\documentclass[prd,twocolumn,floatfix,nofootinbib,superscriptaddress,longbibliography]{revtex4-2}
\usepackage{xpatch}
\xpatchcmd\bibsection{\begingroup}{\vskip 19pt\begingroup}{}{}
\usepackage{slashed}
\usepackage{amsfonts,amssymb,amsmath}
\usepackage{bm}
\usepackage{rotating}
\usepackage{hyperref}
\hypersetup{pdfauthor={me}, colorlinks=true, citecolor=blue, urlcolor=blue, linkcolor=black}
\newcommand{\eps}{\epsilon}
\newcommand{\ord}[1]{{\mathcal{O}}( #1 )}
\begin{document}
\title{Target normal single-spin asymmetry in inclusive electron-nucleon scattering \\
in the $1/N_c$ expansion}
\author{J.~L.~Goity}
\email[ E-mail: ]{goity@jlab.org}
\affiliation{Department of Physics, Hampton University, Hampton, VA 23668, USA}
\affiliation{Theory Center, Jefferson Lab, Newport News, VA 23606, USA}
\author{C.~Weiss}
\email[ E-mail: ]{weiss@jlab.org}
\affiliation{Theory Center, Jefferson Lab, Newport News, VA 23606, USA}
\author{C.~Willemyns}
\email[ E-mail: ]{cintiawillemyns@gmail.com}
\affiliation{Service de Physique Nucl\'eaire et Subnucl\'eaire,
Universit\'e de Mons, UMONS Research Institute for Complex Systems,
Place du Parc 20, 7000 Mons, Belgium}
\begin{abstract}
The target normal single-spin asymmetry in inclusive electron-nucleon scattering is studied in the
low-energy regime that includes the $\Delta$ resonance. The particular interest in the asymmetry
resides in that it is driven by two-photon exchange effects. It probes the spin-dependent absorptive
part of the two-photon exchange amplitude, which is free of infrared and collinear singularities and
represents the most pristine expression of two-photon exchange dynamics. The study presented here
uses the $1/N_c$ expansion of QCD, which combines the $N$ and $\Delta$ through the emergent $SU(4)$
spin-flavor symmetry in the baryon sector and allows for a systematic construction of the transition
EM currents. The analysis includes the first subleading corrections in the $1/N_c$ expansion and
presents results for elastic and inelastic final states. The asymmetry is found to be in the range
$10^{-3}$--$10^{-2}$. The $\Delta$ resonance plays an important role as an intermediate state in the
elastic asymmetry and as a final state in the inclusive asymmetry.
\end{abstract}
\maketitle
%
%
%
%
\section{Introduction}
The electromagnetic interaction is a fundamental tool for the study of hadronic structure and
dynamics. In general, the processes involved have been traditionally analyzed at the leading order
in the EM interaction. In electron-hadron scattering this is $\ord{\alpha_{\rm em}}$, the so called
one-photon exchange approximation (OPE). In hadronic observables there are however important effects
that require the consideration of genuine higher order EM interactions, such as the isospin breaking
in hadronic masses, e.g., the mass difference between the charged and neutral pions that is almost
entirely due to EM, and the important contribution to the proton-neutron mass difference where the
EM contribution is of similar magnitude to the one due to the isospin breaking by the quark
masses. In electron scattering, the subleading EM contributions due to two-photon exchange (TPE)
have been identified as the likely source of the discrepancy observed in the OPE approximation
extraction of the ratio $G_E^p/G_M^p$ from measurements using the Rosenbluth separation versus the
polarization transfer methods \cite{JeffersonLabHallA:1999epl,Guichon:2003qm,Blunden:2003sp}.
Measurements that expose the TPE effects are thus of particular interest. One of them consists in
the comparison of the cross sections of electron and positron scattering on the proton, such as the
recent experiments at DESY \cite{OLYMPUS:2016gso,OLYMPUS:2020dgl} and possible future experiments at
Jefferson Lab \cite{Accardi:2020swt}. In addition, observables in parity-violating electron
scattering receive corrections from TPE \cite{Afanasev:2005ex}. In general the theoretical study of
the TPE effects is affected by significant uncertainties as it requires knowledge of EM hadronic
structure beyond the EM form factors, namely the doubly-virtual-photon Compton amplitudes, and is
thus still a work in progress.

A particularly interesting TPE effect is the transverse target single-spin asymmetry (TSSA) in
electron-nucleon scattering with quasi-two-body final states, i.e. elastic scattering $e + N
\rightarrow e' + N$ or inclusive scattering $e + N\rightarrow e' + X$. If the target nucleon is
polarized transversely to the scattering plane, the cross section generally depends on the scalar
product of the spin vector and the normal vector of the plane.  Due to P and T invariance, such a
spin dependence can arise only from TPE, because it requires a non-zero absorptive part of the
electron-nucleon EM scattering amplitude \cite{Barut:1960zz}.  The spin-dependent cross section
produced in this way is given by on-shell matrix elements of the EM current, is free of collinear
and infrared divergences, and can be considered independently of radiative corrections related to
real photon emission into the final state. These features make the TSSA the most unambiguous TPE
effect in electron scattering. (The same TPE mechanism gives rise to a beam single-spin asymmetry in
the case of transverse electron polarization; this effect is proportional to the electron mass and
generally much smaller than the TSSA; see discussion in Sec.~\ref{sec:discussion}.)

The TSSA in elastic $eN$ scattering has been studied theoretically in
Refs.~\cite{Arafune:1970cx,Guenther:1971gv,Leroy:1972mx,DeRujula:1971nnp}, and more recently in
Refs.~\cite{Pasquini:2004pv,Koshchii:2018bog}, using hadronic physics methods.  This asymmetry is
expected to be of the order $\sim 10^{-2}$ for momentum transfers $Q^2 \lesssim 1$ GeV$^2$.
Experiments performed with recoil polarization in $ep$ elastic scattering obtained values consistent
with zero; see \cite{Powell:1970qt} and references therein. Further tests will become possible with
contemporary elastic scattering experiments.

The TSSA in inclusive $eN$ scattering has been analyzed in deep-inelastic kinematics in
Refs.~\cite{Metz:2006pe,Afanasev:2007ii,Metz:2012ui,Schlegel:2012ve} using a partonic picture and
various assumptions regarding QCD interactions. These calculations predict values in the range
$10^{-4}$--$10^{-3}$, substantially smaller than the elastic TSSA.  Measurements in DIS kinematics
have been performed with a proton target at HERMES at 27.5 GeV beam energy \cite{HERMES:2009hsi},
and with a ${\rm ^3He}$ target at Jefferson Lab Hall A with beam energies 2.4, 3.6, and 5.9 GeV
\cite{Katich:2013atq,Zhang:2015kna,Long:2019iig}. A next-generation measurement with proton target
and electron and positron beams with energies from 2.2 to 6.6 GeV has been proposed at Jefferson Lab
\cite{Grauvogel:2021btg}.

A question of great interest is the behavior of the TSSA in inclusive $eN$ scattering in the first
resonance region, where the $\Delta$ isobar can appear both as an intermediate state in the TPE
amplitude and as a final state in inelastic scattering.  This region lies between the domain of
elastic scattering at low energies and that of deep-inelastic scattering at high energies.  If one
understands the behavior of the TSSA in the resonance region, one could follow its evolution with
energy, connect the elastic and deep-inelastic domains, and explain the different order-of magnitude
predicted for the two regions.  Little is presently known about the inclusive TSSA in the resonance
region from either theory or experiment.  Measurements could be performed in electron scattering
with energies $\sim$ 0.5--1.5 GeV, perhaps at the lower end of the proposed experiment of
Ref.~\cite{Grauvogel:2021btg}, or in future dedicated experiments.

The elastic TSSA in the resonance region can be calculated in terms of the empirical
electroproduction amplitudes extracted from $eN$ scattering data; see Ref.~\cite{Carlson:2007sp} and
references therein. Theoretical uncertainties are significant, as the effect is sensitive to the
phases and arises as a sum over contributions of comparable size and varying sign. The inelastic or
inclusive TSSA in the resonance region is much more difficult to calculate, as it requires also
amplitudes such as $\Delta \rightarrow \Delta$, which cannot be measured in $eN$ scattering.  In
addition to the $\Delta$ it can also receives contributions from nonresonant $\pi N$ final states.
This calls for a theoretical framework that can organize the hadronic intermediate/final states and
predict the EM transition amplitudes.
  
The $1/N_c$ expansion organizes hadron structure and reactions on the basis of the scaling behavior
in the number of colors in QCD \cite{tHooft:1973alw,Witten:1979kh}. It is particularly useful for
baryons and permits a unified description of the $N$ and $\Delta$. In the large-$N_c$ limit the
baryon sector of QCD develops a dynamical spin-flavor symmetry $SU(2N_f)$, with $N_f=2$ the number
of light flavors here \cite{Gervais:1983wq,Gervais:1984rc,Dashen:1993as,Dashen:1993jt,Dashen:1994qi}.
$N$ and $\Delta$ belong to the $SU(4)$ totally symmetric irreducible representation with
$I=S=\frac{1}{2},\cdots,\frac{N_c}{2}$, where $I$ and $S$ are the baryon's isospin and spin. $N
\rightarrow N$, $N \rightarrow \Delta$, and $\Delta \rightarrow \Delta$ transition matrix elements
are thus related by the $SU(4)$ symmetry.  A systematic $1/N_c$ expansion of the EM transition
currents can be performed, including subleading corrections, with all parameters fixed by the
nucleon sector. A parametric distinction between resonant $\Delta$ and nonresonant $\pi N$ states
appears, with the latter relegated to subleading level.  These features of the $1/N_c$ expansion
allow one to develop an efficient framework for the present purpose.

In this work the TSSA in low-energy electron-nucleon scattering with TPE is analyzed using the
$1/N_c$ expansion.  The study covers the energy region below and above the $\Delta$ excitation
threshold and considers the TSSA for both elastic scattering $eN \rightarrow e'N$ and inclusive
scattering $eN \rightarrow e'X$, $X = N, \Delta$.  The application of the $1/N_c$ expansion to the
kinematic variables of electron scattering is discussed, and versions of the expansion appropriate
in the different kinematic regimes are defined. Using the $1/N_c$ expansion of the EM current
operators and their matrix elements between $N$ and $\Delta$ states, the TSSA resulting from TPE is
computed to first subleading order in $1/N_c$. The TSSA is evaluated numerically, and the
contributions of $\Delta$ isobars as intermediate states (in elastic or inclusive scattering) and
final states (in inclusive scattering) are quantified. Possible extensions of the techniques to the
beam spin asymmetry and other observables in low-energy $eN$ scattering are discussed.

In the regime of interest one can identify a low-energy domain below the onset of the $\Delta$
resonance, where only the elastic contribution in the TPE amplitude is present (the low-energy
$\pi N$ continuum contributes only beyond the order in $1/N_c$ considered here); a low-energy domain
above the $\Delta$ resonance, where elastic and inelastic channels are open; and an intermediate
energy domain that extends from the $\Delta$ resonance up to the onset of the higher
resonances. Because the photon virtualities in the TPE amplitude cover a broad range (limited only
by the CM energy of the $eN$ collision), the form factors of the baryon EM currents play an
important role. It is shown that they dramatically affect the contributions of the $\Delta$
resonance to the TSSA.  This underscores the need of a systematic treatment of the transition
currents as provided by the $1/N_c$ expansion.

The article is organized as follows: Section~\ref{sec:methods} summarizes the general methods for
describing the target spin dependence of $eN$ scattering and implementing the $1/N_c$ expansion in
the baryon sector.  Section~\ref{sec:calculation} describes the application of the $1/N_c$ expansion
in the different kinematic regions, construction of the one- and two-photon exchange amplitudes, and
calculation of the TSSA.  Section~\ref{sec:results} presents the numerical results and compares the
contributions of various intermediate/final states.  Section~\ref{sec:discussion} discusses the
significance of the results and possible extensions of the methods.
Appendices~\ref{app:su4}--\ref{app:delta_width} summarize technical material supporting the
calculations, including the $SU(4)$ spin-flavor symmetry, the integrals appearing in the $1/N_c$
expansion of the TSSA, the results for the spin-independent and dependent cross sections, and the
treatment of the $\Delta$ width.
\section{Methods}
\label{sec:methods}
\subsection{Target spin dependence in inclusive electron scattering}
This work considers the process of inclusive scattering of an unpolarized electron on a transversely
polarized nucleon,
\begin{align}
e(k_{\rm i}) + \text{$\mathit{N}$$\uparrow$}(p_{\rm i})
\rightarrow  e(k_{\rm f}) + X,
\label{process}
\end{align}
where $X = N, \pi N, ...$ denotes the hadronic final states accessible at the incident energy. In
the regime to be studied here the inelastic final states are $\pi N$ and dominated by the decay of
the $\Delta$ resonance (as explained below), and the contributions of elastic and inelastic final
states will be analyzed separately in the following.  The 4-momentum transfer is given by
\begin{align}
q \equiv k_{\rm i} - k_{\rm f} = p_{\rm f} - p_{\rm i},
\end{align}
and the process is characterized by the invariants
\begin{align}
s \equiv (k_{\rm i} + p_{\rm i})^2, 
\hspace{1em}
t \equiv q^2,
\hspace{1em}
M_X^2 = (q + p_{\rm i})^2 = p_{\rm f}^2.
\label{KinVars}
\end{align}
The differential cross section can be represented as \cite{Afanasev:2007ii}
\begin{align}
\frac{d\sigma}{d\Gamma_{\rm f}} = \frac{d\sigma_U}{d\Gamma_{\rm f}}
+ e_N^\mu a_\mu\; \frac{d\sigma_N}{d\Gamma_{\rm f}},
\label{dsigma}
\end{align}
where $d\Gamma_{\rm f}$ is the invariant phase space of the final electron. The first term in
Eq.~(\ref{dsigma}) is the unpolarized cross section and the second one results from the effect of
the polarization of the target nucleon.  $a^\mu$ is the spin 4-vector of the target nucleon, and
$e_N^\mu$ is the normalized space-like 4-pseudovector given by
\begin{align}
e_N^\mu &\equiv \frac{N^\mu}{\sqrt{-N^2}},
\hspace{1em}
N^\mu \equiv \epsilon^{\mu\alpha\beta\gamma} p_{{\rm i}\alpha} k_{{\rm i}\beta} k_{{\rm f}\gamma},
\nonumber \\
N^2 &=\frac{t}{4} [s t+(s-m_N^2)(s-M_X^2)].
\label{e_N}
\end{align}

%
%
\begin{figure}[t]
\includegraphics[width=.8\columnwidth]{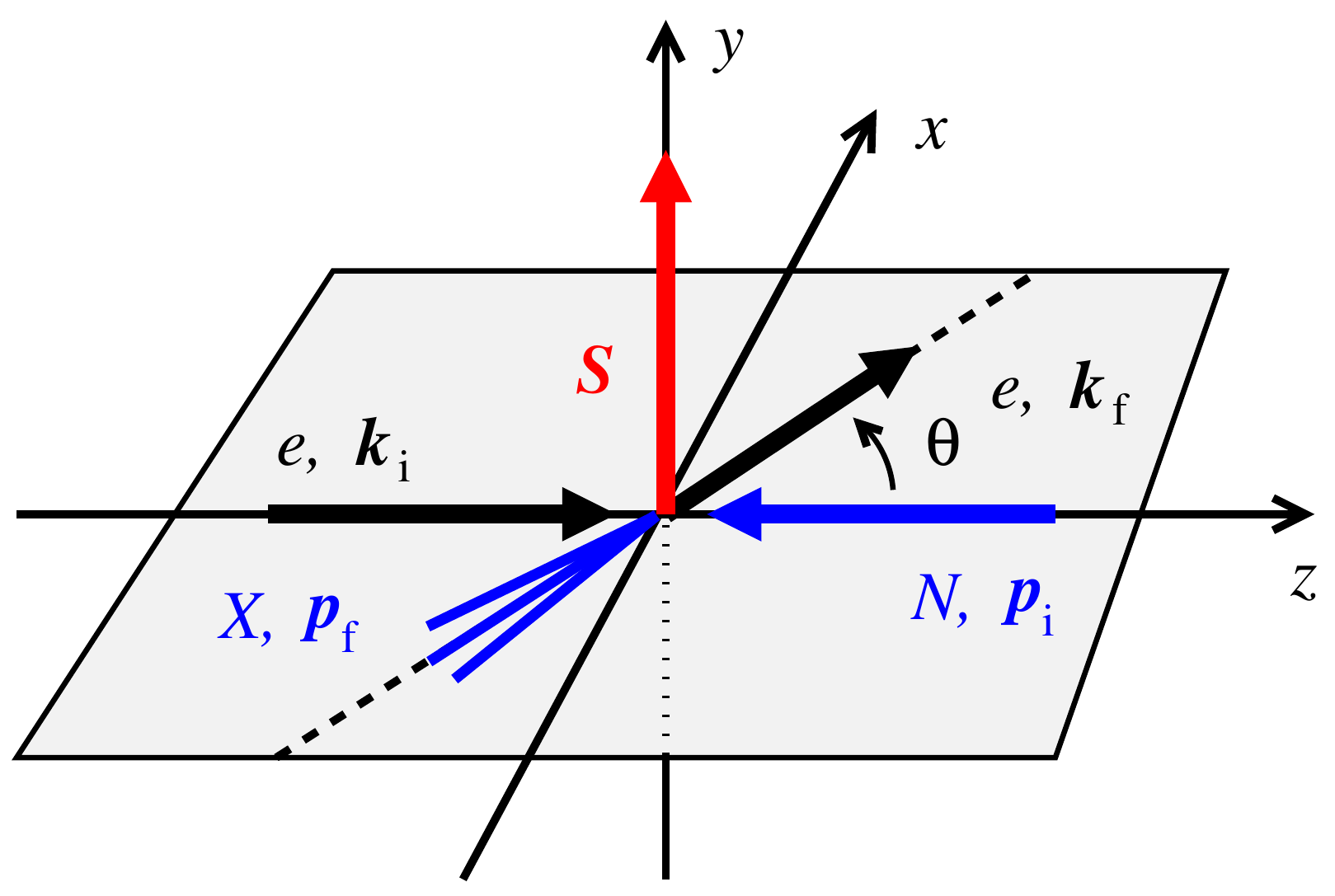}
\caption[]{Inclusive electron-nucleon scattering in the electron-nucleon CM frame.
The nucleon is polarized in the direction normal to the scattering plane.}
\label{fig:cmframe}
\end{figure}
The quasi-two-body scattering process Eq.~(\ref{process}) can be viewed in reference frames where
the 3-momenta (boldface fonts are used for spatial vectors) $\bm{k}_{\rm i}, \bm{k}_{\rm f}$ and
$\bm{p}_{\rm i}$ lie in a plane, e.g. the target nucleon rest frame ($\bm{p}_{\rm i} = 0$), the
electron-nucleon center-of-mass (CM) frame ($\bm{p}_{\rm i} + \bm{k}_{\rm i} = 0$), or the virtual
photon-nucleon CM frame ($\bm{p}_{\rm i} + \bm{k}_{\rm i} - \bm{k}_{\rm f} = 0$).  In such a frame
the vector $\bm{e}_N$ is normal to the scattering plane (see Fig.~\ref{fig:cmframe}) and given by
\begin{align}
e_N = (0, \bm{e}_N),
\hspace{1em}
\bm{e}_N = \frac{\bm{k}_{\rm i} \times \bm{k}_{\rm f}}{|\bm{k}_{\rm i} \times \bm{k}_{\rm f}|},
\end{align}
where $m_N$ is the nucleon mass. The cross section Eq.~(\ref{dsigma}) thus depends on the normal
component of the nucleon spin. The target normal single-spin asymmetry is defined as the ratio
\begin{align}
A_N \equiv \left. \frac{d\sigma_N}{d\Gamma_{\rm f}}
\right/ \frac{d\sigma_U}{d\Gamma_{\rm f}}.
\label{asymmetry_def}
\end{align}
It can be measured either as the asymmetry of the cross sections with the nucleon polarized up and
down for the same scattered electron momentum (up-down asymmetry), or as the asymmetry of the cross
sections with the electron scattered to the left and to the right for the same nucleon polarization
(left-right asymmetry).

The following theoretical analysis uses the electron-nucleon CM frame, where the 3-momenta in the
initial and final states are $\bm{p}_{\rm i} = -\bm{k}_{\rm i}, \bm{p}_{\rm f} = -\bm{k}_{\rm f}$
(see Fig.~\ref{fig:cmframe}).  They are related to the invariants by
\begin{align}
|\bm{k}_{\rm i}| &= \frac{s - m_N^2}{2\sqrt{s}},
\hspace{2em}
|\bm{k}_{\rm f}| = \frac{s - M_X^2}{2\sqrt{s}},
\nonumber \\[1ex]
t &= -2 |\bm{k}_{\rm f}| |\bm{k}_{\rm i}| (1 - \cos \theta ),
\label{invariants_cm}
\end{align}
where $\theta \equiv \textrm{angle} (\bm{k}_{\rm f}, \bm{k}_{\rm i})$ is the scattering angle.  The
expressions in the following do not refer to any specific coordinate system but are formulated in
terms of abstract 3-vector products in this frame.
\subsection{$\mathbf{\it 1/N_c}$ Expansion}
\label{subsec:ncexpansion}
The $1/N_c$ expansion is a powerful method for organizing hadron structure and reactions on the
basis of the scaling behavior in the number of colors in QCD. The expansion needs definition, as it
results from comparing QCD with varying number of degrees of freedom and allows for choices of the
scaling behavior of the parameters of the theory (scale parameter, number of flavors, quark masses).
The commonly adopted version, which works best for the phenomenology of the real world with $N_c=3$
and two or three light flavors, is the 'tHooft expansion, where the number of flavors is fixed and
particular physical observables (e.g., for $N_f = 2$, the $\rho$ and the $\pi$ meson masses) are
used to define the quark masses and the QCD scale. The expansion can furthermore be implemented at
the hadronic level by identifying the $N_c$ scaling of the different quantities. That implementation
can be made into a systematic $1/N_c$ expansion, in particular in the context of effective theories.

The $1/N_c$ expansion is particularly useful in the baryon sector of QCD; see
Ref.~\cite{Jenkins:1998wy} and references therein. The baryon masses are $\ord{N_c}$, and the
$\pi N$ interaction is $\ord{\sqrt{N_c}}$.  The latter requires for consistency that in the
large-$N_c$ limit the baryon sector develops a dynamical contracted spin-flavor symmetry described
by the $SU(2 N_f)$ group, or $SU(4)$ for $N_f = 2$ \cite{Gervais:1983wq,Gervais:1984rc,Dashen:1993as}.
In the rest frame of the baryons, the fifteen generators of $SU(4)$ can be identified with the spin
$\hat{S}^i$, isospin $\hat{I}^a$ and spin-flavor $\hat{G}^{ia}$ operators (see
Appendix~\ref{app:su4}).  In frames where the baryons have momenta $\ord{N_c^0}$, their velocities
are $\ord{N_c^{-1}}$, because the masses are $\ord{N_c}$, and their motion is effectively
nonrelativistic.  Transition matrix elements between baryon states in frames where the momenta are
$\ord{N_c^0}$ can therefore be computed in a non-relativistic expansion, where they are expressed in
terms of the $SU(4)$ generators and the initial/final baryon momenta.

The ground-state baryons belong to the totally symmetric $SU(4)$ multiplet. It consists of states
with isospin/spin $I = S$ and $S = \frac{1}{2},\cdots, \frac{N_c}{2}$, which includes the $N$ and
$\Delta$ states with $I = S = \frac{1}{2}$ and $\frac{3}{2}$. States in the multiple are
characterized by $S$ and the projections $S_3$ and $I_3$ and denoted by $|S S_3 I_3\rangle$.  The
mass splitting between the states is $\ord{N_c^{-1}}$. In this multiplet the generators
$\hat{G}^{ia}$ have matrix elements $\ord{N_c}$ between states with $S=\ord{N_c^0}$, while the
generators $\hat{I}^a$ and $\hat{S}^i$ obviously have matrix elements $\ord{N_c^0}$.
 
This work requires the matrix elements of the EM current operators between baryon states in the
ground state multiplet. The assignment of electric charges to the quarks at arbitrary $N_c$
\cite{Fernando:2019upo} can be made in such a way that the Standard Model gauge and gravitational
anomalies vanish as required for consistency, and such that the charges of the baryons are simply
given by the usual relation $Q=1/2+\hat{I}^3$, independent of $N_c$. The quark charges are then
given by $Q_q=\frac{3}{2N_c}+I_3$. In the following the current is considered for transitions between
baryon states with 3-momenta $\bm{p}, \bm{p}'= \ord{N_c^0}$, and generally different spins $S' \neq
S$, and therefore different masses; the 4-momentum transfer is $q \equiv p' - p$, and its components
are $\bm{q} = \ord{N_c^0}$ and $q^0 = \ord{N_c^{-1}}$.  Including leading and subleading terms in
the $1/N_c$ expansion, the isoscalar ($S$) and isovector ($V$) components of the EM current are
given by \cite{Fernando:2019upo}\footnote{Terms in the currents with higher powers of momenta have
been neglected, such as the isovector contribution to the time component, which stems from a
relativistic correction and is proportional to $\frac{1}{m_N \Lambda} \eps^{0 i j k} q^i p^j
\hat{G}^{ka} $.  Such terms are suppressed except at the upper end of the energy domain considered
here and are subleading in $1/N_c$. The electric quadrupole component of the current, which
mediates $N-\Delta$ transitions, is suppressed by a factor $1/N_c^2$ with respect to the leading
term \cite{Jenkins:2002rj} and thus irrelevant to the present calculation.}
\begin{align}
J_S^{\mu}(q) &= G_E^S(q^2) \frac{1}{2} g^{\mu 0}
- i \frac{1}{2} \frac{G_M^S(q^2)}{\Lambda} \epsilon^{0\mu i j} q^i \hat{S}^j,
\label{current_isoscalar}
\\
J_V^{\mu a}(q) &= G_E^V(q^2) \hat{I}^a g^{\mu 0}
- i\frac{6}{5} \frac{G_M^V(q^2)}{\Lambda} \epsilon^{0\mu i j} q^i \hat{G}^{ja},
\label{current_isovector}
\\[1ex]
J_{\rm EM}^\mu(q)&=J_S^{\mu}(q)+J_V^{\mu 3}(q),
\end{align}
where $G_{E,M}^S$ and $G_{E,M}^V$ are the form factors of the electric and magnetic
components.\footnote{For the sake of convenience in the calculations and without significant
difference the $G_E$ form factor is taken to be equal to the corresponding $F_1$ rather than the
Sachs form factor.}  The currents are expressed in terms of the $SU(4)$ spin-flavor generators and
understood to be evaluated between multiplet states $\langle S'S_3'I_3'| ... |S S_3 I_3\rangle$.
The magnetic terms are written with a generic mass scale $\Lambda = \ord{N_c^0}$, whose value is
identified with the physical nucleon mass (exempt from $N_c$ scaling); this formulation is natural
for the $1/N_c$ expansion and avoids the appearance of spurious powers of $N_c$ that would come from
using the scaling $m_N$ in the denominator. The form factors in Eqs.~(\ref{current_isoscalar}) and
(\ref{current_isovector}) are defined such that they coincide with the physical nucleon form factors
for $\Lambda = m_N(\textrm{physical})$ and $N_c=3$. In particular, the factor $6/5$ in the magnetic
term of the isovector current was introduced such that, for $N_c=3$, $G_M^V$ coincides with the
physical nucleon isovector magnetic form factor.

The currents given by Eqs.~(\ref{current_isoscalar}) and (\ref{current_isovector}) satisfy current
conservation to the necessary accuracy in $1/N_c$. For the magnetic terms (spatial components), this
follows from the vector product structure of the vertices; for the electric terms (time components),
it is realized because $q^0 = \ord{N_c^{-1}}$.

The order in $1/N_c$ of the components of the currents in Eqs.~(\ref{current_isoscalar}) and
(\ref{current_isovector}) is as follows. The isovector magnetic current is $\ord{N_c}$, being
represented by the spin-flavor operator $\hat{G}^{ia}$ that has matrix elements $\ord{N_c}$. This
reflects the fact that the nucleon anomalous magnetic moment is $\ord{N_c}$. (In the quark picture
of baryons, this happens because the magnetic moments of the quarks add up coherently to form the
total magnetic moment of the baryon, see for instance Ref.~\cite{Dai:1995zg}.) The remaining terms
in the current are $\ord{N_c^0}$, being proportional to the operators $\hat 1, \hat S^i$, and $I^a$
that have matrix elements $\ord{N_c^0}$.  At leading order in the $1/N_c$ expansion, the dominant
current component is therefore the isovector magnetic current proportional to the operator
$\hat{G}^{ia}$. Clear evidence of this dominance is the ratio of the isovector and isoscalar
magnetic moments of the nucleon, $G_M^V(0) / G_M^S(0) = \ord{N_c} = 5.34$. The dominant isovector
magnetic current also induces the $M_1$ transitions $N \rightarrow \Delta$; the other current
components only have matrix elements between states with same spin/isospin.

Equations~(\ref{current_isoscalar}) and (\ref{current_isovector}) capture the $1/N_c$ expansion of
the EM currents to the accuracy needed in the present calculation. Higher-order corrections beyond
that accuracy arise from the nonrelativistic expansion of the motion of the baryons.  For momenta
$\ord{N_c^0}$, both the spatial components of the convection current and the time component of the
magnetic currents are $\ord{N_c^{-1}}$. Further higher-order corrections arise from the contribution
of subleading spin-flavor operators, namely $\hat{S}^i \hat{\vec S}^2$ for the isoscalar magnetic
current, and $\{\hat{G}^{i3},\hat S^2\}$ and $\hat{S}^i \hat{I}^3$ for the isovector one.  These
higher-body spin-flavor operators are accompanied by factors $1/N_c^{n-1}$, where $n$ is the number
of spin-flavor generator factors in the composite operator \cite{Dashen:1993jt,Goity:2004pw}.  The
corrections to the magnetic currents are therefore suppressed by $\ord{N_c^{-2}}$ relative to the
dominant isovector magnetic current.  To the accuracy of the present calculation, these higher order
terms in the currents are therefore irrelevant.

%
%
\begin{table*}[t]
\begin{tabular}{l|l|ll|l|l|}
  & Energy regime & \multicolumn{2}{l|}{$1/N_c$ expansion regime} & Channels open &
  Final states possible
\\[1ex]
\hline
I & $m_N < \sqrt{s} < m_\Delta$ & $\sqrt{s} - m_N \sim N_c^{-1}$, & $k \sim N_c^{-1}$ & $N$ & elastic 
\\[1ex]
II & $m_\Delta < \sqrt{s} \ll m_{N\ast}$ & $\sqrt{s} - m_N \sim N_c^{-1}$, & $k \sim N_c^{-1}$ &
   $N, \Delta$ & elastic or inelastic 
\\[1ex]
III & $m_\Delta < \sqrt{s} \lesssim m_{N\ast}$ & $\sqrt{s} - m_N \sim N_c^{0}$, & $k \sim N_c^{0}$ &
$N, \Delta, N^\ast (\text{suppr})$ & elastic or inelastic 
\\[1ex]
\hline
\end{tabular}
\caption{Kinematic regimes in application of the $1/N_c$ expansion to low-energy electron
scattering.}
\label{table:kinematics}
\end{table*}
The momentum dependence of the form factors plays an essential role in the present calculation.  The
scale governing the momentum dependence of the form factors --- the baryon ``size'' in the
large-$N_c$ limit --- is $\ord{N_c^0}$, and the momentum transfer is $t = \ord{N_c^0}$, so that the
functions are evaluated in a region where they differ significantly from their varlues at zero
momentum transfer. The form factors in Eqs.~(\ref{current_isoscalar}) and (\ref{current_isovector})
can be determined by matching the expressions for $N_c = 3$ with the empirical proton and neutron
form factors, which gives
\begin{align}
G_E^{S, V}(t) &= G_E^p(t) \pm G_E^n(t),
\nonumber \\
G_M^{S, V}(t) &= G_M^p(t) \pm G_M^n(t).
\label{formfactor_physical}
\end{align}
In the subsequent calculations, the small contribution of the neutron'e electric form factor is
neglected for simplicity, $G_E^n \equiv 0$, such that $G_E^S = G_E^V = G_E^p$. Furthermore, it is
assumed that the $t$-dependence of all the form factors is of dipole form with a common mass scale
$\Lambda_{\rm EM}^2=0.71\, {\rm GeV}^2$.

The construction of the currents Eqs.~(\ref{current_isoscalar}) and (\ref{current_isovector})
demonstrates the predictive power of the $1/N_c$ expansion. The structure is dictated by the
spin-flavor symmetry in the large-$N_c$ limit. The coefficients are fixed by observables measured in
$N \rightarrow N$ transitions.  Together, this then predicts the matrix elements of the same
operator for $N \rightarrow \Delta$ and $\Delta \rightarrow \Delta$ transitions.
\section{Calculation}
\label{sec:calculation}
\subsection{Kinematic regimes for the $1/N_c$ expansion}
\label{subsec:kinematic}
In this work the $1/N_c$ expansion is used to study the spin dependence of inclusive $eN$ scattering
Eq.~(\ref{process}).  When applying the $1/N_c$ expansion to the scattering process, it is necessary
to specify the parametric order in $1/N_c$ of the kinematic variables -- the scattering energy,
momentum transfer, and final-state mass, Eq.~(\ref{KinVars}).  The physical scales for the
scattering energy and final-state mass are set by the excitation energy of the $\Delta$ and $N^\ast$
baryon resonances, which are of the parametric order
\begin{align}
m_\Delta - m_N & = \ord{N_c^{-1}},
\\ 
m_{N\ast} - m_N & = \ord{N_c^{0}}.
\label{excitation_energy_scales}
\end{align}
Another physical scale arises from the excitation energy of non-resonant $\pi N$ states, namely
$m_{\pi N} - m_N$; this scale permits various choices for the assignment of its $1/N_c$ scaling (see
below). How the scattering energy is chosen relative to the scales
Eq.~(\ref{excitation_energy_scales}) determines what channels are open in the process, and how the
$1/N_c$ expansion is to be applied to the transition currents. Different choices are possible,
leading to different versions of the $1/N_c$ expansion.

The present study considers three kinematic regimes (see Table~\ref{table:kinematics} for a summary):

I) {\it Low-energy elastic regime:} This is the regime of scattering energies below the physical
$\Delta$ threshold, $m_N < \sqrt{s} < m_\Delta$. The $1/N_c$ scaling of the scattering energy and CM
momentum in this regime are $\sqrt{s} - m_N = \ord{N_c^{-1}}$ and $k = \ord{N_c^{-1}}$. This regime
therefore has vanishing extent $\ord{N_c^{-1}}$ in the large-$N_c$ limit. In this regime the only
open channel in the intermediate and final states is the nucleon. Both in this regime (and the
following inelastic regime II) the electric term in the current and the isovector magnetic one
become of the same order. As seen later, in those regimes, the effect of terms in the asymmetry
involving the electric charge become very important for the proton.

II) {\it Low-energy inelastic regime:} This is the regime of scattering energies above the physical
$\Delta$ threshold but significantly below the $N^\ast$ threshold,
$m_\Delta < \sqrt{s} \ll m_{N\ast}$. The $1/N_c$ scaling of the scattering energy and CM momentum in
this regime are $\sqrt{s} - m_N = \ord{N_c^{-1}}$ and $k = \ord{N_c^{-1}}$ (same as I), but the
$\Delta$ channel is now open. This regime can be treated within the low-energy expansion, in which
the momenta are counted as $\ord{N_c^{-1}}$ \cite{CalleCordon:2012xz,Fernando:2019upo}.  Because the
momentum transfer at the vertices is parametrically small, $t = \ord{N_c^{-1}}$, the $t$-dependence
of the form factors is formally suppressed.  In reality one observes significant numerical effects
from the momentum dependence of the form factors already in this regime (see Sec.~\ref{sec:results})

III) {\it Intermediate-energy inelastic regime:} This is the regime where the scattering energy is
above the $\Delta$ threshold and can reach values up to and including the first resonance region,
$m_\Delta < \sqrt{s} \lesssim m_{N\ast}$. The $1/N_c$ scaling of the scattering energy and CM
momentum are now $\sqrt{s} - m_N = \ord{N_c^0}$ and $k = \ord{N_c^0}$, parametrically larger than in
I and II.  Both $\Delta$ and $N^\ast$ states are now accessible as intermediate states (the
amplitude for $N \rightarrow N^\ast$ transitions are suppressed compared to
$N \rightarrow N, \Delta$ transitions by $N_c^{-1/2}$ \cite{Scoccola:2007sn,Goity:2007ft}). This
regime corresponds to the conventional $1/N_c$ expansion of baryon form factors at momentum
transfers $t = \ord{N_c^0}$ and was considered in Ref.~\cite{Goity:2022yro}.  The $t$-dependence of
the baryon form factors plays an essential role in this regime.

Besides the baryon resonances, also non-resonant $\pi N$ states can contribute to the TSSA in
inclusive $eN$ scattering as intermediate and final states.  The importance of these contributions
can be rigorously assessed in the three regimes I--III.  In the low-energy regimes I and II, one can
perform a combined chiral and $1/N_c$ expansion using the $\xi$ power counting
scheme \cite{CalleCordon:2012xz,Fernando:2019upo}, where $k$ and $1/N_c$ are counted as $\ord{\xi}$.
The pion-baryon coupling is given by $\frac{6 g_A}{5 F_\pi} k_\pi^i \hat{G}^{ia}$, where
$g_A = \ord{N_c}$ is the nucleon isovector axial coupling, $k_\pi$ is the pion momentum, and $a$ is
the pion isospin.  The three body phase space brings in a generic suppression factor
$k^2/(32 \pi^2)$.  With these ingredients, and using the spin-flavor algebra, one finds that in the
low-energy regimes I and II the contribution of non-resonant $\pi N$ states to the $eN$ cross
section is suppressed by at least $\ord{\xi^2}$ with respect to the leading order of the present
calculation, and thus it is consistent to neglect it.  In the intermediate-energy regime III, where
the pion momenta are $\ord{N_c^0}$ and not small, the suppression is no longer as effective, and
non-resonant $\pi N$ states can contribute at subleading order of the calculation performed in this
work. If one limits oneself to the final states $N$ and $\Delta$ as in this work, then the
calculation only misses the $\pi N$ continuum in the box diagram, and those are only affecting
subleading contributions in regime III.

The numerical boundaries of these regimes in the $eN$ CM momentum $k$, Eq.~(\ref{invariants_cm}),
are as follows: The $\Delta$ threshold $\sqrt{s} = m_\Delta =$ 1.23 GeV is at $k =$ 0.26 GeV; the
generic $N^\ast$ threshold $\sqrt{s} = m_{N*} \approx $ 1.5 GeV is at $k \approx$ 0.46 GeV.  The
expansion scheme of regime II should be applicable for $0.26 < k \lesssim 0.35$ GeV; that of regime
III for $0.3 \lesssim k \lesssim 0.6$ GeV \cite{Goity:2022yro}.  The quality of the approximation at
upper end of the CM momentum ranges depends on the size of $N^\ast$ contributions, which cannot be
estimated with the present method.
\subsection{Amplitude and cross section}
The scattering amplitude for the process $eN \rightarrow e'B \; (B = N, \Delta)$ in the CM frame of
the $eN$ collision (see Fig.~\ref{fig:cmframe}) is denoted by
\begin{align}
M(\bm{k}_{\rm f}, \bm{k}_{\rm i} | \lambda; S_{\rm f} S_{{\rm f}3}I_{{\rm f}3};
S_{\rm i} S_{{\rm i}3}I_{{\rm i}3}).
\label{amplitude}
\end{align}
Here $\lambda$ is the electron helicity -- the spin projection on $\bm{k}_{\rm i}$ in the initial
state and $\bm{k}_{\rm f}$ in the final state, which is conserved in the scattering process (the
electron mass is neglected).  $S_{\rm i} S_{{\rm i}3}I_{{\rm i}3}$ are the spin-isospin quantum
numbers of the initial nucleon state, where $S_i = \frac{1}{2}$ and $I_{i3} = \pm \frac{1}{2}$ for
proton/neutron.  $S_{\rm f} S_{{\rm f}3}I_{{\rm f}3}$ are the quantum numbers of the final baryon
state, with $S_{\rm f} = \frac{1}{2}$ or $\frac{3}{2}$ for $N$ or $\Delta$, and
$I_{{\rm f}3} = I_{{\rm i}3}$.  The spins of the initial and final baryons are quantized along a
common axis, which can be chosen e.g.\ as the direction of the initial momenta in the CM frame. The
differential cross section for the scattering of unpolarized electrons on polarized nucleons,
Eq.~(\ref{dsigma}), is obtained as\footnote{The amplitude Eq.~(\ref{amplitude}) and the cross
section Eq.~(\ref{cross_section_from_amplitude}) use the relativistic normalization convention for
the electron and baryon momentum states,
$\langle p'|p\rangle = 2 p^0 (2\pi)^3 \delta^{(3)}(\bm{p}' - \bm{p})$.
Reference~\cite{Goity:2022yro} used the nonrelativistic normalization
$\langle p'|p\rangle = (2\pi)^3 \delta^{(3)}(\bm{p}' - \bm{p})$ for the baryon states. The
relativistic convention used here is more transparent for keeping track of kinematic effects caused
by the $N$--$\Delta$ mass difference, which appear in higher orders of the $1/N_c$ expansion.}
\begin{align}
\frac{d\sigma}{d\Omega} &= \frac{{|\bm{k}_{\rm f}|}}{64 \pi^2  |\bm{k}_{\rm i}|  s} \;
\sum_{S_{\rm f} S_{{\rm f}3}} \;
\sum_{\bar S_{{\rm i}3} S_{{\rm i}3}} \rho(S_{{\rm i}3} \bar S_{{\rm i}3}) \;
\frac{1}{2} \sum_{\lambda}
\nonumber
\\ & 
\times 
M^\ast (\bm{k}_{\rm f}, \bm{k}_{\rm i}|\lambda; S_{\rm f} S_{{\rm f}3};
S_{\rm i} \bar S_{{\rm i}3})
\; 
M (\bm{k}_{\rm f}, \bm{k}_{\rm i}|\lambda; S_{\rm f} S_{{\rm f}3};
S_{\rm i} S_{{\rm i}3}).
\label{cross_section_from_amplitude}
\end{align}
The initial nucleon spin projection is averaged over with the spin density matrix
\begin{align}
\rho(S_{{\rm i}3} \bar S_{{\rm i}3}) &= \frac{1}{2}
\left[  \delta (S_{{\rm i}3} \bar S_{{\rm i}3})
+ \bm{a} \cdot \bm{\sigma} (S_{{\rm i}3} \bar S_{{\rm i}3}) \right] ,
\label{density_matrix}
\end{align}
where $\bm{a}$ is nucleon spin 3-vector in Eq.~(\ref{dsigma}) in the CM frame and $\bm{\sigma}$ are
the Pauli matrices. The unpolarized cross section is given by the diagonal sum over initial nucleon
spins.  The polarized cross section for polarization normal to the scattering plane $\bm{a} =
\bm{e}^y$ is given by the non-diagonal sum over nucleon spins with the matrix $\sigma^y$.

Equation~(\ref{cross_section_from_amplitude}) includes the summation over the final baryon spin
$S_{\rm f} = \frac{1}{2}, \frac{3}{2}$ ($N, \Delta$) and represents the cross section for inclusive
scattering. In the following also the individual contributions of $N$ and $\Delta$ final states will
be computed and quoted.

%
%
\begin{figure}[t]
\includegraphics[width=.85\columnwidth]{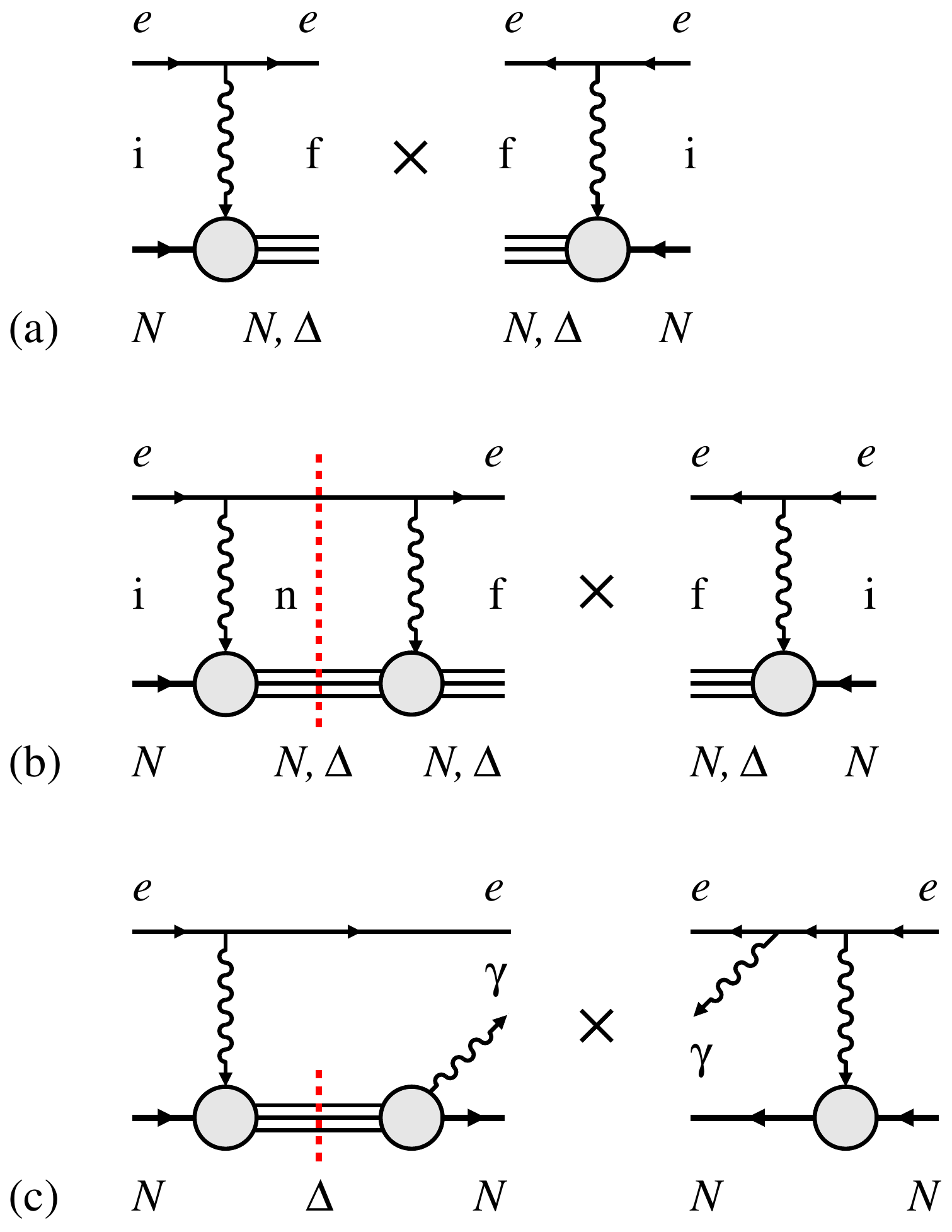}
\caption[]{Inclusive electron-nucleon scattering cross section with $N$ and $\Delta$ final states in
the $1/N_c$ expansion in the kinematic regimes described in Sec.~\ref{subsec:kinematic}.
(a)~Spin-independent cross section from square of $e^2$ amplitudes. The circle
denotes the electromagnetic current matrix element between baryon states. (b)~Spin-dependent cross
section from interference of $e^4$ and $e^2$ amplitudes. (c)~Interference of real photon emission
from electron and baryon (not included in this work).}
\label{fig:diagrams}
\end{figure}
\subsection{Spin-independent cross section}
The $eN$ scattering amplitude is computed as an expansion in the EM coupling,
\begin{align}
M ({\bm k}_{\rm f}, {\bm k}_{\rm i} | \lambda; S_{\rm f} S_{{\rm f}3}; S_{\rm i} S_{{\rm i}3})
\; \equiv \; M_{\rm fi} = M^{(e2)}_{\rm fi} + M^{(e4)}_{\rm fi} + \cdots
\end{align}
The $e^2$ term is the standard OPE amplitude. It is given by the contraction of the electron and
baryon current with the photon propagator,
\begin{align}
M^{(e2)}_{\rm fi} &= -\frac{e^2}{q^2} \bar u (k_{\rm f}) \gamma_\mu u(k_{\rm i}) \;
\langle B_{\rm f} | J_{\rm EM}^\mu (q) | B_{\rm i}\rangle ,
\label{amplitude_ope}
\end{align}
where $B \equiv \{ S S_3 I_3\}$ collectively denotes the baryon spin-flavor quantum numbers. The
squared modulus of this amplitude gives the OPE cross section (see Fig.~\ref{fig:diagrams}a), which
is independent of the target spin.  The explicit form of the OPE cross section generated by the
$1/N_c$-expanded baryon currents, Eqs.~(\ref{current_isoscalar}) and (\ref{current_isovector}), can
be obtained from Eqs.~(\ref{cross_section_from_amplitude}) and (\ref{amplitude_ope}) using standard
techniques.  The result is summarized in Appendix~\ref{app:cross_section}, Eq.~(\ref{OPEXsection}),
for the case of elastic scattering (final $N$, $S_{\rm f} = \frac{1}{2}$).

For the purpose of the present study the spin-independent cross section is needed only as the
denominator of the TSSA Eq.~(\ref{asymmetry_def}) and can be taken at the lowest order in EM
coupling, i.e.\ as the OPE cross section. For simplicity in the following the asymmetry will be
computed with the elastic rather than the inclusive OPE cross section in the denominator; this
choice facilitates the discussion of the behavior near the $\Delta$ threshold.

\subsection{Spin-dependent cross section from two-photon exchange}
The spin-dependent cross section for unpolarized electron scattering is zero at the OPE order because
the amplitude is real (Christ-Lee theorem) \cite{Christ:1966zz}. The first non-zero contribution
appears through the interference between the OPE amplitude and the imaginary (absorptive) part of
the TPE amplitude, which is given by the order $e^4$ box diagram (see Fig.\ref{fig:diagrams}b).
The imaginary part arises from the TPE process with physical (on-shell) intermediate states,
and can be computed by taking the product of the OPE amplitudes of order $e^2$ and integrating
over the phase space of the intermediate states. 

Explicitly, the $e^4$ amplitude resulting from the box diagram is given by
\begin{align}
&M_{\rm fi}^{(e4)}(S_{\rm n})
\nonumber \\
&= -i  e^4 \int\frac{d^4 k_{\rm n}}{(2\pi)^4} \;
\frac{1}{( (k_{\rm i} - k_{\rm n})^2 +i\eps)((k_{\rm n} - k_{\rm f})^2 + i\eps)}\;
\nonumber\\
&\times\bar u(k_{\rm f}) \gamma_\nu \frac{1}{\slashed{k}_{\rm n} - m_e + i\eps}\gamma_\mu
u(k_{\rm i}) \;\;
\nonumber\\
&\times\sum_{S_{3{\rm n}} I_{3{\rm n}}}\langle B_{\rm f}| J_{\rm EM}^\nu(k_{\rm n}-k_{\rm f})
| B_{\rm n}\rangle
\nonumber \\
& \times \frac{2 m_{B_{\rm n}}}{(p_{\rm i} + k_{\rm i} - k_{\rm n})^2-  m_{B_{\rm n}}^2+i\eps}
\langle B_{\rm n} | J_{\rm EM}^\mu(k_{\rm i} - k_{\rm n}) | B_{\rm i}\rangle ,
\label{amplitude_e4}
\end{align}
where $k_n$ is the 4-momentum of the electron in the intermediate state. The amplitude is presented
for a given spin of the baryon $B_n$ in the intermediate state, $S_{\rm n}$; the spin/isospin
projections $S_{3{\rm n}}$ and $I_{3{\rm n}}$ are summed over. In the present case, where the
initial baryon is a nucleon, $S_{\rm i}= \frac{1}{2}$, the intermediate baryons can only be $N$ or
$\Delta$, $S_{\rm n}= \frac{1}{2}$ or $\frac{3}{2}$.  The absorptive part of the amplitude is
obtained by applying the Cutkosky rules.  The interference of the $e^2$ and $e^4$ amplitudes needed
for the spin-dependent cross section then becomes
\begin{align}
& M^{(e2)^*}_{\rm  fi} M^{(e4)}_{\rm fi}(S_{\rm n}) |_{\rm abs} + \textrm{c.c.}
\nonumber \\
&= \frac{e^6 \, m_{B_{\rm n}}}{32 \pi^2\, t \sqrt{s} \;
|\bm{k}_{\rm i}| |\bm{k}_{\rm f}|  |\bm{k}_{\rm n}|}
\nonumber\\
&\times {\rm Im}\left(\int d\Omega_{  \bm {k}_{\rm n}} \right.
\left.\frac{ L_{\mu\nu\rho}(k_{\rm i},k_{\rm f},k_{\rm n})
H_{{\rm fi,n}}^{\mu\nu\rho}(k_{\rm i},k_{\rm f},k_{\rm n})}
{(1-\bm{\hat k}_{\rm i}\cdot\bm{\hat k}_{\rm n})
(1-\bm{\hat k}_{\rm f}\cdot {\bm{\hat k}_{\rm n}})} \right) .
\label{interfterm}
\end{align}
Here the momenta are in the CM frame. $\bm{\hat k}_{\rm i}, \bm{\hat k}_{\rm f}$ and
$\bm{\hat k}_{\rm n}$ are the unit vectors along the initial, final and intermediate electron
momenta; the moduli $|{\bm k}_{\rm i}|$ and $|{\bm k}_{\rm i}|$ are given by
Eq.~(\ref{invariants_cm}), amd $|{\bm k}_{\rm n}|$ is given by the same expression with the
intermediate baryon mass $m_{B_{\rm n}}$. The spin-dependent cross section resulting from TPE is
then given by
\begin{align}
& e_N^\mu a_\mu \frac{{d\sigma_N}}{d\Omega}(I_{3{\rm i}}, S_{\rm f}, S_{\rm n})
\nonumber \\
&=\frac{\alpha^3}{16\pi} \frac{|\bm{k}_{\rm f}|}{|\bm{k}_{\rm i}|}
\frac{m_N m_{B_{\rm f}}m_{B_{\rm n}}}{t s^{3/2} |\bm{k}_{\rm i}| |\bm{k}_{\rm f}| |\bm{k}_{\rm n}| }
\nonumber \\
&\times {\rm Im}\left(\int d\Omega_{ \bm{\hat k}_{\rm n}} \right. \left.
\frac{ L_{\mu\nu\rho}(k_{\rm i},k_{\rm f},k_{\rm n})
H_{{\rm fi,n}}^{\mu\nu\rho}(k_{\rm i},k_{\rm f},k_{\rm n})}
{(1- \bm{\hat k}_{\rm i}\cdot \bm{\hat k}_{\rm n})
(1-\bm{\hat k}_{\rm f}\cdot \bm{\hat k}_{\rm n})} \right) .
\label{interfXsection}
\end{align}
The leptonic and hadronic tensors in the above expressions are given by
\begin{align}
& L^{\mu\nu\rho}(k_{\rm i}, k_{\rm f}, k_{\rm n})
\nonumber \\
&= {\rm Tr}(\slashed{k}_{\rm i} \gamma^\mu \slashed{k}_{\rm f} \gamma^\nu
\slashed{k}_{\rm n} \gamma^\rho)
\nonumber\\[1ex]
&= 4 \left( k_{\rm i}^\mu k_{\rm n}^\rho k_{\rm f}^\nu
+ k_{\rm i}^\mu k_{\rm n}^\nu k_{\rm f}^\rho \right.
\nonumber \\[1ex]
&+ k_{\rm i}^\rho
\left( k_{\rm n}^\nu k_{\rm f}^\mu + k_{\rm n}^\mu k_{\rm f}^\nu
- k_{\rm n}\!\cdot\! k_{\rm f}\; g^{\mu\nu} \right)
\nonumber \\[1ex]
&+ k_{\rm i}^\nu \left( k_{\rm n}^\rho k_{\rm f}^\mu - k_{\rm n}^\mu k_{\rm f}^\rho
+ k_{\rm n}\!\cdot\! k_{\rm f} \; g^{\mu\rho} \right)
\nonumber \\[1ex]
&- \left( k_{\rm i}^\mu \, k_{\rm n}\!\cdot\! k_{\rm f}
- k_{\rm i}\!\cdot\! k_{\rm f} \; k_{\rm n}^\mu +
k_{\rm i}\!\cdot\! k_{\rm n}\; k_{\rm f}^\mu \right) g^{\nu\rho}
\nonumber \\[1ex]
&-  k_{\rm i}\!\cdot\! k_{\rm f} \; k_{\rm n}^\rho \, g^{\mu\nu}
- k_{\rm i}\!\cdot\! k_{\rm f}\; k_{\rm n}^\nu \, g^{\mu\rho}
\nonumber \\[1ex]
& \left. -k_{\rm i} \!\cdot\! k_{\rm n}\; k_{\rm f}^\nu \, g^{\mu\rho}
+ k_{\rm i}\!\cdot\! k_{\rm n} \; k_{\rm f}^\rho \, g^{\mu\nu} \right) ,
\label{Leptonic_tensor}
\\[2ex]
& H_{{\rm fi},{\rm n}}^{\mu\nu\rho} (k_{\rm i}, k_{\rm f}, k_{\rm n})
\nonumber \\
&= \sum_{\bar S_{{\rm i}3} S_{{\rm i}3}}
\frac{1}{2} \bm{a} \cdot \bm{\sigma} (S_{{\rm i}3} \bar S_{{\rm i}3})
\sum_{S_{{\rm f}3} I_{{\rm f}3}}
\sum_{S_{{\rm n}3} I_{{\rm n}3}}
\nonumber \\[1ex]
&\times \langle B_{\rm i} | (J_{\rm EM}^\mu (k_{\rm i} - k_{\rm f}))^\dagger | B_{\rm f} \rangle
\nonumber \\[2ex]
&\times\langle B_{\rm f} | J_{\rm EM}^\nu (k_{\rm n} - k_{\rm f}) | B_{\rm n}\rangle
\;
\langle B_{\rm n} | J_{\rm EM}^\rho (k_{\rm i} - k_{\rm n}) | B_{\rm i}\rangle .
\label{Hadronic_tensor}
\end{align}
Equation~(\ref{interfXsection}) presents the cross section depending on the isospin projection of
the initial nucleon $I_{{\rm i}3} = \pm\frac{1}{2}$, the spin of the final baryon
$S_{\rm n} = \frac{1}{2}, \frac{3}{2}$; and the spin of the intermediate baryon in the box diagram
$S_{\rm n} = \frac{1}{2}, \frac{3}{2}$; the contributions of the different final and intermediate
states will be discussed below.

The $1/N_c$ expansion is now implemented for the hadronic tensor. The method makes use of the
t-channel spin and isospin of the tensor, $J$ and $I$, which can be viewed as the quantum numbers of
an operator connecting the nucleon states. In the spin-dependent cross section, only the $J=1$
component of the tensor is needed, and because it is a forward matrix element between the initial
nucleon state and its conjugate, only the total $I =$ 0 or 1 components of the tensor can
contribute.  Thus, in the end those components of the hadronic tensor will reduce to the operators
$\hat{S}^i$ and $\hat{S}^i \hat{I}^3$. The spin-flavor reduction of the hadronic tensor can be
carried out for general $N_c$ making use of the $SU(4)$ algebra. The sketch of the calculation is as
follows: starting with the general structure of the product of currents
\begin{align}
&(J_{\rm EM}^\mu (k_{\rm i} - k_{\rm f}))^\dagger | B_{\rm f} \rangle
\nonumber \\[1ex]
& \times 
\langle B_{\rm f} | J_{\rm EM}^\nu (k_{\rm n} - k_{\rm f}) | B_{\rm n}\rangle
\;
\langle B_{\rm n} | J_{\rm EM}^\rho (k_{\rm i} - k_{\rm n}) ,
\end{align}
the spatial and the time components of the currents, as well as the isoscalar and the isovector
components need separate consideration along with the projections onto the final and the
intermediate baryon states.  The product of currents is decomposed in two steps, namely the currents
in the box diagram are first coupled to t-channel $(J_1,I_1)$, and then the result is coupled to the
(conjugate) current of the one-photon exchange to total $(J=1,I=0,1)$ as needed here. At each stage
the resulting composite spin-flavor operators are decomposed into the basis of spin-flavor
operators.  An advantage of this procedure is that one obtains explicitly the results for generic
$N_c$, making possible a detailed organization in powers of $1/N_c$ of the different combinations of
the EM current components, with even more details such as the individual contributions of the
different $(J_1,I_1)$ and $(J,I)$ projections.

The integrals over the intermediate momentum direction $\bm{\hat k}_{\rm n}$ in
Eq.~(\ref{interfterm}) are reduced to cases where the numerator is a tensor product of
$\bm{\hat k}_{\rm n}$ multiplied by powers of $\bm{\hat k}_{\rm i} \cdot \bm{\hat k}_{\rm n}$ and
$\bm{\hat k}_{\rm f} \cdot \bm{\hat k}_{\rm n}$ (see Appendix~\ref{app:integrals}). As explained in
Sec.~\ref{subsec:ncexpansion}, the momentum dependence of the form factors needs to be included in
the integral, and a common dipole form is chosen for the form factors of all components of the EM
current. The integrations with these form factors are performed analytically in
Appendix~\ref{app:integrals}.

In general, the individual integrals show IR or collinear divergencies resulting from one of the
photons in the box diagram becoming soft or real within the integration domain. The collinear
singularities occur for a photon coupling to a current making a transition between $N$ and $\Delta$,
where a real photon with energy equal to the mass difference is possible, the other cases are IR
singularities.  Those divergencies are regulated by including an infinitesimal photon mass whose
effect is represented by the parameter $\eps = 0^+$ in Appendix~\ref{app:integrals}. The end
result is, for both cases with and without the inclusion of form factors \footnote{The IR and
collinear divergencies of individual integrals do depend on the form factors, thus additional
cancellations occur in this case. }, that those divergencies of the individual contributions cancel
in the imaginary part of the spin-dependent part of the integral in Eq.~(\ref{interfXsection}). It
is important to emphasize that the cancellations of the divergencies only occur for the precise
on-shell kinematics.  The divergencies cancel individually for the different final and intermediate
baryon states, and for each possible t-channel $(J=1,I=0,1)$ and $(J_1,I_1)$ projection of the box
diagram, as far as in the TPE absorptive amplitude EM gauge invariant combinations of the two
hadronic currents are considered, i.e., for terms with two different components of the EM current
the two possible orderings must be added up.  In particular, those cancellations serve as one useful
check of the calculations.

The explicit calculation shows that for a stable $\Delta$ the interference differential cross
section has a finite discontinuity at the $\Delta$ threshold. This discontinuity is only present in
the elastic asymmetry, i.e., nucleon final state. It is explained as follows: the leptonic tensor is
proportional to the energy $E_{\rm n}$ of the electron in the box, the absorptive part of the
diagram has a phase space proportional to $E_{\rm n}$, and each photon propagator gives a factor
$1/E_{\rm n}$, so that there is a finite contribution in the limit $E_{\rm n}\to 0$, which is at the
threshold for the $\Delta$. This finite threshold enhancement is thus understood as the two photons
becoming real and collinear with $\bm {k}_{\rm i}$ and $\bm {k}_{\rm f}$.

Although the $\Delta$ width is $\ord{N_c^{-2}}$, it is then necessary to include it to reproduce the
realistic behavior at the onset of its contribution, resulting in a smoothing of the mentioned
discontinuity. In the present calculation the width is implemented by a Breit-Wigner form, as shown
in Appendix~\ref{app:delta_width}. The effect of the width is reduced to a smearing of the $\Delta$
mass in the calculation at zero width using Eq.~(\ref{smearing}).
\section{Results}
\label{sec:results}
%
%
\begin{figure*}[t]
\begin{tabular}{ll}
\includegraphics[width=8cm]{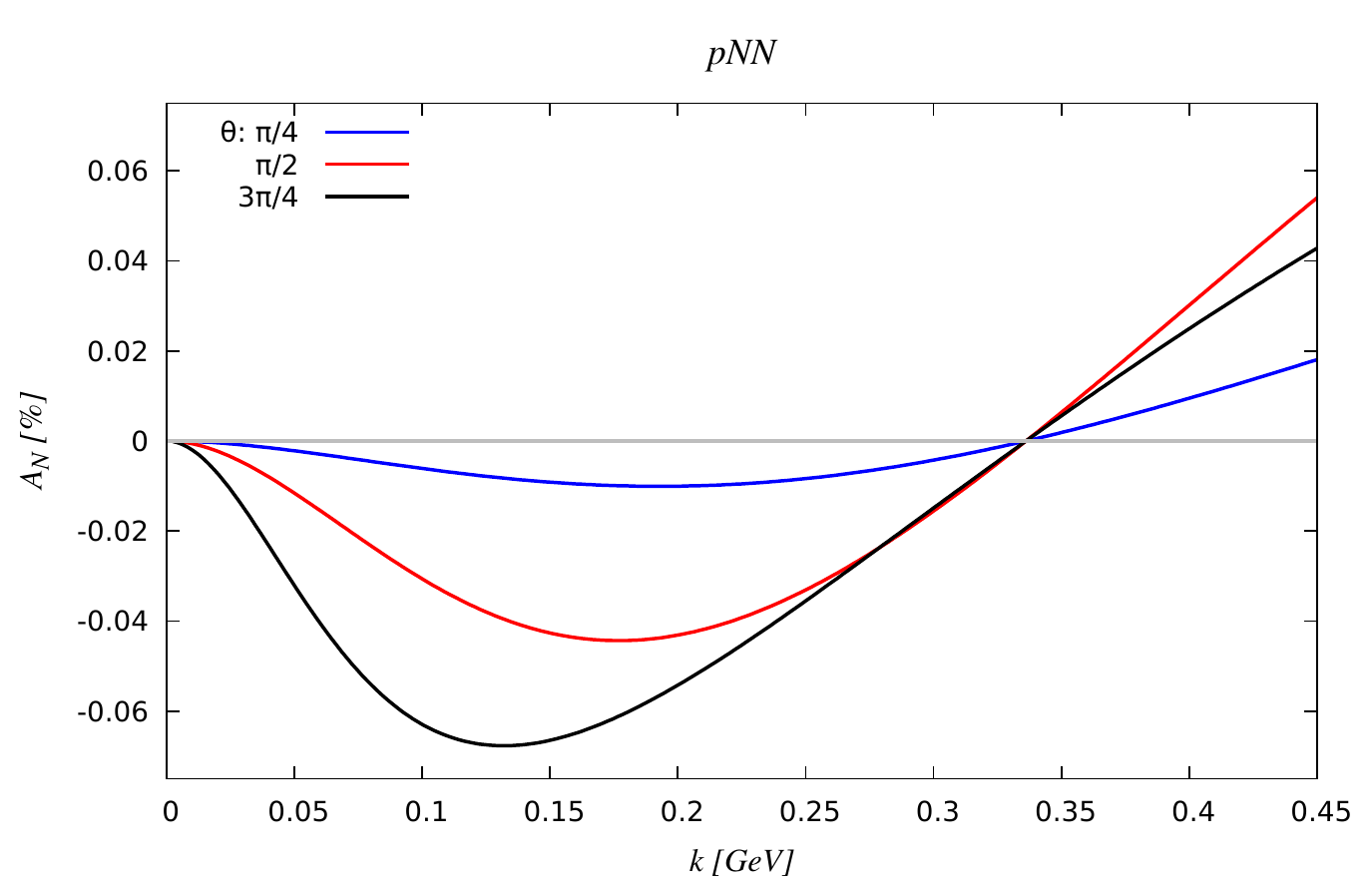} &
\includegraphics[width=8cm]{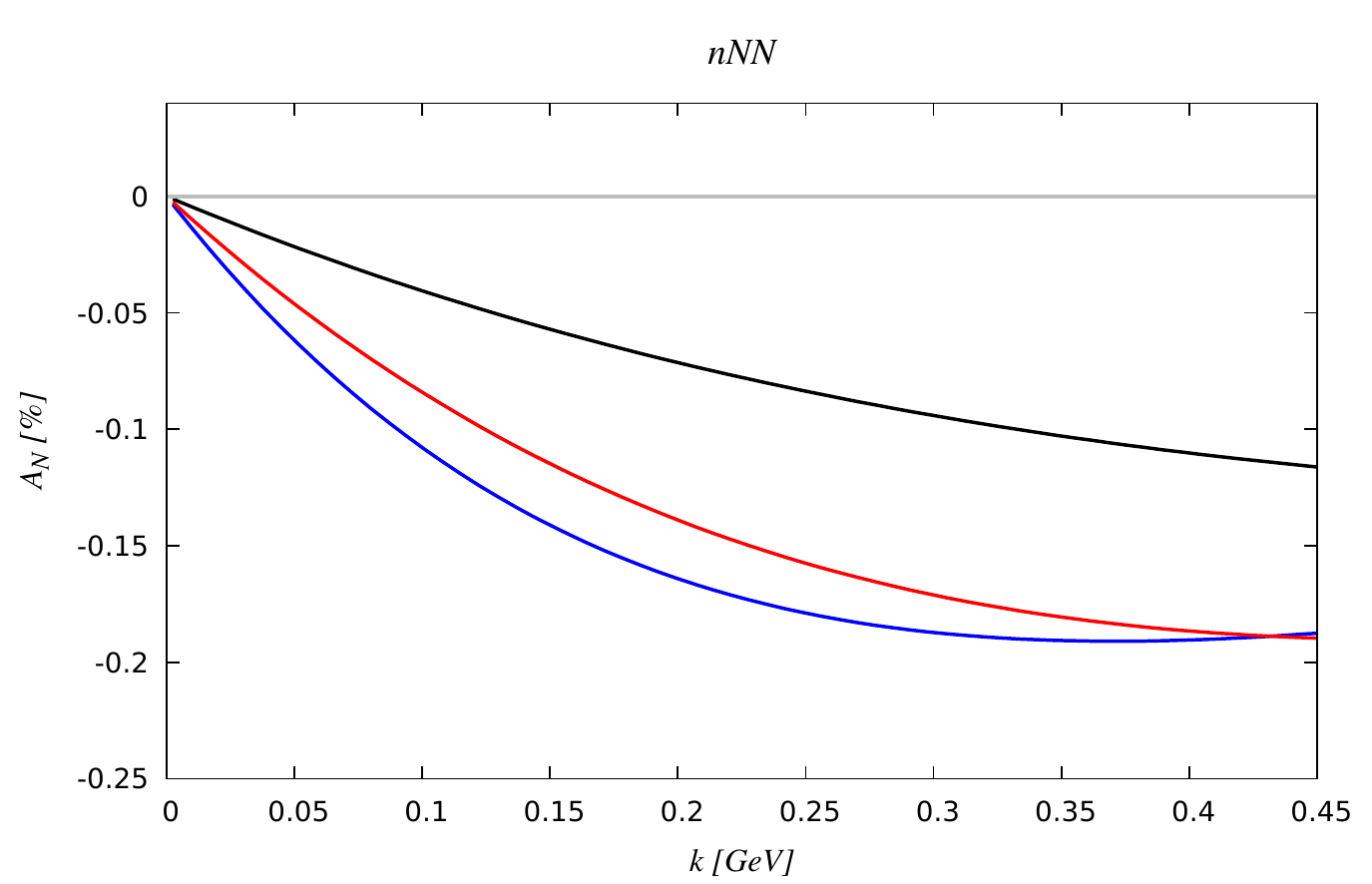}
\\
\hspace{-1.7em}
\includegraphics[width=8.6cm]{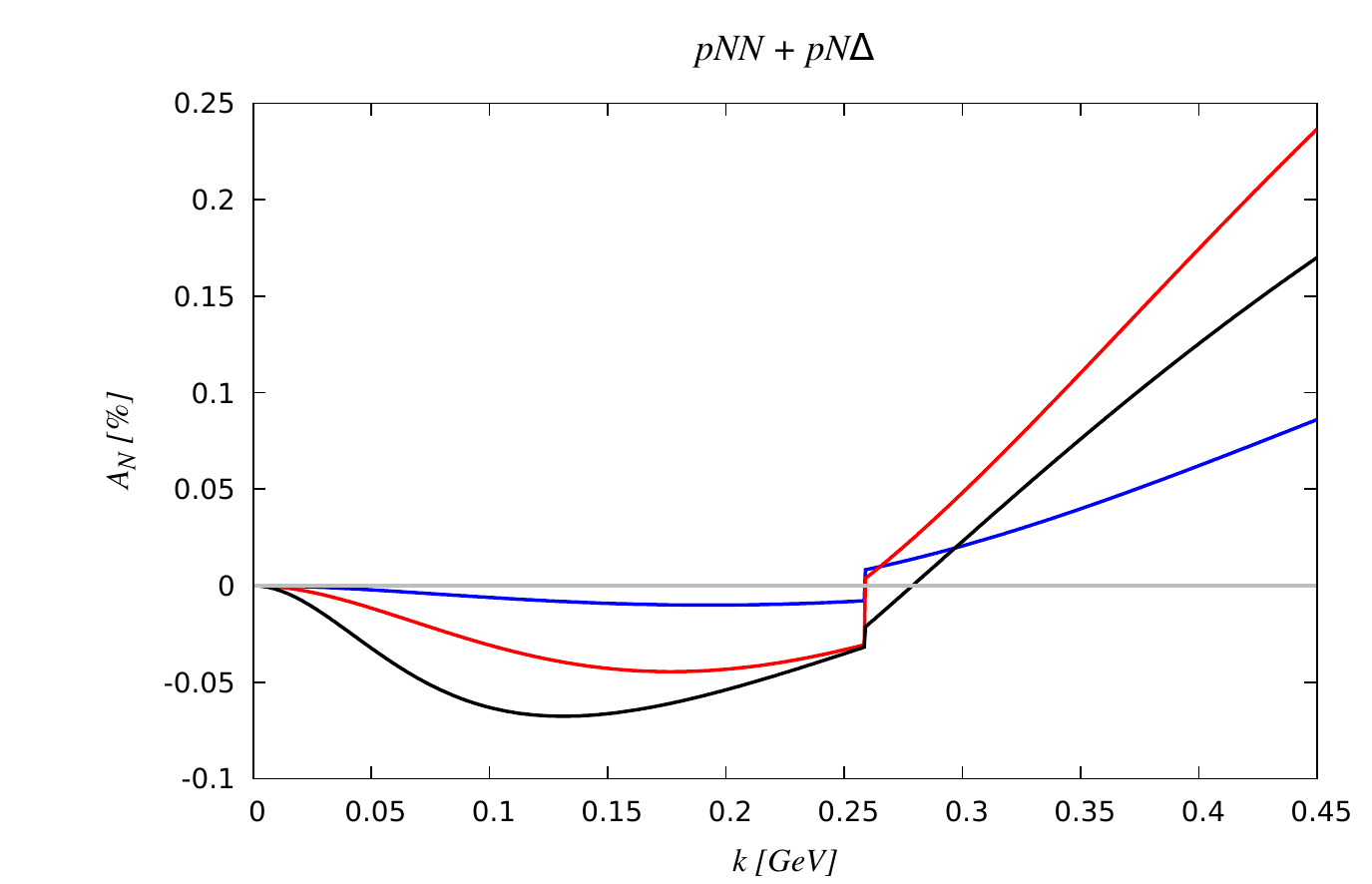} &
\includegraphics[width=8.6cm]{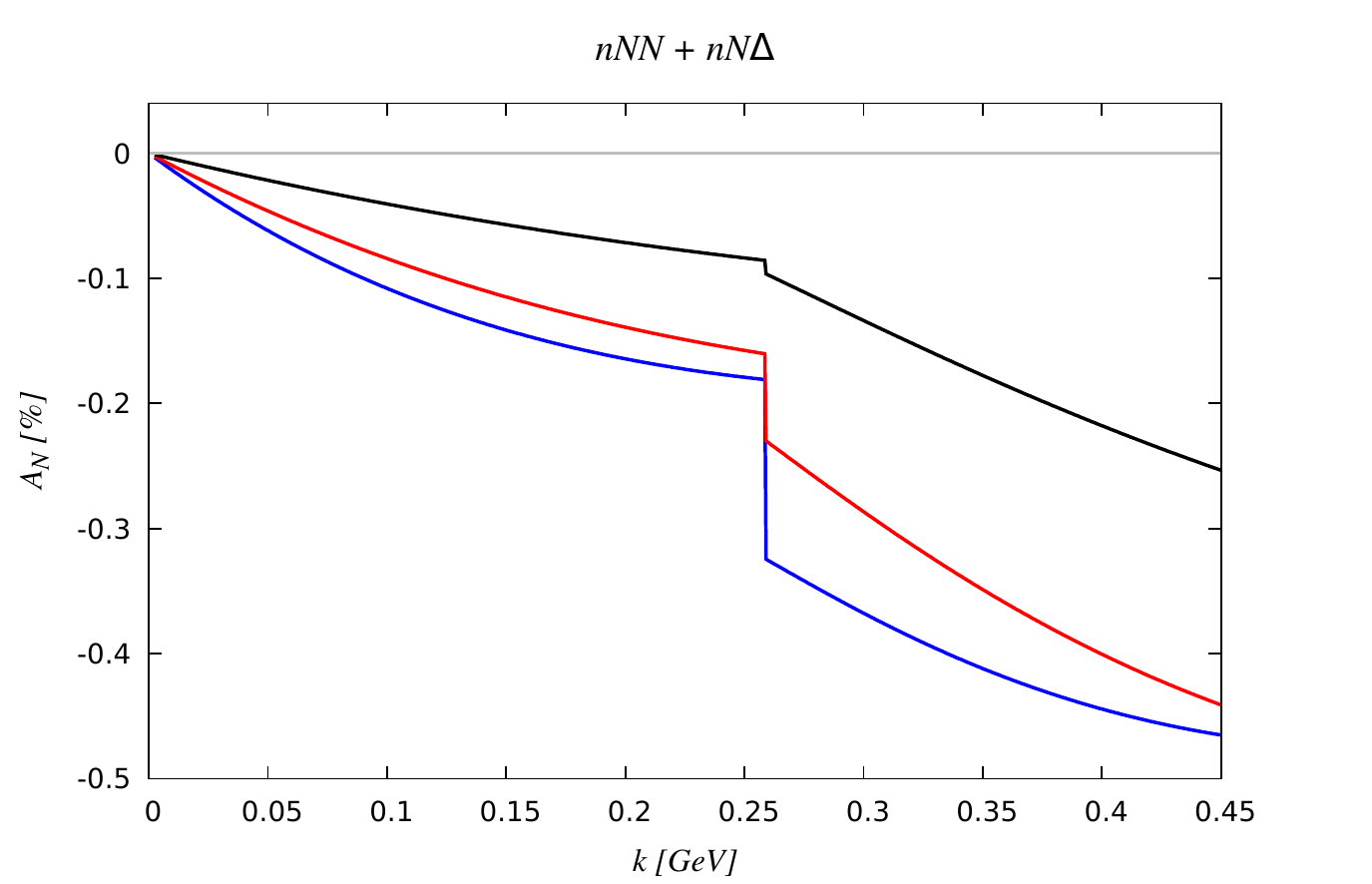}
\\
\includegraphics[width=8cm,height=5.6cm]{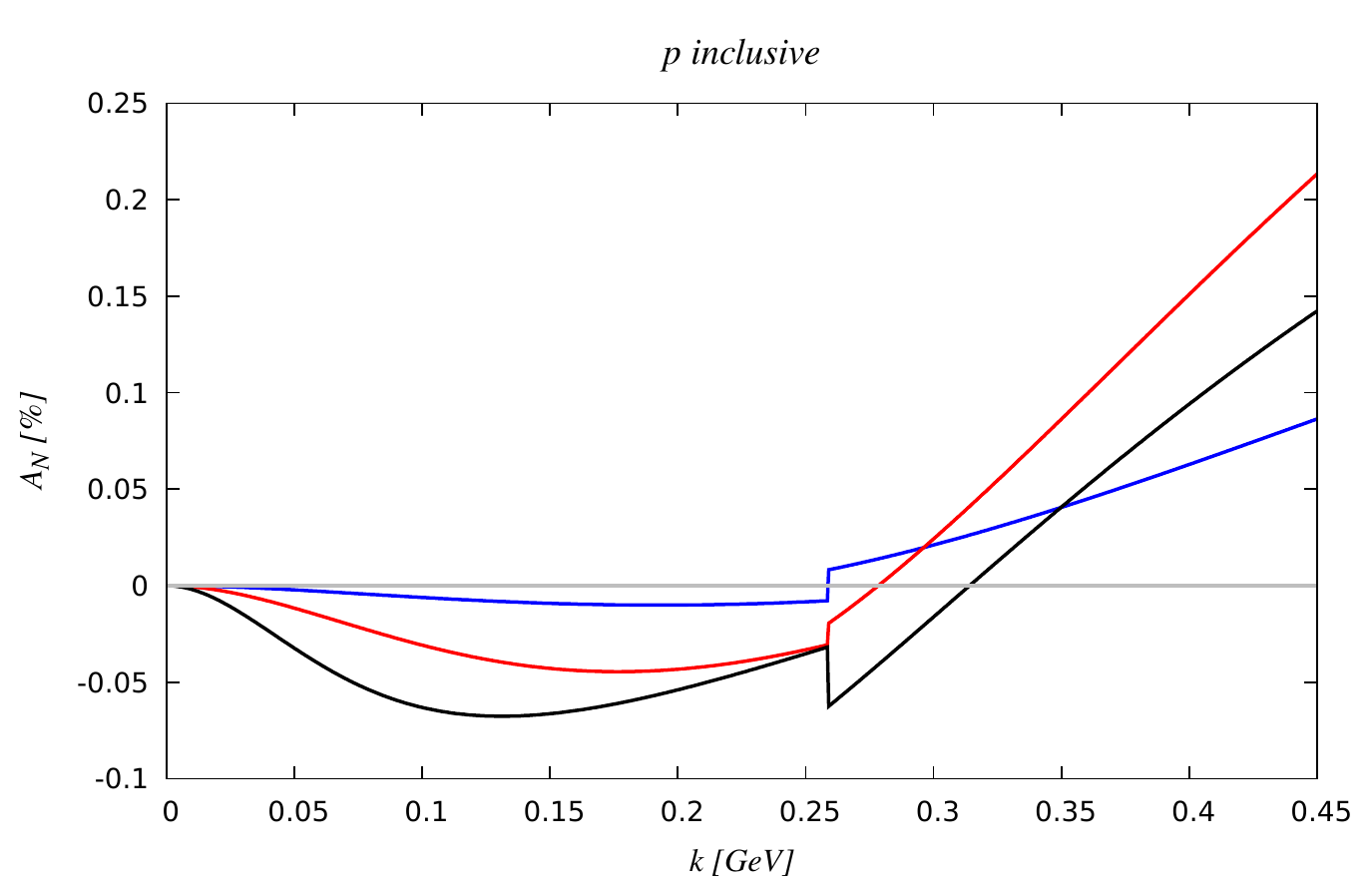} &
\includegraphics[width=8cm,height=5.6cm]{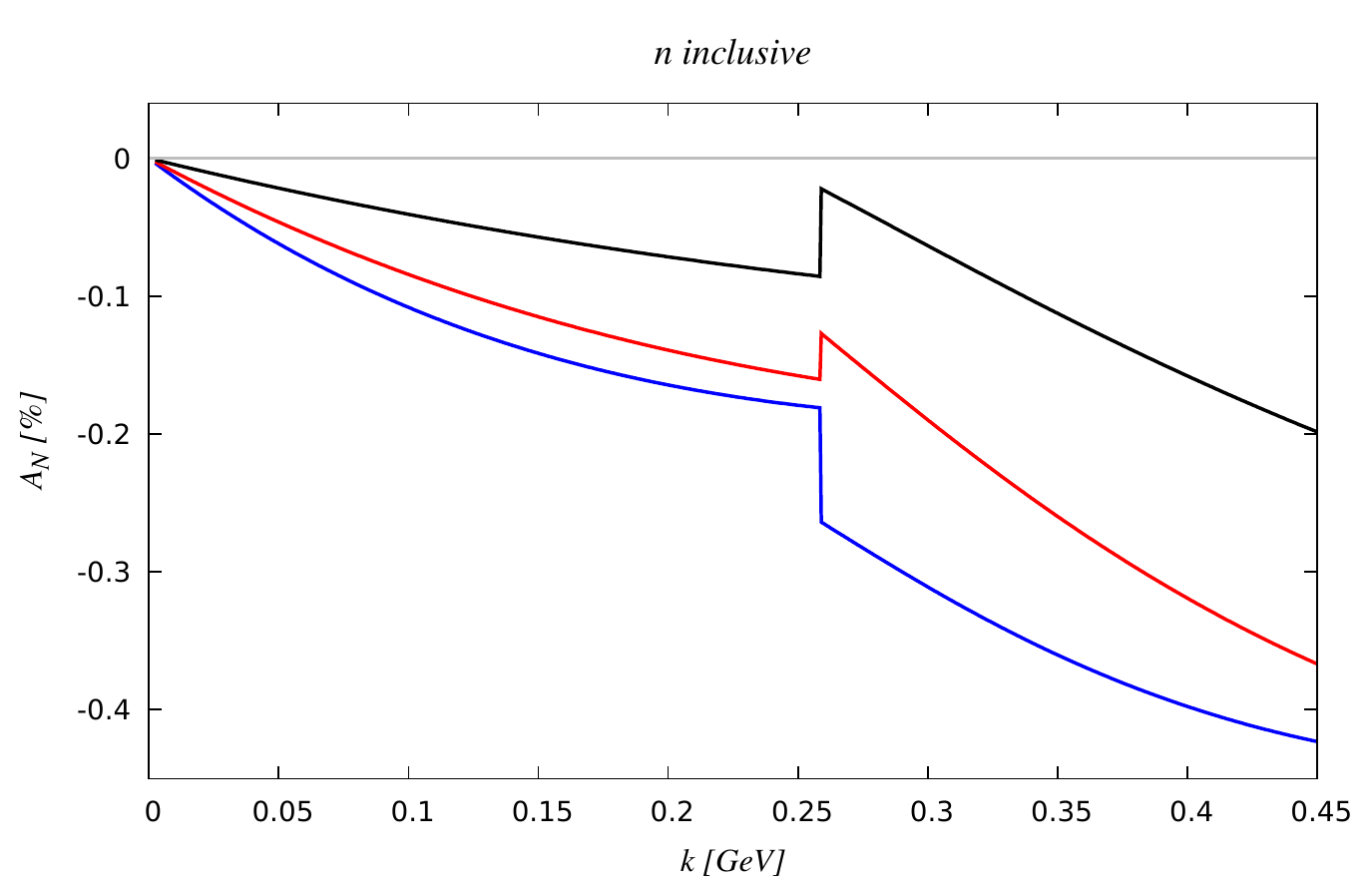}
\end{tabular}
\caption[]{$A_N$ vs $k$ with no form factors and stable $\Delta$. Left/right column: Proton/neutron
target.  Top row: Elastic $A_N$ with only nucleon in the TPE amplitude. Middle row: Elastic $A_N$
with nucleon and $\Delta$ in TPE amplitude. Bottom row: Inclusive $A_N$.}  \label{fig:TSSAElastic}
\end{figure*}
\subsection{Evaluation and validation}
The TSSA is now evaluated numerically, using the expressions obtained from the $1/N_c$ expansion of
the hadronic tensor.  The results cover the parametric regions I--III identified in
Sec.~\ref{subsec:kinematic} and are accurate to subleading order in $1/N_c$.  In regions I and II
the present results are new and predict the behavior of the TSSA below and above the $\Delta$
threshold. In region III the present results can be matched with the leading-order $1/N_c$ expansion
results of the previous publication \cite{Goity:2022yro} but include also the subleading $1/N_c$
corrections, which improve the predictions and illustrate the theoretical uncertainty.

The results shown here have been validated with two independent tests: (i)~Comparison with the
leading-order $1/N_c$ expansion results \cite{Goity:2022yro}, which were evaluated using an
independent algebraic method.  (ii)~Comparison of the nucleon-only contribution (i.e., nucleon in
intermediate and final state) with the well-known result of the relativistic Feynman diagram
calculation, expanded such as to match the $1/N_c$ expansion calculation.

\subsection{Role of form factors and electric/magnetic currents}
It is instructive to first display the results when the $t$ dependence of the form factors is
neglected, as this gives a rough idea of the role of the different components of the EM current, and
also serves as a reference point for the calculation with form factors. As indicated earlier, the
TSSA $A_N$ is defined in the following with respect to the unpolarized elastic cross section
Eq.~(\ref{OPEXsection}).  The results for the separate contributions to the interference term of the
cross section $d\sigma_{N}$ by the nucleon and $\Delta$ in the intermediate and final states are
given in Appendix~\ref{app:sigma_N} Eq.~(\ref{dsigmaSSA}). The contributions are at most linear in
the electric form factors\footnote{A non-relativistic expansion of the cross section starting with
the relativistic one gives terms that are proportional to $(G_E/m_N)^2$; such terms are of higher
order in $1/N_c$ and are not captured by the present expansion. They involve the contributions from
the spatial components of the convection EM current, which to the order of the present calculation
are irrelevant. }, and in the strict non-relativistic limit, independent of the nucleon mass, as one
would expect.  As mentioned earlier, neglecting the width of the $\Delta$ leads to a finite
discontinuity in the interference differential cross section in the case of a final nucleon at the
$\Delta$ threshold.

Figure~\ref{fig:TSSAElastic} shows the TSSA $A_N$ evaluated without form factors (here and in the
following $k \equiv |\bm{k}_i|$). It is observed that the $\Delta$ intermediate state in the box
amplitude makes a large contribution to the elastic asymmetry. On the other hand, the $\Delta $
final state makes a very small contribution to the inclusive asymmetry. This was observed already in
the leading-order $1/N_c$ expansion in the kinematic region III in Ref.~\cite{Goity:2022yro}. As
shown below, the inclusion of the $t$ dependence in the form factors profoundly affects the
suppression of the $\Delta$ state in the inelastic asymmetry. For the proton the behavior of the
asymmetry is very much affected in the kinematic domains I and II by the terms proportional to
$G_E$, which are of opposite sign to the purely magnetic ones and larger, leading to the cross-over
to negative values shown in Fig.~\ref{fig:TSSAElastic}.
 
\begin{figure*}[!]
\includegraphics[width=8cm,height=5.6cm]{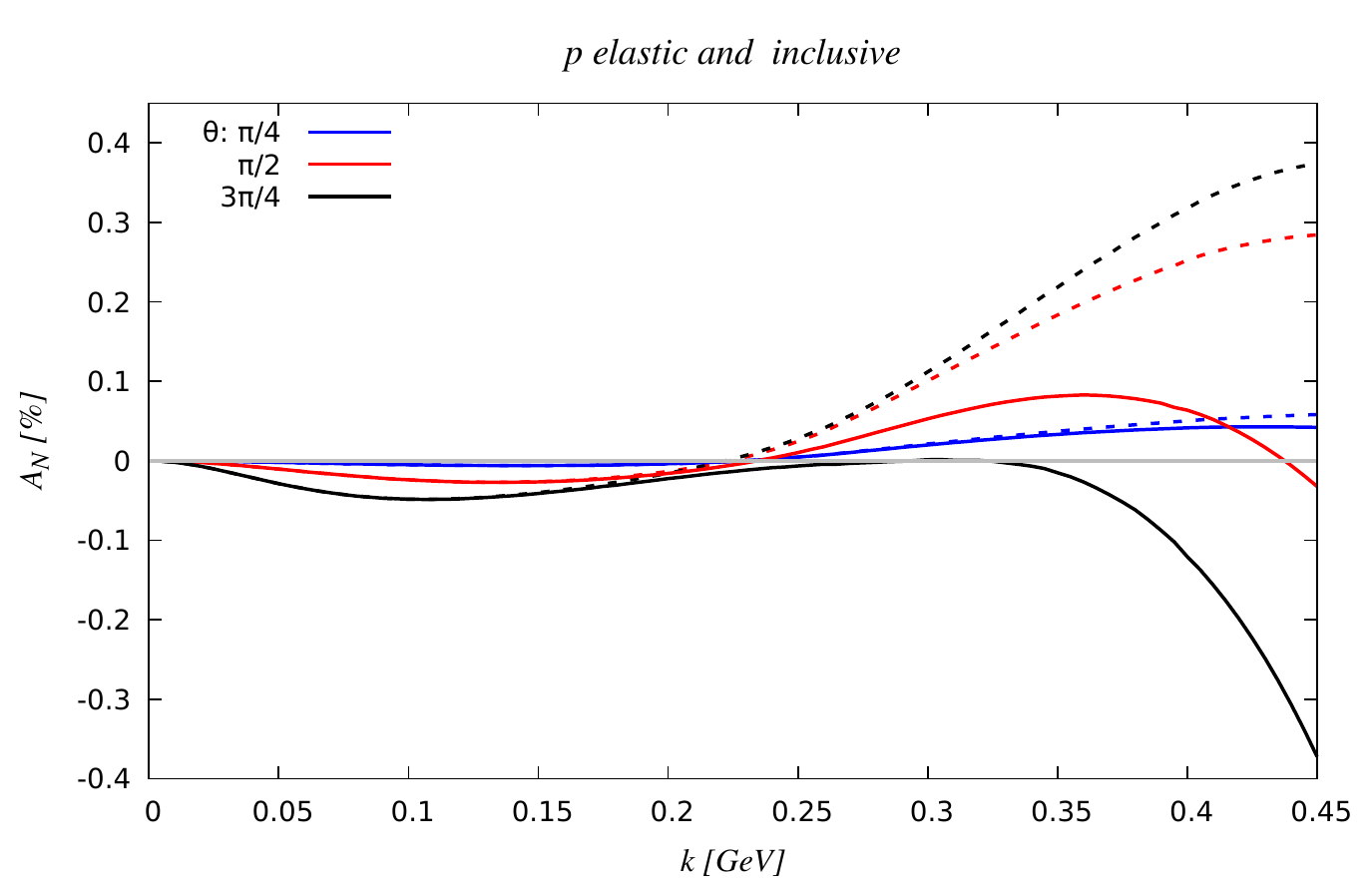}
\includegraphics[width=8cm,height=5.6cm]{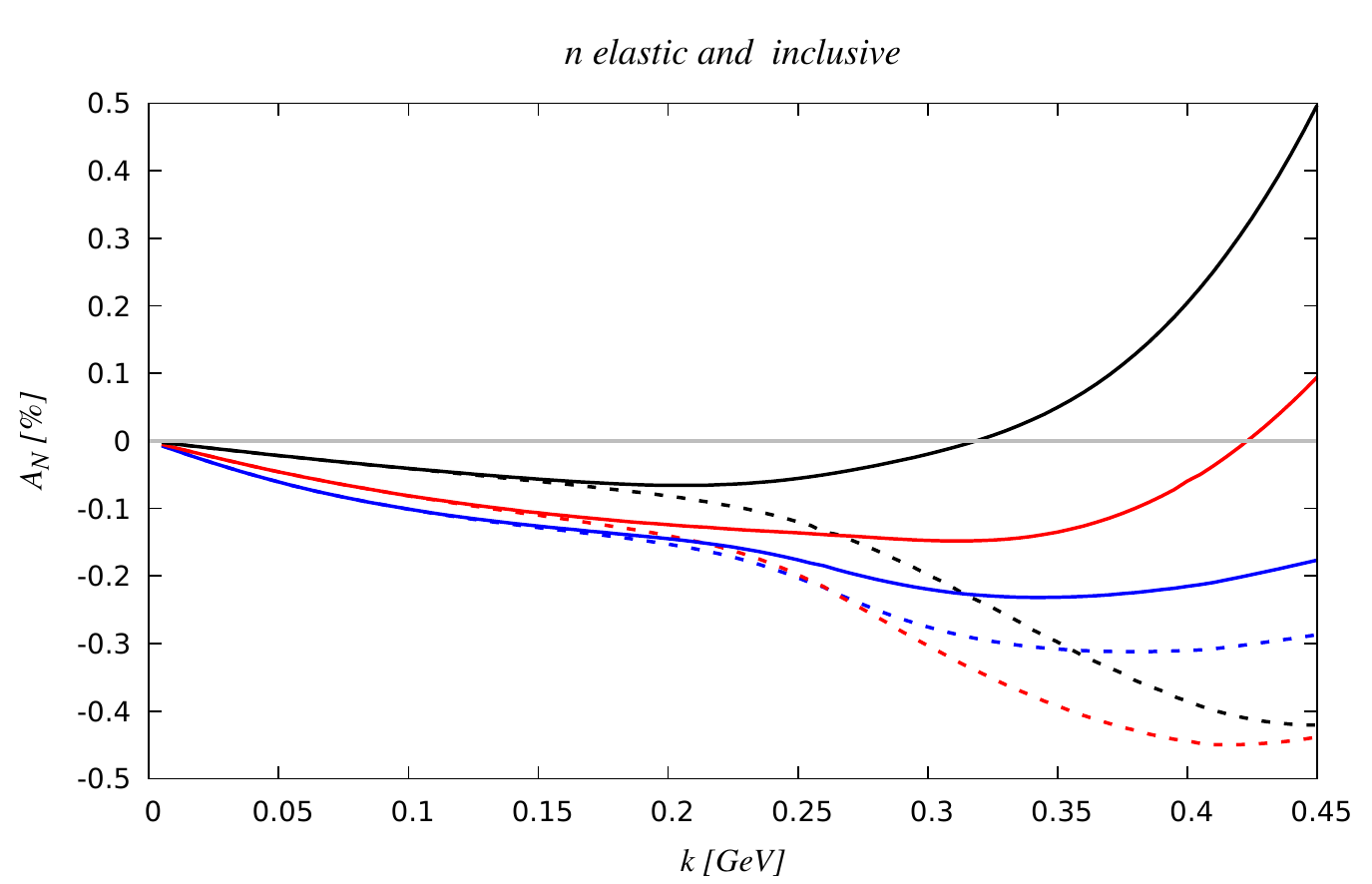}
\\
\includegraphics[width=8cm,height=5.6cm]{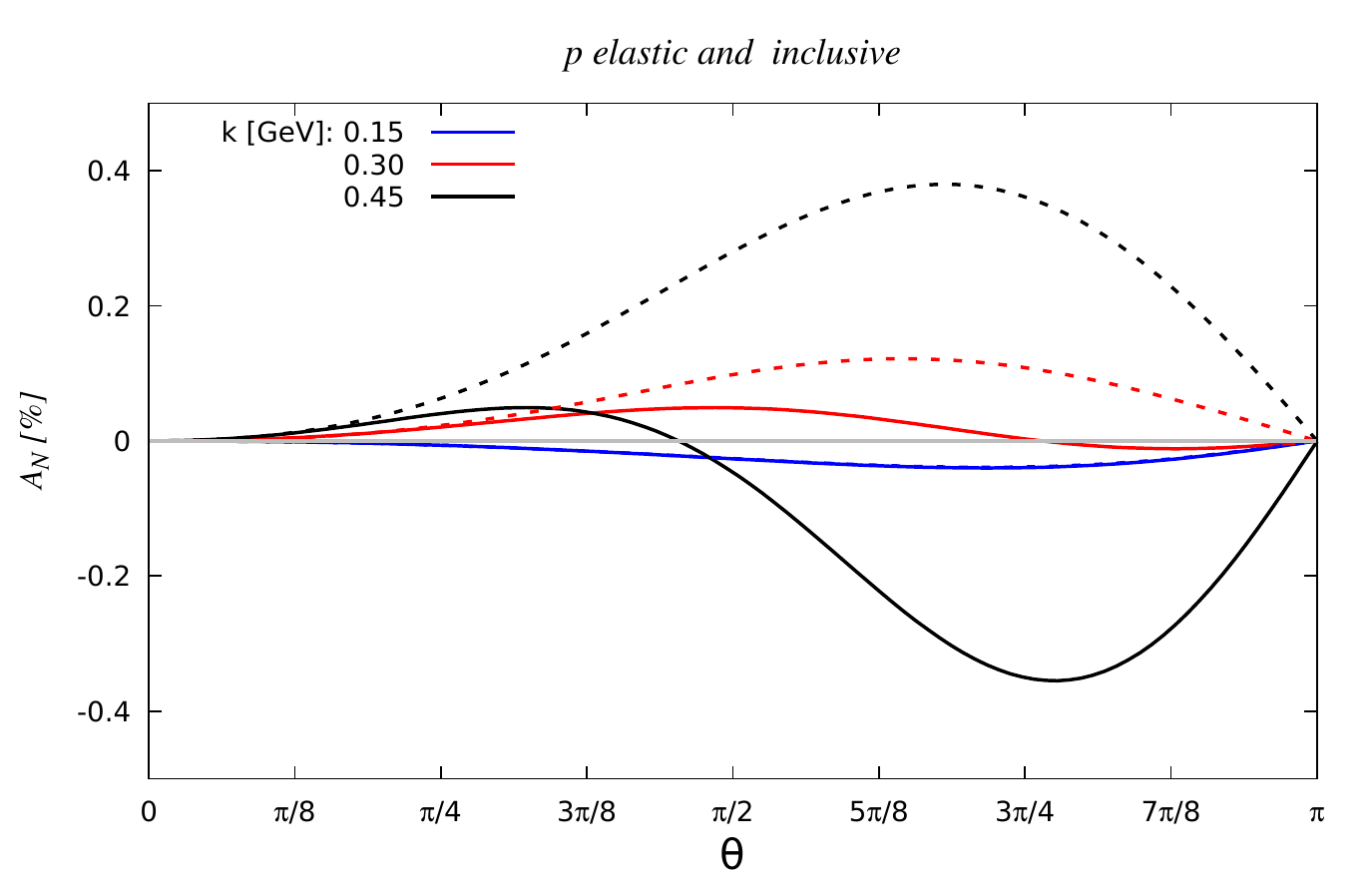}
\includegraphics[width=8cm,height=5.6cm]{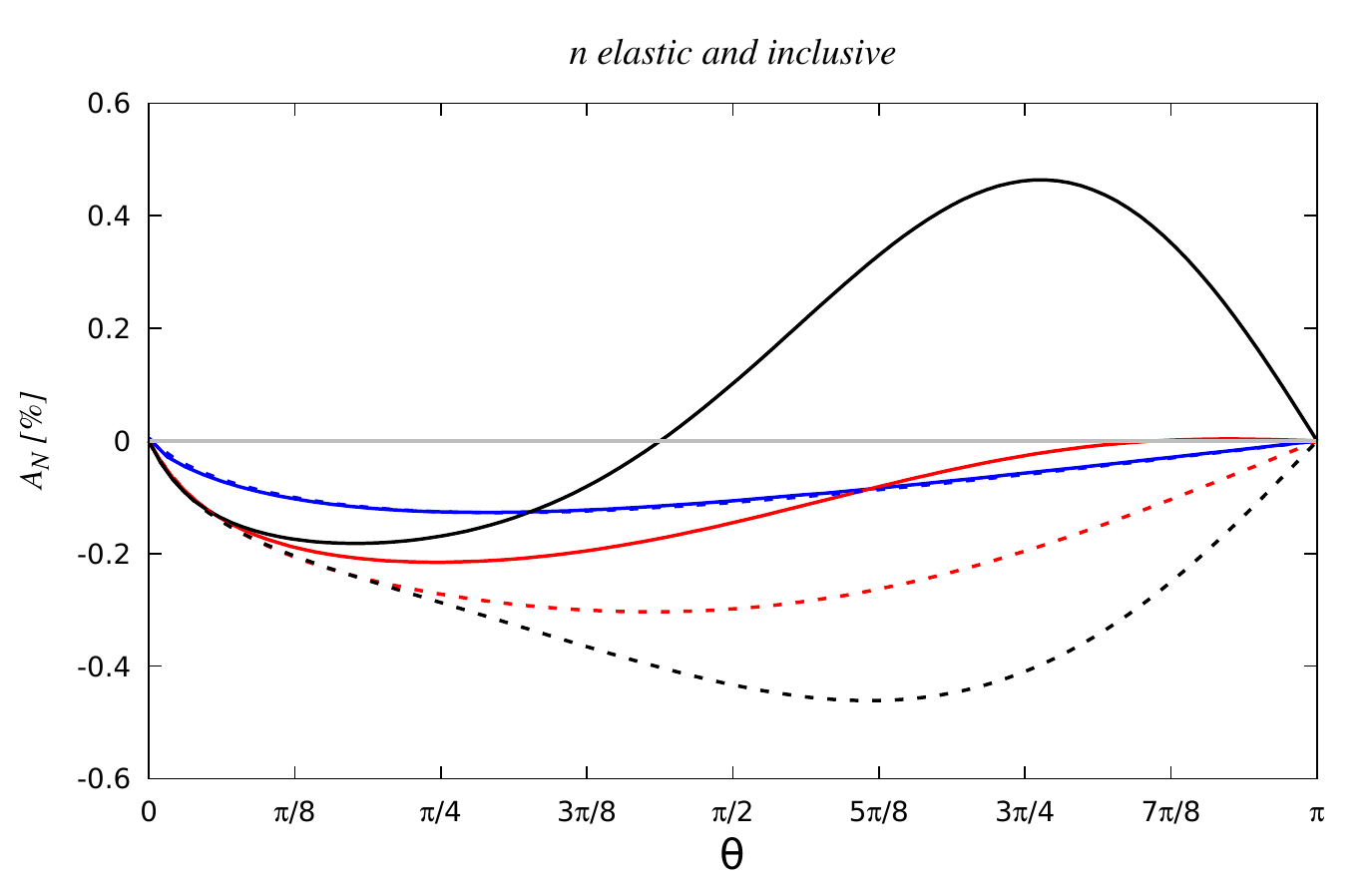}
\caption[]{$A_{N}$ vs $k$ (top row) and $A_{N}$ vs $\theta$ (bottom row) with inclusion of form
factors and $\Delta$ width. Proton target (left column) and neutron target (right column). Elastic
(dashed lines) and inclusive (solid lines).}
\label{fig:TSSAFF}
\end{figure*}

\begin{figure*}[!]
\includegraphics[width=8cm,height=5.6cm]{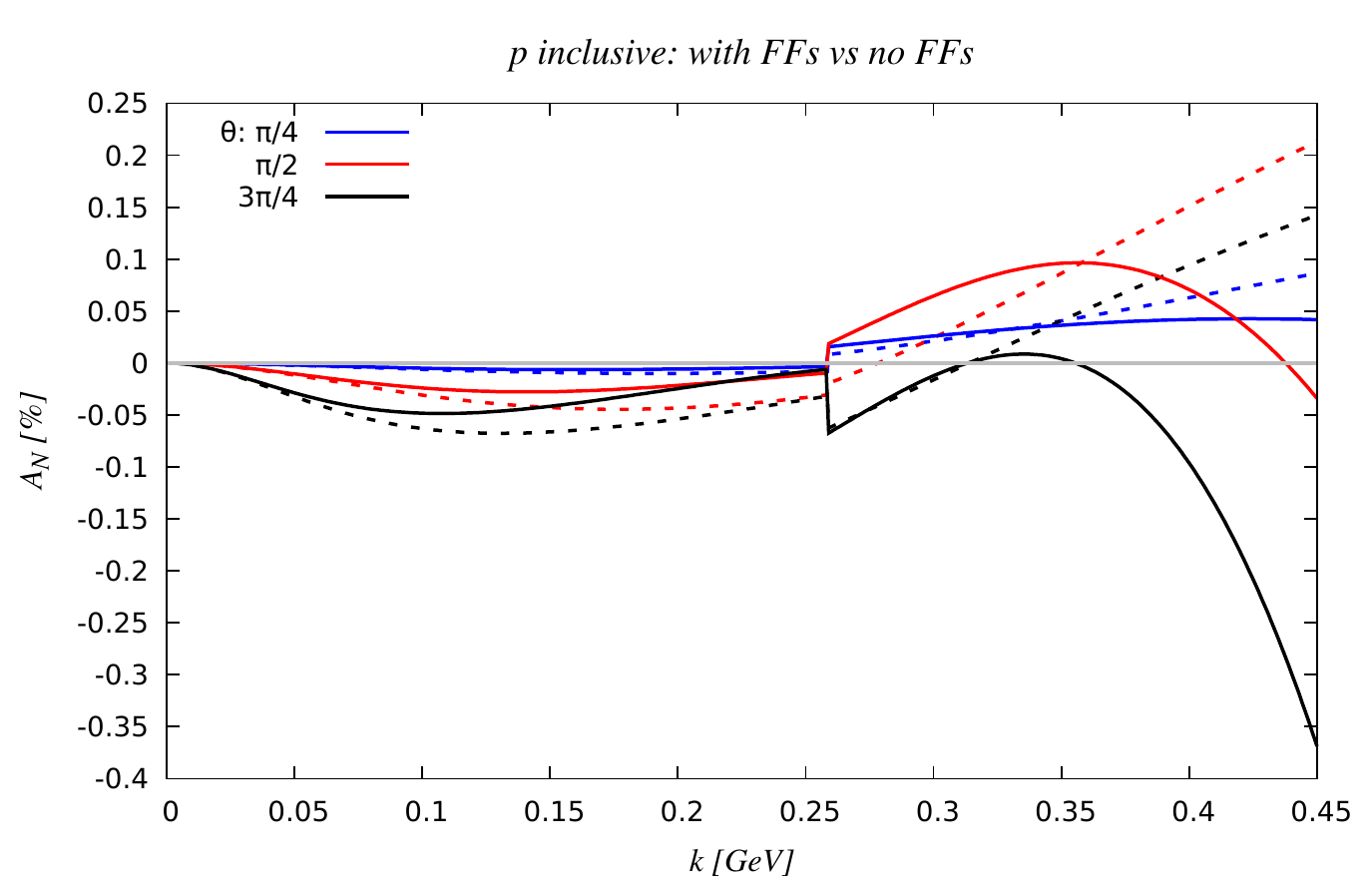}
\includegraphics[width=8cm,height=5.6cm]{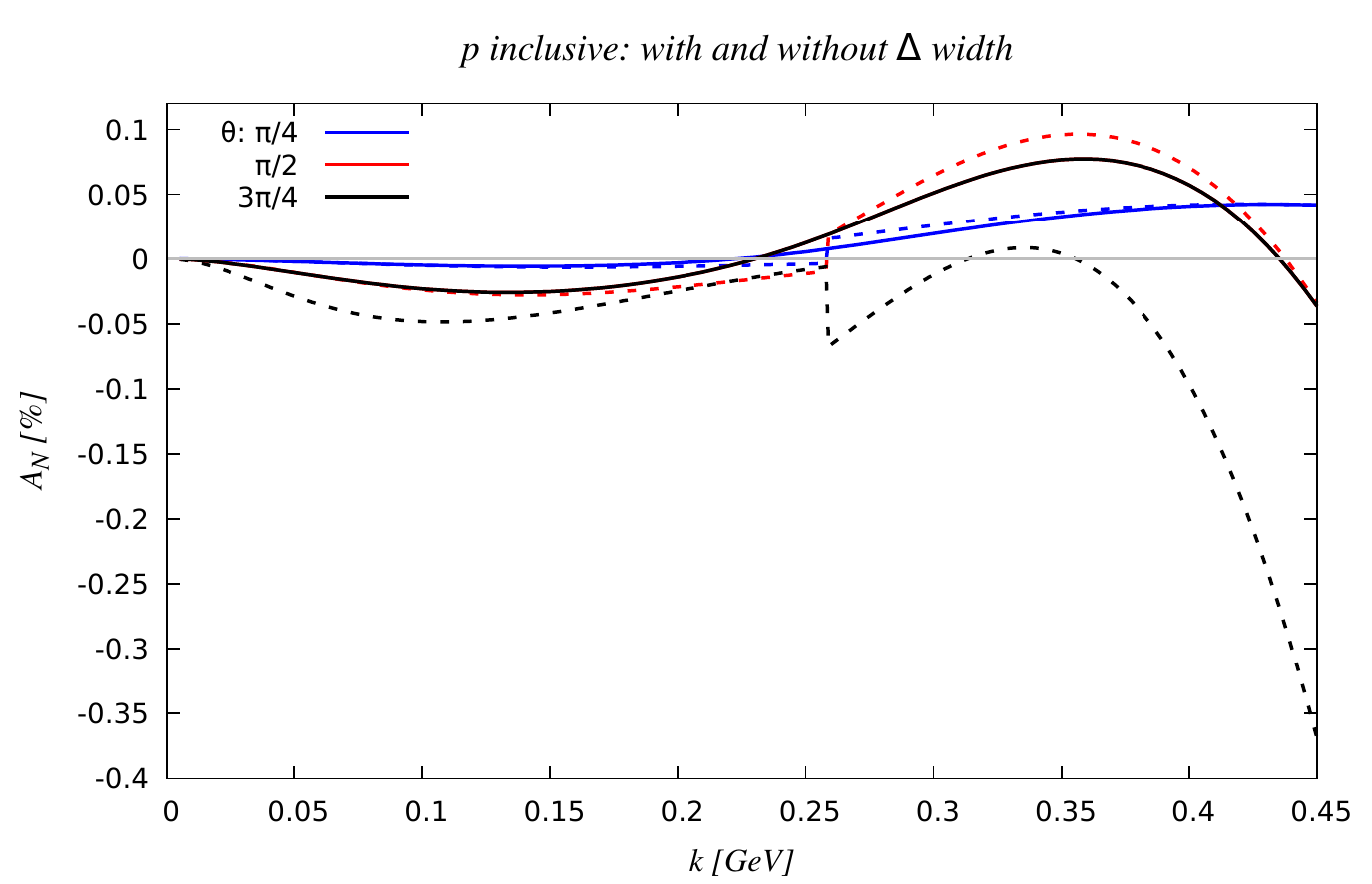}
\caption[]{Inclusive $A_{N}$ for proton target. Left panel: Comparison of results without form
factors (dashed lines) and with form factors (solid lines). Right panel: Comparison of results
without $\Delta$ width (dashed) and with $\Delta$ width (solid); both are with form factors.}
\label{fig:TSSA-FF-noFF}
\end{figure*}

\begin{figure*}[!]
\includegraphics[width=8cm,height=5.6cm]{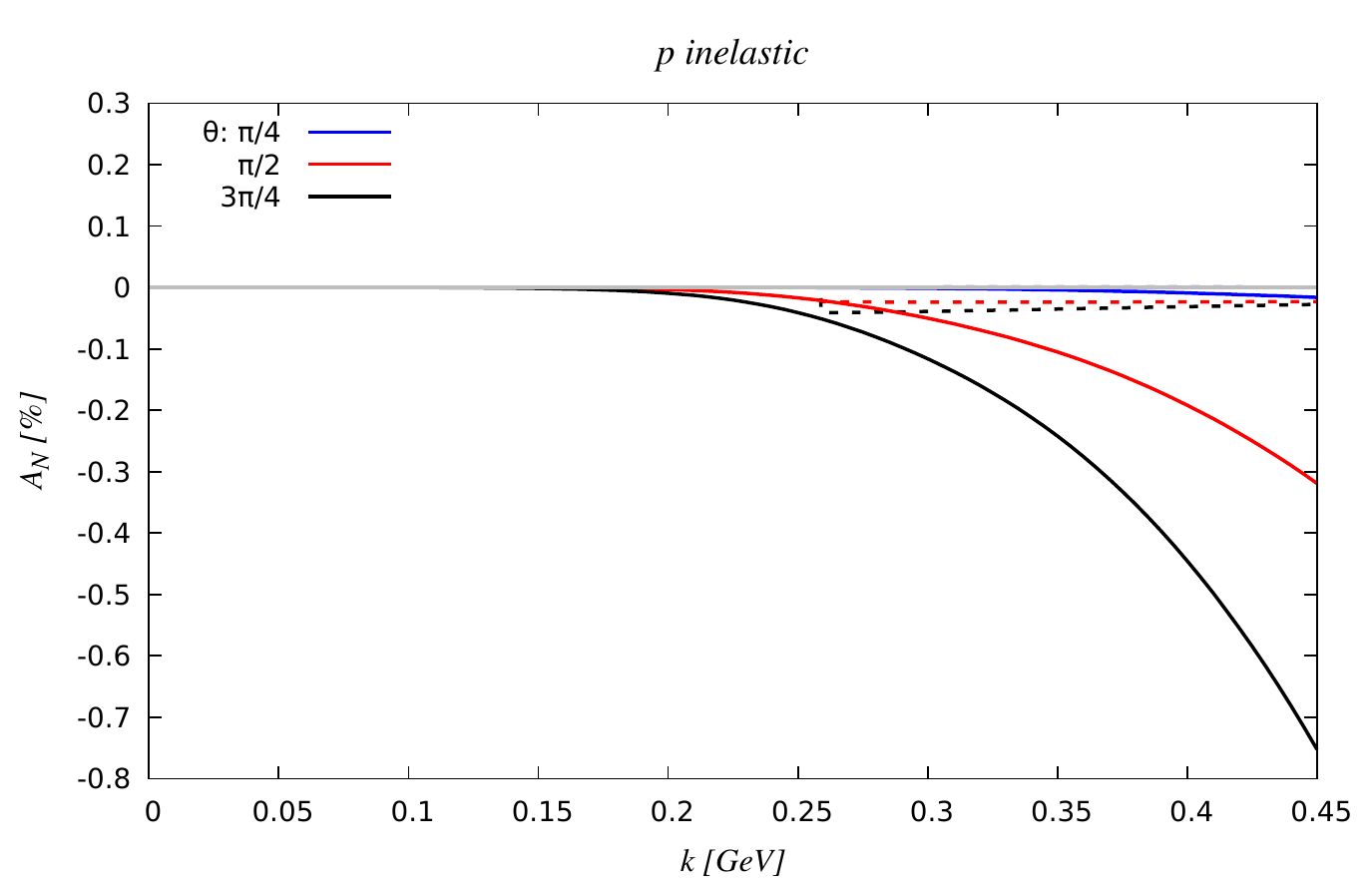}
\includegraphics[width=8cm,height=5.6cm]{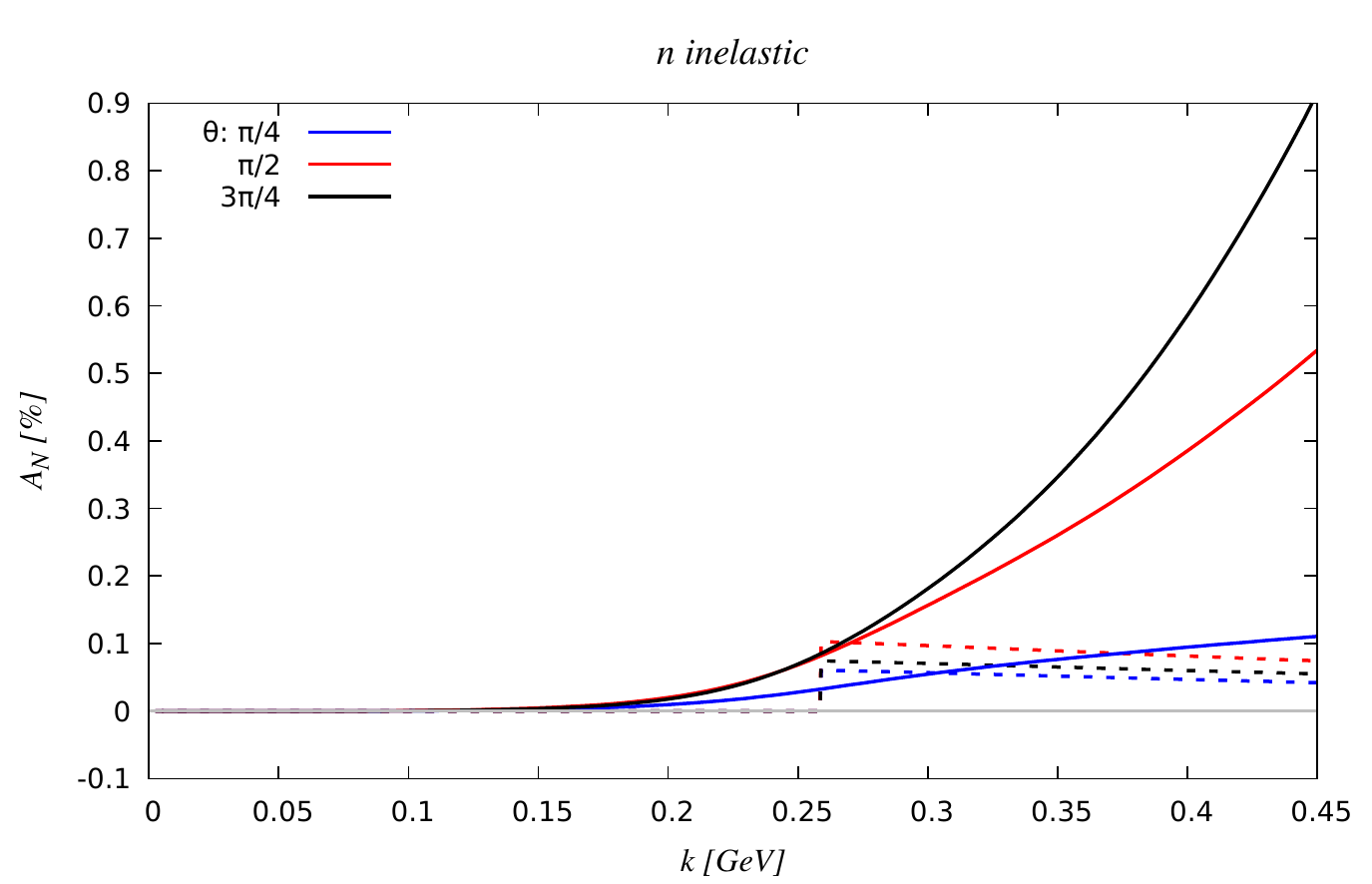}
\caption[]{Inelastic $A_{N}$ for proton (left panel) and neutron target (right panel). Comparison
between results without form factors (dashed lines) and with form factors (solid lines).}
\label{fig:TSSA-Inelastic-FF-noFF}
\end{figure*}

Figure~\ref{fig:TSSAFF} shows the results obtained with inclusion of the form factors and the
$\Delta$ width (these represent the final numerical results and will be discussed further below).
The width is implemented using $\Gamma_\Delta=0.125$ GeV and $Q=0.2$ GeV. Comparing with
Fig.~\ref{fig:TSSAElastic} one observes that the form factors have only a moderate effect on the
elastic asymmetries (dashed curves in Fig.~\ref{fig:TSSAFF}). However, they have a dramatic effect
on the inelastic asymmetry ($\Delta$ final state). This is further illustrated by
Figs.~\ref{fig:TSSA-FF-noFF} and \ref{fig:TSSA-Inelastic-FF-noFF}, which directly compare the
results with and without form factors for the inelastic and inclusive (elastic + inelastic)
asymmetries. In fact, for energies above the $\Delta$ threshold, the inelastic asymmetry has
opposite sign to that of the elastic one and becomes increasingly dominant with energy.  This effect
of the form factors was observed in the LO $1/N_c$ expansion \cite{Goity:2022yro}.

There is no simple argument explaining the effects of the form factors in the absorptive part of the
box diagram observed here. However, some insight can be gained from considering the large-$N_c$
limit, where the leading-order $1/N_c$ expansion result becomes exact. One finds that logarithmic
terms $\propto \log \sin^2\frac{\theta}{2}$ are important in the interference cross section for elastic
and inelastic final states. In the inelastic case there is a strong cancellation between these
logarithmic terms and polynomial terms in $\sin^2\frac{\theta}{2}$ when the form factors are neglected,
giving the small inelastic interference cross section. This cancellation is upset when the form
factors are included, resulting in the strong sensitivity of the inelastic TSSA to the form
factors. In the strict large $N_c$ limit, the contribution to the elastic asymmetry by the $\Delta$
intermediate state in the box is twice that of the $N$. In the physical case, with the subleading
terms in the EM current included, one finds a similar result for the case of the neutron, while the
contribution of the $\Delta$ is further enhanced for the proton.  On the other hand, for $\Delta$ in
the final state, in the strict large $N_c$ limit, the contribution of the $N$ in the box is five
times as large as that of the $\Delta$. This remains roughly the same in the physical case for both
proton and neutron.
%
%

%
%

%
%

%
%
\begin{figure}[!]
\includegraphics[width=8cm,height=5.6cm]{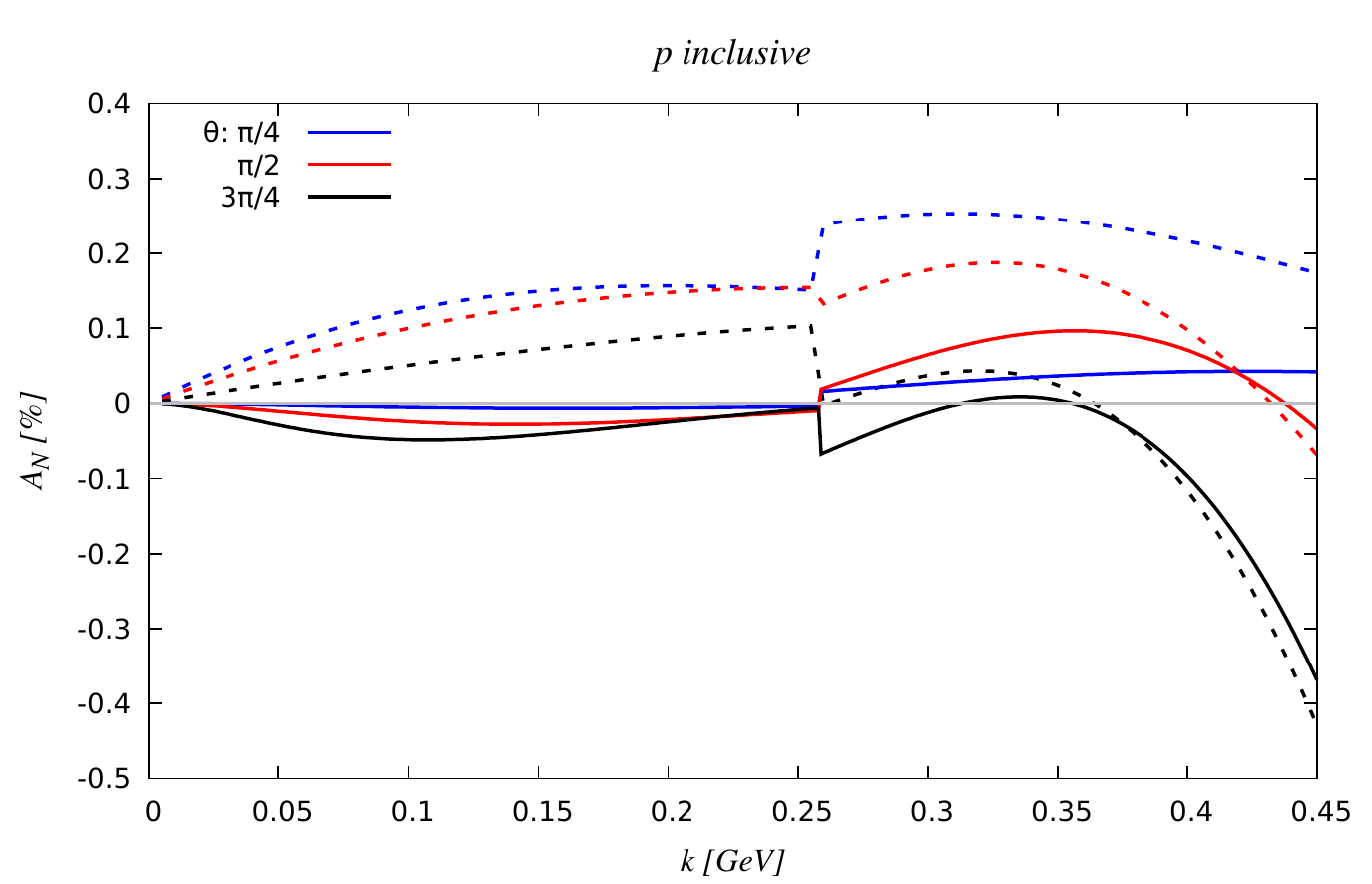}
\includegraphics[width=8cm,height=5.6cm]{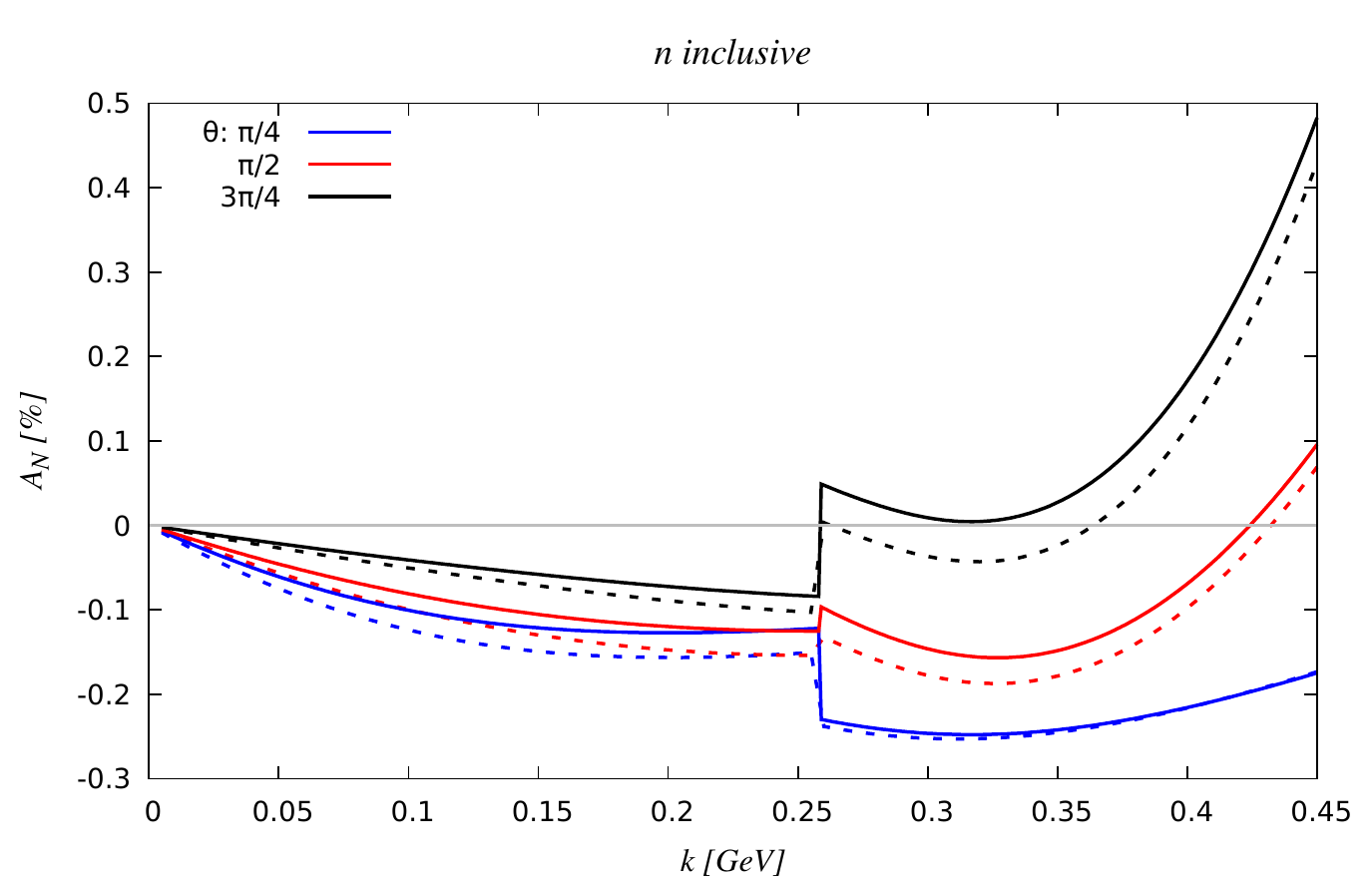}
\caption[]{Inclusive $A_{N}$ for proton (left panel) and neutron target (right panel). Comparison
between the LO in $1/N_c$ (dashed lines) and the results of this work (solid lines). Form factors
are included, and the physical phase space is used in the LO result.}
\label{fig:TSSA-LO-vs-NLO}
\end{figure}

\subsection{Size of $1/N_c$ corrections}
It is interesting to investigate the size of $1/N_c$ corrections in the TSSA. This serves to
illustrate the convergence of the parametric expansion and quantify the numerical uncertainty.

The LO $1/N_c$ expansion in kinematic region III, $k = \ord{N_c^0}$, was considered in
Ref.~\cite{Goity:2022yro}.  In this order the $N$ and $\Delta$ are degenerate, and only the
isovector magnetic component of the EM current contributes.  The LO contributions to the TSSA are
$\ord{\alpha N_c}$ and arise from the hadronic currents in the box being coupled to $I_1=J_1$, which
can only be $I_1=J_1=0$ if the final state is $N$ (elastic) and $I_1=J_1=2$ if the final state is
$\Delta$ (inelastic). Using the results of the present calculation, it is now possible to compute
the $1/N_c$ corrections in region III. They arise from the LO components of the two currents in the
box diagram coupled to $I_1\neq J_1$, and from the subleading components of the EM current. At this
order also the mass difference $m_\Delta-m_N$ must be included. Furthermore, it is possible to
compute the size of $1/N_c$ corrections in regions I and II, $k = \ord{N_c^{-1}}$, which is also
covered by the present expressions.

Figure~\ref{fig:TSSA-LO-vs-NLO} shows the comparison of the LO and NLO results. Here the correct
phase space, with the finite $N$--$\Delta$ mass splitting, is used for the LO result. For the
neutron one sees that the LO result is close to the NLO one, which is easy to understand as the
contributions are purely magnetic, and the only difference is the disregard of the isoscalar
magnetic term at LO. On the other hand, for the proton the effect of the electric term in the
current, which is not present at LO, leads to a big difference at NLO. As mentioned earlier, the
modified power counting implied in the kinematic regions I and II shows the relevance of the
electric contributions, especially at the smaller angles. At larger energies and scattering angle
the purely magnetic contributions become dominant, and the LO approximation is remarkably good
(black curve in Fig.~\ref{fig:TSSA-LO-vs-NLO}).

\subsection{Final results}
The results of Fig.~\ref{fig:TSSAFF} represent the final numerical estimate of the TSSA and should
be used to discuss its kinematic dependencies and potential measurements. It is worth noting the
following features: (i)~The elastic and inclusive asymmetries are of the order few $\times 10^{-3}$
for $k \lesssim$ 0.5 GeV, for both proton and neutron.  (ii)~The inelastic contribution to the
asymmetry above the $\Delta$ threshold has opposite sign to the elastic one, and at large angles and
energy $k>0.35$ GeV is about a factor two larger in magnitude than the elastic one.  (iii)~As a
function of the scattering angle, the elastic TSSA has its maximum magnitude at increasing angles
for increasing energy, for both proton and neutron. The elastic asymmetries do not change sign,
while the inclusive ones do.

\section{Discussion}
\label{sec:discussion}
The TSSA for electron nucleon scattering was evaluated in the energy range below the second
resonance region employing a method based on he $1/N_c$ expansion. The method makes use of the
dynamical constraints that the large $N_c$ limit imposes in the baryon sector, which result in a
spin-flavor approximate symmetry broken by subleading corrections that are organized in a $1/N_c$
expansion. That symmetry in particular unifies the treatment of the nucleon and $\Delta$ resonance,
allowing for the systematic analysis carried out here that includes the first subleading terms in
the $1/N_c$ expansion.  The analysis gives results for the elastic and inelastic asymmetry, and also
provides details on the separate $N$ and $\Delta$ contributions in the absorptive part of the TPE
scattering amplitude.

It is found that form factors play a crucial role, in particular in enhancing the inelastic
asymmetry. The latter turns out to have, for CM scattering angles larger than $90^\circ$, opposite
sign to that of the elastic one and significantly larger in magnitude. For electron CM momenta below
0.5 GeV the TSSA is found to be in the range $10^{-3}-10^{-2}$.  If experiments in this energy
domain could be performed, measurements of the TSSA should be feasible.

Some comments are in order regarding the accuracy of the estimates based on the $1/N_c$ expansion.
At the NLO accuracy of the present calculation, all the structures in the baryon EM currents are
nonzero (magnetic and electric).  NNLO terms do not bring in any new structures but only modify the
coefficients of the NLO result.  The NNLO terms neglected in the calculation should therefore have
natural size, and their relative effect can be estimated as $\sim 1/N_c^2|_{N_c = 3} = 1/9$, times a
coefficient of order unity. Such an accuracy was observed in previous implementations of the $1/N_c$
expansion as a low-energy expansion with
$k = \ord{N_c^{-1}}$ \cite{CalleCordon:2012xz,Fernando:2019upo}. In the kinematic regions I and II,
where the $1/N_c$ expansion of the TSSA is implemented as a low-energy expansion, the accuracy of
the estimates should be of this order. In kinematic region III, where the $1/N_c$ expansion is
implemented with $k = \ord{N_c^0}$, the accuracy should be similar as long as the CM energy remains
below the $N^\ast$ threshold, $\sqrt{s} \lesssim$ 1.5 GeV or $k \lesssim$ 0.5 GeV.

When the energy rises above the threshold, the $N^\ast$'s can appear as an intermediate state in the
TPE amplitude.  Generically, the transition matrix elements of the EM current from ground state
baryons to higher resonances carry an additional suppression factor
$1/\sqrt{N_c}$ \cite{Goity:2004pw,Goity:2007ft,Scoccola:2007sn}.  The contribution of individual
$N^\ast$ resonances to the TPE amplitude are therefore suppressed by a factor $\ord{N_c^{-1}}$
relative to the leading order of the calculation in kinematic region III; however, they might be
numerically large.  The inclusion of $N^\ast$ intermediate states in the present calculation of the
TSSA in kinematic region III would be an interesting future extension of the present study.

The transition of the TSSA to the high-energy regime will involve the cumulative contributions of
many resonances, as intermediate states in the TPE amplitude and as final states in the cross
section. In this situation the $\ord{N_c^{-1}}$ suppression of individual $N^\ast$ contributions is
no longer effective, and the accounting changes.  It is expected that both the elastic and the
inclusive TSSA in this regime are generated by TPE amplitudes at the quark level.  Different
arguments have been put forward regarding the dominance of scattering from same quark or different
quarks.  The duality of the descriptions as cumulative resonance contributions and scattering from
quarks is an interesting theoretical problem. Measurements of the TSSA in the resonance region could
provide valuable material for further studies.

The cross section for inclusive $eN$ scattering at $\ord{\alpha^3}$ includes also real photon
emission into the final state, $e + N \rightarrow e' + \gamma + X'$. A TSSA can appear from the
interference of the amplitudes of real photon emission by the nucleon and by the electron --- the
so-called virtual Compton scattering and Bethe-Heitler processes (Fig.~\ref{fig:diagrams}c). It
requires that the amplitude of real photon emission from the nucleon have an imaginary
part \cite{Barut:1960zz}. In the low-energy regime considered here, this is possible if the
intermediate state is a $\Delta$. This contribution to the TSSA can be analyzed in the $1/N_c$
expansion approach in the same manner as the TPE contribution (Fig.~\ref{fig:diagrams}b).  In the
kinematic region III, where $k = \ord{N_c^0}$, parametric analysis shows that the real photon
emission contribution is suppressed at least by a factor $1/N_c$ compared to the TPE contribution.
This happens because the energy of the emitted photon in the CM frame is of the order of the
$N$--$\Delta$ mass difference $m_\Delta - m = \ord{N_c^{-1}}$; its momentum is therefore
$k_\gamma = \ord{N_c^{-1}}$; and its coupling to the dominant isovector magnetic vertex is
suppressed by $1/N_c$. Here the $1/N_c$ expansion reproduces the well-known result from ``soft
photons'' physics in QED, that such photons couple only to the charge of the colliding particles but
not to their spin \cite{Low:1958sn}. The calculation of real photon emission in kinematic region III
to leading non-vanishing order, and the extension of the above analysis to region II, remain
interesting problems for future study.

TPE also gives rise to a transverse beam spin asymmetry in $eN$ scattering. It is proportional to
the electron mass and expected to be several orders of magnitude smaller than the target spin
asymmetry \cite{Gorchtein:2004ac,Pasquini:2004pv,Afanasev:2004pu,Gorchtein:2005yz,Carlson:2017lys,Koshchii:2019mgv}.
The transverse beam spin asymmetry can be measured in electron scattering experiments with high beam
polarization quality as used for parity-violating scattering; it represents an important background
to the longitudinal beam spin asymmetry caused by weak interactions parity-violating electron
scattering. It could also be measured in $\mu N$ scattering, where it is enhanced by the muon mass
\cite{Koshchii:2019mgv}.  The $1/N_c$ expansion method developed here can be extended to calculate
the beam spin asymmetry in elastic or inclusive $eN$ scattering in the resonance region. Work on
this extension is in progress.

\section*{Acknowledgments}
This material is based upon work supported by the U.S.~Department of Energy, Office of Science,
Office of Nuclear Physics under contract DE-AC05-06OR23177 (JLG and CWe), by the National Science
Foundation, Grant Number PHY 1913562 (JLG) and by the Fonds de la Recherche Scientifique (FNRS)
(Belgium), Grant Number 4.45.10.08 (CWi).

\appendix
\section{$SU(4)$ algebra}
\label{app:su4}
This appendix summarizes properties of the $SU(4)$ spin-flavor symmetry group used in the present
analysis.  The algebra of $SU(4)$ contains fifteen generators: the spin generators $\hat{S}^i$, the
isospin generators $\hat{I}^a$, and the spin-flavor generators $\hat{G}^{ia}$, where $i$ and $a$ run
from 1 to 3. The commutation relations are
\begin{align}
& [\hat{S}^i, \hat{S}^j ] = i \eps^{ijk}\hat{S}^k, \hspace{1em}
[ \hat{I}^a, \hat{I}^b ] = i \eps^{abc} \hat{I}^c , \hspace{1em}
[ \hat{I}^a, \hat{S}^i ] = 0, &
\nonumber \\[1ex]    
& [\hat{S}^i, \hat{G}^{ja} ]=i  \eps^{ijk} \hat{G}^{ka}, \hspace{1em}
[\hat{I}^a, \hat{G}^{ib} ] = i \eps^{abc} \hat{G}^{ic}, &
\nonumber \\   
& [\hat{G}^{ia}, \hat{G}^{jb} ]
= \frac{i}{4}\delta^{ij}\eps^{abc}\hat{I}^c + \frac{i}{4} \delta^{ab} \eps^{ijk} \hat{S}^k. &
\label{eq:commutation-relations}
\end{align}
The ground state baryon states fill the $SU(4)$ representation formed by totally symmetric tensors
with $N_c$ indices.  These states have spin/isospin $S=I=\frac{1}{2}\cdots \frac{N_c}{2}$ and are
denoted by $|S S_3 I_3\rangle$.  The matrix elements of the $SU(4)$ generators in these states are
\begin{align}
&\langle S' S'_3 I'_3 | \hat{S}^i | S S_3 I_3\rangle
\nonumber \\
&= \sqrt{S(S + 1)} \delta_{S'S} \delta_{I_3' I_3} \langle S S_3, 1 i| S'S_3' \rangle ,
\\[1ex]
&\langle S' S'_3 I'_3 | \hat{I}^a | S S_3 I_3\rangle
\nonumber \\
&= \sqrt{S(S + 1)} \delta_{S'S} \delta_{I_3' I_3}
\langle S I_3, 1 a| S'I_3' \rangle ,
\\[1ex]
&\langle S' S'_3 I'_3 | \hat{G}^{ia} | S S_3 I_3\rangle
\nonumber \\
&=\frac{1}{4}
\sqrt{\frac{2S+1}{2S'+1}}\sqrt{(N_c+2)^2-(S-S')^2(S+S'+1)^2}\nonumber\\
 &\times\langle S S_3, 1 i | S' S'_3\rangle\langle S I_3, 1 a | S' I'_3\rangle .
\end{align}
$\hat{S}^i$ and $\hat{I}^a$ have matrix elements $\ord{N_c^0}$ and connect only states with $S' =
S$; $\hat{G}^{ia}$ have matrix elements $\ord{N_c}$ and can connect states with $S' = S$ or
$S\pm 1$.
 	
\section{Phase space integrals}
\label{app:integrals}
This appendix describes the phase space integrals arising in the calculation of the absorptive part
of the box diagram in Eqs.~(\ref{interfterm}) and (\ref{interfXsection}). Giving explicit analytic
results is important because individual integrals present infrared and/or collinear divergencies due
to the photon propagators in the box, which cancel in the final result. The divergencies of the
individual integrals are regulated by including an infinitesimal photon mass, provided by the
regulator $\eps=0^+$ below.
 
\subsection{Integrals without form factors}
The first set of integrals is for the case where no form factors are included. In the following the
unit vector ${\hat {\bm K}}$ is the integration variable [the intermediate electron direction
${\hat {\bm k}}_{\rm n}$ in Eq.~(\ref{interfterm})], and ${\hat {\bm k}}$ and ${\hat {\bm k}}'$ are
external unit vectors on which the integral depends [the initial/final electron direction
${\hat {\bm k}}_{\rm f, i}$ in Eq.~(\ref{interfterm})].  The integrals with a single denominator
arising in
the calculation are:
\begin{align}
J(n)&=\int d\Omega_{K} \frac{( \bm{\hat   k}\cdot \bm{\hat   K})^n}{1- \bm{\hat k}
\cdot \bm{\hat   K}+\eps}
\nonumber \\
&=-2\pi\left(\log \frac{\eps}{2} + 2\sum_{m=0}^{[\frac{n-1}{2}]}\frac{1}{2m+1}\right),
\nonumber\\[2ex]
J^i( \bm{\hat   k},n)&=\int d\Omega_{K}
\frac{ \bm{\hat   K}^i\hat ( \bm{\hat   k}\cdot \bm{\hat   K})^n}
{1 - \bm{\hat k}\cdot \bm{\hat K} + \eps}
\nonumber\\[1ex]
&= \bm{\hat k}^i J(n+1),
\nonumber\\[2ex]
J^{ij}( \bm{\hat k},n)&=\int d\Omega_{K}
\frac{ \bm{\hat K}^i  \bm{\hat K}^j\hat ( \bm{\hat k}\cdot \bm{\hat K})^n}
{1- \bm{\hat k}\cdot \bm{\hat K}+\eps}
\nonumber\\
&=\frac{1}{2} \left\{ \delta^{ij} [J(n)-J(n+2)] \phantom{{\hat k}^i} \right.
\nonumber\\
& \left. \hspace{1em} + \; \bm{\hat   k}^i \bm{\hat   k}^j [3 J(n+2)-J(n)] \right\} .
\label{masterint1}
\end{align} 
The integrals with a double denominator are:
\begin{align}
&J( \bm{\hat k}, \bm{\hat k}', n, n')
\nonumber\\
&=\int d\Omega_{K}
\frac{( \bm{\hat   k}\cdot \bm{\hat K})^n( \bm{\hat k}'\cdot \bm{\hat K})^{n'}}
{(1- \bm{\hat k}\cdot \bm{\hat K} + \eps)(1- \bm{\hat k}'\cdot \bm{\hat K} + \eps)} ,
\nonumber\\[2ex]
&J^i( \bm{\hat k}, \bm{\hat k}', n, n')
\nonumber\\
&= \int d\Omega_{K}\frac{ \bm{\hat K}^i( \bm{\hat k}\cdot \bm{\hat K})^n
(\bm{\hat k}'\cdot \bm{\hat K})^{n'}}
{(1- \bm{\hat k}\cdot \bm{\hat K} + \eps)(1- \bm{\hat   k}'\cdot \bm{\hat K} + \eps)} ,
\nonumber\\[2ex]
&J^{ij}( \bm{\hat k}, \bm{\hat k}', n, n')
\nonumber\\
&=\int d\Omega_{K} \frac{ \bm{\hat K}^i \bm{\hat K}^j( \bm{\hat k}\cdot \bm{\hat K})^n
(\bm{\hat k}'\cdot \bm{\hat K})^{n'}}
{(1 - \bm{\hat k}\cdot \bm{\hat K}+\eps)(1 - \bm{\hat k}'\cdot \bm{\hat   K}+\eps)} .
\label{masterint2}
\end{align} 
The results for these integrals as needed in the present work are given in the following tables.
Here $ \bm{\hat k}\cdot \bm{\hat k}'=\cos\theta$, where $\theta$ is the scattering angle.
\begin{align} 	
\renewcommand{\arraystretch}{1.75}
\begin{array}{cc|l}
n & n' &  J( \bm{\hat   k}, \bm{\hat   k}',n,n')
\\[1ex]
\hline
0 &0 & \displaystyle \frac{4\pi}{1-\cos\theta}
\left( \log\sin^2\frac{\theta}{2} - \log\frac{\eps}{2} \right)
\\[1ex]
1 &0 &  J( \bm{\hat   k}, \bm{\hat   k}',0,0) - J(0)
\\[1ex]
1 &1 & J( \bm{\hat   k}, \bm{\hat   k}',1,0) - J(1)
\\[1ex]
2 & 0 & \bm{\hat   k}^i J^i( \bm{\hat   k}, \bm{\hat   k}',0,0) - \cos\theta\; J(1)
\end{array}
\nonumber
\end{align}

\begin{align} 	
\renewcommand{\arraystretch}{1.75}
\begin{array}{cc|l}
n & n' & J^i( \bm{\hat   k}, \bm{\hat   k}',n,n')
\\[1ex]
\hline
0 & 0 & \displaystyle
-2\pi\frac{ \bm{\hat k}^i+ \bm{\hat k}'^i}{\sin^2\theta}
\left[ (1+\cos\theta)\log\frac{\eps}{2}-2\;\log\sin^2\frac{\theta}{2} \right]
\\[1ex]
1 & 0 & J^i (\bm{\hat k}, \bm{\hat k}', 0, 0) - \bm{\hat k}'^i J(1)
\\[1ex]
1 & 1 & J^i (\bm{\hat k}, \bm{\hat k}', 0, 0) - (\bm{\hat k}^i + \bm{\hat k}'^i) J(1)
\\[1ex]
2 & 0 & J^i (\bm{\hat k}, \bm{\hat k}', 1, 0) - \bm{\hat k}^j J^{ij} (\bm{\hat k}', 0)
\end{array}
\nonumber
\end{align}

\begin{align} 	
\renewcommand{\arraystretch}{1.75}
\begin{array}{cc|l}
n & n' & J^{ij}( \bm{\hat   k}, \bm{\hat   k}',n,n') 
\\[1ex]
\hline
0 & 0  &  \displaystyle
\frac{2\pi}{ \cos^2\frac{\theta}{2} \sin^2\theta}
\left[ - \delta^{ij} \sin^2\theta \;\log\sin^2\frac{\theta}{2} \right.
\\[1ex]
& & \displaystyle +2( \bm{\hat k}^i \bm{\hat k}^j + \bm{\hat k}'^i \bm{\hat k}'^j)
\left( \log\sin^2\frac{\theta}{2}-\cos\theta\cos^2\frac{\theta}{2} \right.
\\[1ex]
& & \hspace{1em} \left. \displaystyle -\cos^4\frac{\theta}{2}\;\log\frac{\eps}{2} \right) \\ 
  & & \displaystyle \left. +2 ( \bm{\hat k}^i \bm{\hat k}'^j+ \bm{\hat k}^j \bm{\hat k}'^i)
  \left( \cos^2\frac{\theta}{2} + \sin^2\frac{\theta}{2}\;\log\sin^2\frac{\theta}{2}\right) \right]
\\[2ex]
1 & 0 & J^{ij}( \bm{\hat k}, \bm{\hat k}', 0, 0) - J^{ij}( \bm{\hat k}', 0)
\end{array}
\end{align}
\subsection{Integrals with form factors}
The second set of integrals is for the case where form factors are included. The choice is a common
form factor for all components of the current with the dipole form,
\begin{align}
F(t) = \frac{\Lambda_{\rm EM}^4}{(\Lambda_{\rm EM}^2-t)^2} .
\end{align}
The integrals can be given analytically, rendering very large expressions. They are obtained through
the following steps.  One first expresses the form factor as the derivative of a monopole,
\begin{align}
F(t)= \left. -\Lambda_{\rm EM}^4\frac{\partial}{\partial a}
\frac{1}{a-t}\right|_{a\to\Lambda_{\rm EM}^2}.
\end{align}
The momentum transfer at the EM vertices in the box diagram are
$t = -2 k K (1 - \bm{\hat k}\cdot \bm{\hat K})$ and
$t' = -2 k' K (1 - \bm{\hat k}'\cdot \bm{\hat K})$, where $k$ and $K$, and $k'$ and $K$, are the
moduli of the electron 3-momenta entering in the respective vertices. The box integrals involving
the form factors are of the general form, where Pol indicates polynomial in the arguments:
\begin{widetext}
\begin{align}
&\int d\Omega_{K} \frac{\Lambda_{\rm EM}^8
\text{Pol} (\bm{\hat k} \cdot \bm{\hat K}, \bm{\hat k}'\cdot \bm{\hat K}, \bm{\hat K}^i)}
{(1 - \bm{\hat k}\cdot \bm{\hat K}+\eps)
[\Lambda_{\rm EM}^2+2 k K(1- \bm{\hat k}\cdot \bm{\hat K} )]^2
(1 - \bm{\hat k}'\cdot \bm{\hat K})
[\Lambda_{\rm EM}^2+2 k' K(1 - \bm{\hat k}'\cdot \bm{\hat K})]^2}
\\[2ex]
= \; & \frac{\Lambda_{\rm EM}^8}{(4 k k' K^2)^2}\frac{\partial}{\partial a}
\frac{\partial}{\partial a'}\int d\Omega_{K}
\left.
\frac{\text{Pol}( \bm{\hat k}\cdot \bm{\hat K}, \bm{\hat k}'\cdot \bm{\hat K}, \bm{\hat K}^i)}
{(1 - \bm{\hat k}\cdot \bm{\hat K}+\eps) (1 - \bm{\hat k}'\cdot \bm{\hat K}+\eps)
(a - \bm{\hat k}\cdot \bm{\hat K}) (a' - \bm{\hat k}'\cdot \bm{\hat K})} \, 
\right|_{a^{(')}\rightarrow 1 + \frac{\Lambda_{\rm EM}^2}{2k^{(')} K}} .
\nonumber 
\end{align}
\end{widetext}
Using partial fractions
\begin{align}
&\frac{1}{(1 - \bm{\hat k}\cdot \bm{\hat K} + \eps) (a - \bm{\hat k}\cdot \bm{\hat K})}
\nonumber \\
&= \frac{1}{a-1}\left(\frac{1}{1 - \bm{\hat k}\cdot \bm{\hat K} + \eps}
+ \frac{1}{a - \bm{\hat k}\cdot \bm{\hat K}}\right)
\end{align}
reduces the integrals to be calculated to the general form
\begin{align}
\int d\Omega_{K}
\frac{\text{Pol}( \bm{\hat k}\cdot \bm{\hat K}, \bm{\hat k}'\cdot \bm{\hat K}, \bm{\hat K}^i)}
{(a - \bm{\hat k}\cdot \bm{\hat K} + \eps)(a' - \bm{\hat k}'\cdot \bm{\hat K} + \eps)},
\end{align}
where $a, a' \geq1$. By expanding the numerator these integrals can be reduced to integrals with
single or double denominators.  The integrals with single denominators are:
\begin{align}
& {\bm J}(a,n)
\nonumber \\
&= \int d\Omega_{K} \frac{( \bm{\hat k}\cdot \bm{\hat K})^n}
{a - \bm{\hat k}\cdot \bm{\hat K} + \eps}
\nonumber \\
&= 2\pi\left( a^n \log(\frac{a+1}{a-1+\eps})
- 2\sum_{m=0}^{[\frac{n-1}{2}]} \frac{a^{n-1-2m}}{2m+1} \right),
\nonumber \\[2ex]
&{\bm J}^i( \bm{\hat k}, a, n)
\nonumber \\
&= \int d\Omega_{K} \frac{\bm{\hat K}^i (\bm{\hat k}\cdot \bm{\hat K})^n}
{a - \bm{\hat k}\cdot \bm{\hat K} + \eps}
\nonumber \\[2ex]
&= \bm{\hat k}^i {\bm J}(a, n),
\end{align}
\begin{align}
& {\bm J}^{ij}( \bm{\hat k}, a, n)
\nonumber \\
&= \int d\Omega_{K}
\frac{ \bm{\hat K}^i \bm{\hat K}^j (\bm{\hat k}\cdot \bm{\hat K})^n}
{a - \bm{\hat k}\cdot \bm{\hat K} + \eps}
\nonumber \\[1ex]
&= \frac{1}{2} \left\{ \delta^{ij} \left[ {\bm J}(a,n) - {\bm J}(a, n+2) \right]
\phantom{\hat{k}^i} \right. 
\nonumber \\[1ex]
& \left. + \; \bm{\hat k}^i \bm{\hat k}^j \left[ 3 {\bm J}(a, n+2) - {\bm J}(a, n) \right] \right\} ,
\label{masterint3}
\end{align}
The integrals with double denominators are:
\begin{align}
& {\bm J}( \bm{\hat k}, \bm{\hat k}', a, a', n, n')
\nonumber \\
&=\int d\Omega_{K}
\frac{( \bm{\hat k}\cdot \bm{\hat K})^n( \bm{\hat k}'\cdot \bm{\hat K})^{n'}}
{(a - \bm{\hat k}\cdot \bm{\hat K} + \eps)(a' - \bm{\hat k}'\cdot \bm{\hat K} + \eps)} ,
\end{align}
\begin{align}
& {\bm J}^i( \bm{\hat k}, \bm{\hat k}', a, a', n, n')
\nonumber \\
&= \int d\Omega_{K}
\frac{ \bm{\hat K}^i (\bm{\hat k}\cdot \bm{\hat K})^n (\bm{\hat k}'\cdot \bm{\hat K})^{n'}}
{(a - \bm{\hat k}\cdot \bm{\hat K} + \eps)(a' - \bm{\hat k}'\cdot \bm{\hat K} + \eps)} ,
\end{align}
\begin{align}
& {\bm J}^{ij}( \bm{\hat   k}, \bm{\hat   k}',a,a',n,n')
\nonumber \\
&=\int d\Omega_{K}
\frac{\bm{\hat K}^i \bm{\hat K}^j (\bm{\hat k}\cdot \bm{\hat K})^n
(\bm{\hat k}'\cdot \bm{\hat K})^{n'}}
{(a - \bm{\hat k}\cdot \bm{\hat K} + \eps)(a' - \bm{\hat k}'\cdot \bm{\hat K} + \eps)} .
\label{masterint4}
\end{align} 

Explicit results are shown in the following tables, where
$A(a, a', \theta) \equiv \sqrt{a^2 a'^2 - 2 a a' \cos\theta-\sin ^2\theta}$:
\begin{widetext}
\begin{align}
\renewcommand{\arraystretch}{1.75}
\begin{array}{cc|l}
n & n' & {\bm J}( \bm{\hat   k}, \bm{\hat   k}',a,a',n,n')
\\
\hline
0 & 0 & \displaystyle \frac{2\pi}{A(a, a', \theta)}
\left( \log\frac{2\sin^2\frac{\theta}{2} - a(a - a') + A(a, a', \theta)}
{2\sin^2\frac{\theta}{2} - a(a - a') - A(a, a', \theta)} +a \leftrightarrow a' \right)
\\[3ex]
1 & 0 & a\,{\bm J}( \bm{\hat k}, \bm{\hat k}', a, a', 0, 0) - {\bm J}(a', 0)
\\[2ex]
1 & 1 & 4\pi + a a' \,{\bm J}( \bm{\hat k}, \bm{\hat k}', a, a', 0, 0)
- a \,{\bm J}(a, 0) - a'\,{\bm J}(a', 0)
\\[2ex]
2 & 0 & a  \, {\bm J}( \bm{\hat k}, \bm{\hat k}', a, a', 1, 0) - \cos\theta\; {\bm J}
(a',0)
\end{array}
\end{align}
\begin{align}
& {\bm J}^i (\bm{\hat k}, \bm{\hat k}', a, a', n, n')
\nonumber \\
&= -\csc^2 \theta \; \left[ (\cos \theta \; \bm{\hat k}^i
- \bm{\hat k}^{\prime i}) {\bm J}( \bm{\hat k}, \bm{\hat k}', a, a', n, n' + 1)
+ \; ( \cos \theta \; \bm{\hat k}^{\prime i} - \bm{\hat k}^i ) {\bm J}(k, k', a, a', n + 1, n')
\right] ,
\\[2ex]
& {\bm J}^{ij} (\bm{\hat k}, \bm{\hat k}', a, a', n, n')
\nonumber \\
& = \delta^{ij} \left\{ {\bm J} (\bm{\hat k}, \bm{\hat k}', a, a', n, n') \right.
\nonumber \\
& - \csc^2\theta \; \left[
- 2 \cos\theta\; {\bm J} (\bm{\hat k}, \bm{\hat k}', a, a', n + 1, n' + 1) 
+  \left. {\bm J}( \bm{\hat k}, \bm{\hat k}', a, a', n, n'+2)
+ {\bm J}(\bm{\hat k}, \bm{\hat k}', a, a', n+2, n') \right] \right\}
\nonumber\\
&+\csc^4\theta\; \left\{ \bm{\hat k}^i  \bm{\hat k}^j
\left[ -\sin^2\theta\; {\bm J}( \bm{\hat k}, \bm{\hat k}', a, a', n, n')
+ (\cos^2\theta +1) {\bm J}( \bm{\hat k}, \bm{\hat k}', a, a', n, n' + 2) \right. \right.
\nonumber\\
&\left.\left. -4 \cos\theta\; {\bm J}( \bm{\hat k}, \bm{\hat k}', a, a', n + 1, n' + 1)
+ 2 {\bm J}( \bm{\hat k}, \bm{\hat k}', a, a', n+2, n') \right]
+ \{ k, a, n \} \leftrightarrow \{ k', a',n' \} \right\}
\nonumber\\
& +\csc^4 \theta \;
\left( \bm{\hat k}^{\prime i}  \bm{\hat k}^j + \bm{\hat k}^i \bm{\hat k}^{\prime j} \right)
\left\{ (3 \cos^2\theta + 1 ) {\bm J}( \bm{\hat k}, \bm{\hat k}', a, a', n + 1, n' + 1)
\right.
+\cos\theta\; \sin ^2\theta \; {\bm J}( \bm{\hat k}, \bm{\hat k}', a, a', n, n')
\nonumber\\
&\left. -2 \cos\theta \left[ {\bm J}( \bm{\hat k}, \bm{\hat k}', a, a', n, n' + 2) +
{\bm J}( \bm{\hat k}, \bm{\hat k}', a, a', n + 2, n') \right] \right\} .
\end{align}
\vspace{10ex}
\end{widetext}
\section{Spin-independent OPE cross section}
\label{app:cross_section}
This appendix presents the explicit expression of the OPE cross section for elastic scattering
$eN \rightarrow eN$, which is used as denominator in the calculation of the TSSA in this work. The
OPE cross section is given by\footnote{This is the cross section in terms of the form factors
$G_E=F_1$ and $G_M=F_1+F_2$.}
\begin{align}
\frac{d\sigma_U}{d\Omega} &= \frac{\alpha ^2 }{4 m_N^2 s t^2}
\left\{ \left[ G_E^2 \left(4 m_N^2-t\right) +\frac{2}{\Lambda} {G_E} G_M m_N t\right]
\right.
\nonumber \\[1ex]
& \times \left[ (s-m_N^2)^2+s t\right]
\nonumber\\
&-\left. \frac{1}{\Lambda^2}G_M^2 m_N^2 t \left[ (s-m_N^2)^2+s t-2 m_N^2 t\right]\right\} ,
\label{OPEXsection}
\end{align}
where $\Lambda$ is the generic mass scale accompanying the magnetic form factors introduced in
Sec.~\ref{subsec:ncexpansion}.  When used in the context of the $1/N_c$ expansion,
Eq.~(\ref{OPEXsection}) is expanded to the corresponding order in the non-relativistic expansion The
cross section for general $N_c$ is obtained by replacing in Eq.~(\ref{OPEXsection})
\begin{align}
G_E &\equiv G_E(I_3) = G_E^S + 2 I_3 G_E^V,
\nonumber
\\[1ex]
G_M &\equiv G_M(I_3) = G_M^S + \frac{2}{5} I_3 (N_c+2) G_M^V,
\label{formfactors_isospin}
\end{align}
where $G_E^S$, etc. are the isoscalar and isovector physical form factors,
Eq.~(\ref{formfactor_physical}), and $I_3 = \pm\frac{1}{2}$ is the isospin projection of the initial
nucleon state.  Then, the strict $1/N_c$ expansion is performed with the scaling assignments
$m_N=\ord{N_c}$, $\Lambda=\ord{N_c^0}$, $\sqrt{s}-m_N=\ord{N_c^0}$, and $t=\ord{N_c^0}$.
\section{Spin-dependent TPE cross section}
\label{app:sigma_N}
This appendix presents the results for the spin-dependent interference cross section the CM frame,
Eq.~(\ref{interfXsection}), for the case where the form factors and the $\Delta$ width are ignored.
The expressions display separately the contributions where the final and intermediate baryon states
are $N$ or $\Delta$. The notation $(N_{\rm i},B_{\rm f},B_{\rm n})$ indicates the initial nucleon
$N_{\rm i} = p, n$ with isospin projection $I_3 = \pm \frac{1}{2}$, the final baryon
$B_{\rm f} = N, \Delta$ and the intermediate baryon $B_{\rm n} = N, \Delta$.
\begin{widetext}
\begin{align}
\frac{d\sigma_N}{d\Omega}  (N_{\rm i},N,N) &= \frac{ \alpha ^3 k^2 m_N^3 }{4000\, \Lambda ^3\,
s^{3/2}\, t \,(1+x)} \left[ 2 (1+x)-(x+3) \log  \frac{1-x}{2}\right]
\nonumber\\[1ex]
&\times
\left[ (N_c-3) G_M(-I_3)-(N_c+7) G_M(I_3)\right]^2
\nonumber\\[2ex]
&\times \left\{ 10
\Lambda  G_E(I_3) +k \left[ (N_c-3)
G_M(-I_3)-(N_c+7) G_M(I_3) \right]\right\} ,
\nonumber
\end{align}
\begin{align}
\frac{d\sigma_N}{d\Omega}  (N_{\rm i},N,\Delta)
&= \frac{\Theta (k_\Delta)\alpha ^3 m_N^2 m_\Delta }{2000 \,\Lambda ^3\, s^{3/2}\, t\, (1+x)}
(N_c-1) (N_c+5) \left[ G_M(-I_3)-G_M(I_3) \right]^2
\nonumber\\
&\times \left\{ 2
k k_\Delta\, (1+x)-\log  \frac{1-x}{2}\;\, \left[ k^2 (1+x)+2 k_\Delta^2 \right] \right\}
\nonumber\\[1ex]
&\times \left\{ k \left[ (N_c-3)
G_M(-I_3)-(N_c+7) G_M(I_3) \right] -5 \Lambda G_E(I_3) \right\} ,
\nonumber
\end{align}
\begin{align}
\frac{d\sigma_N}{d\Omega}  (N_{\rm i},\Delta,N)
&= \frac{\Theta (k_\Delta)\alpha ^3 k_\Delta m_N^2 m_\Delta }
{16000\, \Lambda ^3 \,s^{3/2}\, t \,(1+x)}
(N_c-1) (N_c+5) \left[ G_M(-I_3)-G_M(I_3)\right]^2
\nonumber\\
&\times\left( 2 \log \frac{1-x}{2}\;\, \left\{ 20 \Lambda  G_E(I_3)
\left[ 2 k-k_\Delta (1+x) \right]   \phantom{0^0} \right. \right.
\nonumber\\[1ex]
&\left. - \left[ (N_c-3) G_M(-I_3) -(N_c+7) G_M(I_3)\right]
\left[ 2 k^2-3 k k_\Delta\, (x-1)-2  k_\Delta ^2 (x-2)\right] \right\}
\nonumber\\[1ex]
&-(1+x) \left\{ \left( 11 k^2-k k_\Delta\,+4 k_\Delta^2\right)
\left[ (N_c-3) G_M(-I_3)-(N_c+7) G_M(I_3) \right] \right.
\nonumber\\[0ex]
&\left.\left. -40 \,\Lambda \, G_E(I_3) (k- k_\Delta ) \right\}
\phantom{\frac{0}{0}}\hspace{-1em} \right) ,
\nonumber
\end{align}
\begin{align}
\frac{d\sigma_N}{d\Omega}  (N_{\rm i},\Delta,\Delta)
&= \frac{\Theta (k_\Delta)\alpha ^3 k_\Delta^2 m_N^2 m_\Delta }
{80000\, k \,\Lambda ^3\, s^{3/2} \,t \,(1+x)}
(N_c-1) (N_c+5) \left[ G_M(-I_3)-G_M(I_3)\right]^2
\nonumber\\
&\times
\left( 200 \,\Lambda \, G_E(I_3) \left\{ (1+x) (k - k_\Delta)+\log \frac{1-x}{2}\;
\left[ k (1+x)-2 k_\Delta\right] \right\} \right.
\nonumber\\
&+ \left[ (N_c-23) G_M(-I_3) - (N_c+27) G_M(I_3) \right] 
\nonumber\\
&\times \left. \left\{ 2 \log \frac{1-x}{2}  \left[ -6 k^2+k k_\Delta\, (5 x+3)-6 k_\Delta^2\right]
+ k_\Delta (1+x) (9 k-23 k_\Delta ) \right\} \right)  .
\label{dsigmaSSA}
\end{align}
\end{widetext}
Here $x \equiv \cos\theta$; $k$ and $k_\Delta$ are the CM momenta in $eN$ and $e\Delta$ states given
by Eq.~(\ref{invariants_cm}), and $t = -2 k_{\rm i} k_{\rm f} (1-x)$ in each of the above
expressions, with $k_{\rm i} = k$ and $k_{\rm f} = k$ or $k_\Delta$ depending on the final baryon
state.  $G_{M,E}(I_3)$ are the form factors for initial nucleon isospin projection $I_3$ given by
Eq.~(\ref{formfactors_isospin}).  Since the above expressions are for the case where the momentum
dependence of the form factors is neglected, $G_E^p=1$ and $G_E^n=0$.

The expansion in $1/N_c$ of the cross sections Eq.~(\ref{dsigmaSSA}) is easily performed using the
scaling of the masses as $m_N=\ord{N_c}$, $m_\Delta-m_N=\ord{N_c^{-1}}$, and $\Lambda=\ord{N_c^0}$
(to be chosen equal to the physical nucleon mass).  For the scaling of the CM momentum $k$ there are
the three distinguished regimes described in Sec.~\ref{subsec:kinematic}: the low-energy elastic
regime, where the energy is below the $\Delta$ threshold and $k = \ord{N_c^{-1}}$; the low-energy
inelastic regime, where the energy is above the $\Delta$ threshold and $k = \ord{N_c^{-1}}$; and the
intermediate-energy inelastic regime where $k=\ord{N_c^0}$. The expressions Eq.~(\ref{dsigmaSSA})
cover all three regimes and can be expanded further with the appropriate scaling assignment for the
momenta in each regime.
\section{Implementation of $\Delta$ width} 
\label{app:delta_width}
This appendix describes the implementation of the $\Delta$ width in the context of the present
approach based on the $1/N_c$ expansion. The decay width of the $\Delta$ is a quantity
$\ord{N_c^{-2}}$, it however plays an important role in the shape of the asymmetry as the electron
energy is near te excitation energy of the $\Delta$. A Breit-Wigner form is used, which leads to the
following convolution (smearing) in the calculation of the absorptive part of the box diagram. At
vanishing width the integrals in the absorptive part are of the general form:
\begin{align}
&  \int \frac{d^4K}{(2\pi)^4} \delta^+(K^2) \delta(p^0+q^0-\Delta m) \frac{\text{Pol}(K)}{q^2 q'^2} ,
\end{align}
where $q$, $q'$ are the photon momenta in the box, and $\Delta m$ the $N$-$\Delta$ mass difference.
With finite width $\Gamma$ the corresponding integral becomes:
\begin{align}
& \frac{1}{4 \arctan\frac{2Q}{\Gamma}} \int_{-Q}^Q d\mu \frac{\Gamma}{\mu^2+\frac{\Gamma^2}{4}}
\nonumber \\
\times & \int \frac{d^4K}{(2\pi)^4} \delta^+(K^2)  \delta(p^0+q^0-(\Delta m-\mu))
\frac{\text{Pol}(K)}{q^2 q'^2} .
\label{smearing}
\end{align}
The domain of integration in $\mu$ must be limited by the scale $Q$, as otherwise the result
diverges for large $|\mu|$. It is also logical that $|\mu|<\Delta m$. Results have little
sensitivity to $Q$ as far as $\Gamma< Q <\Delta m$.  The factor in front of the above expression
provides the proper normalization for the convolution.
 
The end result is that the implementation of the $\Delta$ width in the interference cross section is
simply given by a smearing of the cross section as follows:
\begin{align}
& d\sigma_N(N,B_{\rm f},B_{\rm n})\to
\nonumber \\
& \frac{1}{4 \arctan\frac{2Q}{\Gamma}} \int_{-Q}^Q d\mu \frac{\Gamma}{\mu^2+\frac{\Gamma^2}{4}}
d\sigma_N(N,B_{\rm f},B_{\rm n}) |_{m_\Delta\to m_\Delta-\mu} .
 \end{align}
\bibliography{ssa_largen.bib}{}

\begin{thebibliography}{48}%
\makeatletter
\providecommand \@ifxundefined [1]{%
 \@ifx{#1\undefined}
}%
\providecommand \@ifnum [1]{%
 \ifnum #1\expandafter \@firstoftwo
 \else \expandafter \@secondoftwo
 \fi
}%
\providecommand \@ifx [1]{%
 \ifx #1\expandafter \@firstoftwo
 \else \expandafter \@secondoftwo
 \fi
}%
\providecommand \natexlab [1]{#1}%
\providecommand \enquote  [1]{``#1''}%
\providecommand \bibnamefont  [1]{#1}%
\providecommand \bibfnamefont [1]{#1}%
\providecommand \citenamefont [1]{#1}%
\providecommand \href@noop [0]{\@secondoftwo}%
\providecommand \href [0]{\begingroup \@sanitize@url \@href}%
\providecommand \@href[1]{\@@startlink{#1}\@@href}%
\providecommand \@@href[1]{\endgroup#1\@@endlink}%
\providecommand \@sanitize@url [0]{\catcode `\\12\catcode `\$12\catcode
  `\&12\catcode `\#12\catcode `\^12\catcode `\_12\catcode `\%12\relax}%
\providecommand \@@startlink[1]{}%
\providecommand \@@endlink[0]{}%
\providecommand \url  [0]{\begingroup\@sanitize@url \@url }%
\providecommand \@url [1]{\endgroup\@href {#1}{\urlprefix }}%
\providecommand \urlprefix  [0]{URL }%
\providecommand \Eprint [0]{\href }%
\providecommand \doibase [0]{https://doi.org/}%
\providecommand \selectlanguage [0]{\@gobble}%
\providecommand \bibinfo  [0]{\@secondoftwo}%
\providecommand \bibfield  [0]{\@secondoftwo}%
\providecommand \translation [1]{[#1]}%
\providecommand \BibitemOpen [0]{}%
\providecommand \bibitemStop [0]{}%
\providecommand \bibitemNoStop [0]{.\EOS\space}%
\providecommand \EOS [0]{\spacefactor3000\relax}%
\providecommand \BibitemShut  [1]{\csname bibitem#1\endcsname}%
\let\auto@bib@innerbib\@empty
\bibitem [{\citenamefont {Jones}\ \emph {et~al.}(2000)\citenamefont {Jones}
  \emph {et~al.}}]{JeffersonLabHallA:1999epl}%
  \BibitemOpen
  \bibfield  {author} {\bibinfo {author} {\bibfnamefont {M.~K.}\ \bibnamefont
  {Jones}} \emph {et~al.} (\bibinfo {collaboration} {Jefferson Lab Hall A}),\
  }\bibfield  {title} {\bibinfo {title} {{$G_{Ep} / G_{Mp}$ ratio by
  polarization transfer in $\vec{e} p \rightarrow e \vec{p}$}},\ }\href
  {https://doi.org/10.1103/PhysRevLett.84.1398} {\bibfield  {journal} {\bibinfo
   {journal} {Phys. Rev. Lett.}\ }\textbf {\bibinfo {volume} {84}},\ \bibinfo
  {pages} {1398} (\bibinfo {year} {2000})},\ \Eprint
  {https://arxiv.org/abs/nucl-ex/9910005} {arXiv:nucl-ex/9910005} \BibitemShut
  {NoStop}%
\bibitem [{\citenamefont {Guichon}\ and\ \citenamefont
  {Vanderhaeghen}(2003)}]{Guichon:2003qm}%
  \BibitemOpen
  \bibfield  {author} {\bibinfo {author} {\bibfnamefont {P.~A.~M.}\
  \bibnamefont {Guichon}}\ and\ \bibinfo {author} {\bibfnamefont
  {M.}~\bibnamefont {Vanderhaeghen}},\ }\bibfield  {title} {\bibinfo {title}
  {{How to reconcile the Rosenbluth and the polarization transfer method in the
  measurement of the proton form-factors}},\ }\href
  {https://doi.org/10.1103/PhysRevLett.91.142303} {\bibfield  {journal}
  {\bibinfo  {journal} {Phys. Rev. Lett.}\ }\textbf {\bibinfo {volume} {91}},\
  \bibinfo {pages} {142303} (\bibinfo {year} {2003})},\ \Eprint
  {https://arxiv.org/abs/hep-ph/0306007} {arXiv:hep-ph/0306007} \BibitemShut
  {NoStop}%
\bibitem [{\citenamefont {Blunden}\ \emph {et~al.}(2003)\citenamefont
  {Blunden}, \citenamefont {Melnitchouk},\ and\ \citenamefont
  {Tjon}}]{Blunden:2003sp}%
  \BibitemOpen
  \bibfield  {author} {\bibinfo {author} {\bibfnamefont {P.~G.}\ \bibnamefont
  {Blunden}}, \bibinfo {author} {\bibfnamefont {W.}~\bibnamefont
  {Melnitchouk}},\ and\ \bibinfo {author} {\bibfnamefont {J.~A.}\ \bibnamefont
  {Tjon}},\ }\bibfield  {title} {\bibinfo {title} {{Two photon exchange and
  elastic electron proton scattering}},\ }\href
  {https://doi.org/10.1103/PhysRevLett.91.142304} {\bibfield  {journal}
  {\bibinfo  {journal} {Phys. Rev. Lett.}\ }\textbf {\bibinfo {volume} {91}},\
  \bibinfo {pages} {142304} (\bibinfo {year} {2003})},\ \Eprint
  {https://arxiv.org/abs/nucl-th/0306076} {arXiv:nucl-th/0306076} \BibitemShut
  {NoStop}%
\bibitem [{\citenamefont {Henderson}\ \emph {et~al.}(2017)\citenamefont
  {Henderson} \emph {et~al.}}]{OLYMPUS:2016gso}%
  \BibitemOpen
  \bibfield  {author} {\bibinfo {author} {\bibfnamefont {B.~S.}\ \bibnamefont
  {Henderson}} \emph {et~al.} (\bibinfo {collaboration} {OLYMPUS}),\ }\bibfield
   {title} {\bibinfo {title} {{Hard Two-Photon Contribution to Elastic
  Lepton-Proton Scattering Determined by the OLYMPUS Experiment}},\ }\href
  {https://doi.org/10.1103/PhysRevLett.118.092501} {\bibfield  {journal}
  {\bibinfo  {journal} {Phys. Rev. Lett.}\ }\textbf {\bibinfo {volume} {118}},\
  \bibinfo {pages} {092501} (\bibinfo {year} {2017})},\ \Eprint
  {https://arxiv.org/abs/1611.04685} {arXiv:1611.04685 [nucl-ex]} \BibitemShut
  {NoStop}%
\bibitem [{\citenamefont {Bernauer}\ \emph {et~al.}(2021)\citenamefont
  {Bernauer} \emph {et~al.}}]{OLYMPUS:2020dgl}%
  \BibitemOpen
  \bibfield  {author} {\bibinfo {author} {\bibfnamefont {J.~C.}\ \bibnamefont
  {Bernauer}} \emph {et~al.} (\bibinfo {collaboration} {OLYMPUS}),\ }\bibfield
  {title} {\bibinfo {title} {{Measurement of the Charge-Averaged Elastic
  Lepton-Proton Scattering Cross Section by the OLYMPUS Experiment}},\ }\href
  {https://doi.org/10.1103/PhysRevLett.126.162501} {\bibfield  {journal}
  {\bibinfo  {journal} {Phys. Rev. Lett.}\ }\textbf {\bibinfo {volume} {126}},\
  \bibinfo {pages} {162501} (\bibinfo {year} {2021})},\ \Eprint
  {https://arxiv.org/abs/2008.05349} {arXiv:2008.05349 [nucl-ex]} \BibitemShut
  {NoStop}%
\bibitem [{\citenamefont {Accardi}\ \emph {et~al.}(2021)\citenamefont {Accardi}
  \emph {et~al.}}]{Accardi:2020swt}%
  \BibitemOpen
  \bibfield  {author} {\bibinfo {author} {\bibfnamefont {A.}~\bibnamefont
  {Accardi}} \emph {et~al.},\ }\bibfield  {title} {\bibinfo {title} {{An
  experimental program with high duty-cycle polarized and unpolarized positron
  beams at Jefferson Lab}},\ }\href
  {https://doi.org/10.1140/epja/s10050-021-00564-y} {\bibfield  {journal}
  {\bibinfo  {journal} {Eur. Phys. J. A}\ }\textbf {\bibinfo {volume} {57}},\
  \bibinfo {pages} {261} (\bibinfo {year} {2021})},\ \Eprint
  {https://arxiv.org/abs/2007.15081} {arXiv:2007.15081 [nucl-ex]} \BibitemShut
  {NoStop}%
\bibitem [{\citenamefont {Afanasev}\ and\ \citenamefont
  {Carlson}(2005)}]{Afanasev:2005ex}%
  \BibitemOpen
  \bibfield  {author} {\bibinfo {author} {\bibfnamefont {A.~V.}\ \bibnamefont
  {Afanasev}}\ and\ \bibinfo {author} {\bibfnamefont {C.~E.}\ \bibnamefont
  {Carlson}},\ }\bibfield  {title} {\bibinfo {title} {{Two-photon-exchange
  correction to parity-violating elastic electron-proton scattering}},\ }\href
  {https://doi.org/10.1103/PhysRevLett.94.212301} {\bibfield  {journal}
  {\bibinfo  {journal} {Phys. Rev. Lett.}\ }\textbf {\bibinfo {volume} {94}},\
  \bibinfo {pages} {212301} (\bibinfo {year} {2005})},\ \Eprint
  {https://arxiv.org/abs/hep-ph/0502128} {arXiv:hep-ph/0502128} \BibitemShut
  {NoStop}%
\bibitem [{\citenamefont {Barut}\ and\ \citenamefont
  {Fronsdal}(1960)}]{Barut:1960zz}%
  \BibitemOpen
  \bibfield  {author} {\bibinfo {author} {\bibfnamefont {A.~O.}\ \bibnamefont
  {Barut}}\ and\ \bibinfo {author} {\bibfnamefont {C.}~\bibnamefont
  {Fronsdal}},\ }\bibfield  {title} {\bibinfo {title} {{Spin-Orbit Correlations
  in $\mu - e$ and $e^- - e^-$ Scattering}},\ }\href
  {https://doi.org/10.1103/PhysRev.120.1871} {\bibfield  {journal} {\bibinfo
  {journal} {Phys. Rev.}\ }\textbf {\bibinfo {volume} {120}},\ \bibinfo {pages}
  {1871} (\bibinfo {year} {1960})}\BibitemShut {NoStop}%
\bibitem [{\citenamefont {Arafune}\ and\ \citenamefont
  {Shimizu}(1970)}]{Arafune:1970cx}%
  \BibitemOpen
  \bibfield  {author} {\bibinfo {author} {\bibfnamefont {J.}~\bibnamefont
  {Arafune}}\ and\ \bibinfo {author} {\bibfnamefont {Y.}~\bibnamefont
  {Shimizu}},\ }\bibfield  {title} {\bibinfo {title} {{Proton polarization in
  elastic electron-proton scattering}},\ }\href
  {https://doi.org/10.1103/PhysRevD.1.3094} {\bibfield  {journal} {\bibinfo
  {journal} {Phys. Rev. D}\ }\textbf {\bibinfo {volume} {1}},\ \bibinfo {pages}
  {3094} (\bibinfo {year} {1970})}\BibitemShut {NoStop}%
\bibitem [{\citenamefont {Guenther}\ and\ \citenamefont
  {Rodenberg}(1971)}]{Guenther:1971gv}%
  \BibitemOpen
  \bibfield  {author} {\bibinfo {author} {\bibfnamefont {U.}~\bibnamefont
  {Guenther}}\ and\ \bibinfo {author} {\bibfnamefont {R.}~\bibnamefont
  {Rodenberg}},\ }\bibfield  {title} {\bibinfo {title} {{Polarization of the
  recoil protons in the elastic electron-proton scattering process}},\ }\href
  {https://doi.org/10.1007/BF02723986} {\bibfield  {journal} {\bibinfo
  {journal} {Nuovo Cim. A}\ }\textbf {\bibinfo {volume} {2}},\ \bibinfo {pages}
  {25} (\bibinfo {year} {1971})}\BibitemShut {NoStop}%
\bibitem [{\citenamefont {Leroy}\ and\ \citenamefont
  {Piketty}(1972)}]{Leroy:1972mx}%
  \BibitemOpen
  \bibfield  {author} {\bibinfo {author} {\bibfnamefont {J.~P.}\ \bibnamefont
  {Leroy}}\ and\ \bibinfo {author} {\bibfnamefont {C.~A.}\ \bibnamefont
  {Piketty}},\ }\bibfield  {title} {\bibinfo {title} {{Asymmetry in inelastic
  electron-polarized proton scattering}},\ }\href
  {https://doi.org/10.1016/0550-3213(72)90320-3} {\bibfield  {journal}
  {\bibinfo  {journal} {Nucl. Phys. B}\ }\textbf {\bibinfo {volume} {38}},\
  \bibinfo {pages} {466} (\bibinfo {year} {1972})}\BibitemShut {NoStop}%
\bibitem [{\citenamefont {De~Rujula}\ \emph {et~al.}(1971)\citenamefont
  {De~Rujula}, \citenamefont {Kaplan},\ and\ \citenamefont
  {De~Rafael}}]{DeRujula:1971nnp}%
  \BibitemOpen
  \bibfield  {author} {\bibinfo {author} {\bibfnamefont {A.}~\bibnamefont
  {De~Rujula}}, \bibinfo {author} {\bibfnamefont {J.~M.}\ \bibnamefont
  {Kaplan}},\ and\ \bibinfo {author} {\bibfnamefont {E.}~\bibnamefont
  {De~Rafael}},\ }\bibfield  {title} {\bibinfo {title} {{Elastic scattering of
  electrons from polarized protons and inelastic electron scattering
  experiments}},\ }\href {https://doi.org/10.1016/0550-3213(71)90460-3}
  {\bibfield  {journal} {\bibinfo  {journal} {Nucl. Phys. B}\ }\textbf
  {\bibinfo {volume} {35}},\ \bibinfo {pages} {365} (\bibinfo {year}
  {1971})}\BibitemShut {NoStop}%
\bibitem [{\citenamefont {Pasquini}\ and\ \citenamefont
  {Vanderhaeghen}(2004)}]{Pasquini:2004pv}%
  \BibitemOpen
  \bibfield  {author} {\bibinfo {author} {\bibfnamefont {B.}~\bibnamefont
  {Pasquini}}\ and\ \bibinfo {author} {\bibfnamefont {M.}~\bibnamefont
  {Vanderhaeghen}},\ }\bibfield  {title} {\bibinfo {title} {{Resonance
  estimates for single spin asymmetries in elastic electron-nucleon
  scattering}},\ }\href {https://doi.org/10.1103/PhysRevC.70.045206} {\bibfield
   {journal} {\bibinfo  {journal} {Phys. Rev. C}\ }\textbf {\bibinfo {volume}
  {70}},\ \bibinfo {pages} {045206} (\bibinfo {year} {2004})},\ \Eprint
  {https://arxiv.org/abs/hep-ph/0405303} {arXiv:hep-ph/0405303} \BibitemShut
  {NoStop}%
\bibitem [{\citenamefont {Koshchii}\ and\ \citenamefont
  {Afanasev}(2018)}]{Koshchii:2018bog}%
  \BibitemOpen
  \bibfield  {author} {\bibinfo {author} {\bibfnamefont {O.}~\bibnamefont
  {Koshchii}}\ and\ \bibinfo {author} {\bibfnamefont {A.}~\bibnamefont
  {Afanasev}},\ }\bibfield  {title} {\bibinfo {title} {{Target-normal
  single-spin asymmetry in elastic electron-nucleon scattering}},\ }\href
  {https://doi.org/10.1103/PhysRevD.98.056007} {\bibfield  {journal} {\bibinfo
  {journal} {Phys. Rev. D}\ }\textbf {\bibinfo {volume} {98}},\ \bibinfo
  {pages} {056007} (\bibinfo {year} {2018})},\ \Eprint
  {https://arxiv.org/abs/1803.04004} {arXiv:1803.04004 [hep-ph]} \BibitemShut
  {NoStop}%
\bibitem [{\citenamefont {Powell}\ \emph {et~al.}(1970)\citenamefont {Powell}
  \emph {et~al.}}]{Powell:1970qt}%
  \BibitemOpen
  \bibfield  {author} {\bibinfo {author} {\bibfnamefont {T.}~\bibnamefont
  {Powell}} \emph {et~al.},\ }\bibfield  {title} {\bibinfo {title}
  {{Measurement of the polarization in elastic electron-proton scattering}},\
  }\href {https://doi.org/10.1103/PhysRevLett.24.753} {\bibfield  {journal}
  {\bibinfo  {journal} {Phys. Rev. Lett.}\ }\textbf {\bibinfo {volume} {24}},\
  \bibinfo {pages} {753} (\bibinfo {year} {1970})}\BibitemShut {NoStop}%
\bibitem [{\citenamefont {Metz}\ \emph {et~al.}(2006)\citenamefont {Metz},
  \citenamefont {Schlegel},\ and\ \citenamefont {Goeke}}]{Metz:2006pe}%
  \BibitemOpen
  \bibfield  {author} {\bibinfo {author} {\bibfnamefont {A.}~\bibnamefont
  {Metz}}, \bibinfo {author} {\bibfnamefont {M.}~\bibnamefont {Schlegel}},\
  and\ \bibinfo {author} {\bibfnamefont {K.}~\bibnamefont {Goeke}},\ }\bibfield
   {title} {\bibinfo {title} {{Transverse single spin asymmetries in inclusive
  deep-inelastic scattering}},\ }\href
  {https://doi.org/10.1016/j.physletb.2006.11.009} {\bibfield  {journal}
  {\bibinfo  {journal} {Phys. Lett. B}\ }\textbf {\bibinfo {volume} {643}},\
  \bibinfo {pages} {319} (\bibinfo {year} {2006})},\ \Eprint
  {https://arxiv.org/abs/hep-ph/0610112} {arXiv:hep-ph/0610112} \BibitemShut
  {NoStop}%
\bibitem [{\citenamefont {Afanasev}\ \emph {et~al.}(2008)\citenamefont
  {Afanasev}, \citenamefont {Strikman},\ and\ \citenamefont
  {Weiss}}]{Afanasev:2007ii}%
  \BibitemOpen
  \bibfield  {author} {\bibinfo {author} {\bibfnamefont {A.}~\bibnamefont
  {Afanasev}}, \bibinfo {author} {\bibfnamefont {M.}~\bibnamefont {Strikman}},\
  and\ \bibinfo {author} {\bibfnamefont {C.}~\bibnamefont {Weiss}},\ }\bibfield
   {title} {\bibinfo {title} {{Transverse target spin asymmetry in inclusive
  DIS with two-photon exchange}},\ }\href
  {https://doi.org/10.1103/PhysRevD.77.014028} {\bibfield  {journal} {\bibinfo
  {journal} {Phys. Rev. D}\ }\textbf {\bibinfo {volume} {77}},\ \bibinfo
  {pages} {014028} (\bibinfo {year} {2008})},\ \Eprint
  {https://arxiv.org/abs/0709.0901} {arXiv:0709.0901 [hep-ph]} \BibitemShut
  {NoStop}%
\bibitem [{\citenamefont {Metz}\ \emph {et~al.}(2012)\citenamefont {Metz},
  \citenamefont {Pitonyak}, \citenamefont {Schafer}, \citenamefont {Schlegel},
  \citenamefont {Vogelsang},\ and\ \citenamefont {Zhou}}]{Metz:2012ui}%
  \BibitemOpen
  \bibfield  {author} {\bibinfo {author} {\bibfnamefont {A.}~\bibnamefont
  {Metz}}, \bibinfo {author} {\bibfnamefont {D.}~\bibnamefont {Pitonyak}},
  \bibinfo {author} {\bibfnamefont {A.}~\bibnamefont {Schafer}}, \bibinfo
  {author} {\bibfnamefont {M.}~\bibnamefont {Schlegel}}, \bibinfo {author}
  {\bibfnamefont {W.}~\bibnamefont {Vogelsang}},\ and\ \bibinfo {author}
  {\bibfnamefont {J.}~\bibnamefont {Zhou}},\ }\bibfield  {title} {\bibinfo
  {title} {{Single-spin asymmetries in inclusive deep inelastic scattering and
  multiparton correlations in the nucleon}},\ }\href
  {https://doi.org/10.1103/PhysRevD.86.094039} {\bibfield  {journal} {\bibinfo
  {journal} {Phys. Rev. D}\ }\textbf {\bibinfo {volume} {86}},\ \bibinfo
  {pages} {094039} (\bibinfo {year} {2012})},\ \Eprint
  {https://arxiv.org/abs/1209.3138} {arXiv:1209.3138 [hep-ph]} \BibitemShut
  {NoStop}%
\bibitem [{\citenamefont {Schlegel}(2013)}]{Schlegel:2012ve}%
  \BibitemOpen
  \bibfield  {author} {\bibinfo {author} {\bibfnamefont {M.}~\bibnamefont
  {Schlegel}},\ }\bibfield  {title} {\bibinfo {title} {{Partonic description of
  the transverse target single-spin asymmetry in inclusive deep-inelastic
  scattering}},\ }\href {https://doi.org/10.1103/PhysRevD.87.034006} {\bibfield
   {journal} {\bibinfo  {journal} {Phys. Rev. D}\ }\textbf {\bibinfo {volume}
  {87}},\ \bibinfo {pages} {034006} (\bibinfo {year} {2013})},\ \Eprint
  {https://arxiv.org/abs/1211.3579} {arXiv:1211.3579 [hep-ph]} \BibitemShut
  {NoStop}%
\bibitem [{\citenamefont {Airapetian}\ \emph {et~al.}(2010)\citenamefont
  {Airapetian} \emph {et~al.}}]{HERMES:2009hsi}%
  \BibitemOpen
  \bibfield  {author} {\bibinfo {author} {\bibfnamefont {A.}~\bibnamefont
  {Airapetian}} \emph {et~al.} (\bibinfo {collaboration} {HERMES}),\ }\bibfield
   {title} {\bibinfo {title} {{Search for a Two-Photon Exchange Contribution to
  Inclusive Deep-Inelastic Scattering}},\ }\href
  {https://doi.org/10.1016/j.physletb.2009.11.041} {\bibfield  {journal}
  {\bibinfo  {journal} {Phys. Lett. B}\ }\textbf {\bibinfo {volume} {682}},\
  \bibinfo {pages} {351} (\bibinfo {year} {2010})},\ \Eprint
  {https://arxiv.org/abs/0907.5369} {arXiv:0907.5369 [hep-ex]} \BibitemShut
  {NoStop}%
\bibitem [{\citenamefont {Katich}\ \emph {et~al.}(2014)\citenamefont {Katich}
  \emph {et~al.}}]{Katich:2013atq}%
  \BibitemOpen
  \bibfield  {author} {\bibinfo {author} {\bibfnamefont {J.}~\bibnamefont
  {Katich}} \emph {et~al.},\ }\bibfield  {title} {\bibinfo {title}
  {{Measurement of the Target-Normal Single-Spin Asymmetry in Deep-Inelastic
  Scattering from the Reaction $^{3}\mathrm{He}^{\uparrow}(e,e')X$}},\ }\href
  {https://doi.org/10.1103/PhysRevLett.113.022502} {\bibfield  {journal}
  {\bibinfo  {journal} {Phys. Rev. Lett.}\ }\textbf {\bibinfo {volume} {113}},\
  \bibinfo {pages} {022502} (\bibinfo {year} {2014})},\ \Eprint
  {https://arxiv.org/abs/1311.0197} {arXiv:1311.0197 [nucl-ex]} \BibitemShut
  {NoStop}%
\bibitem [{\citenamefont {Zhang}\ \emph {et~al.}(2015)\citenamefont {Zhang}
  \emph {et~al.}}]{Zhang:2015kna}%
  \BibitemOpen
  \bibfield  {author} {\bibinfo {author} {\bibfnamefont {Y.~W.}\ \bibnamefont
  {Zhang}} \emph {et~al.},\ }\bibfield  {title} {\bibinfo {title} {{Measurement
  of the Target-Normal Single-Spin Asymmetry in Quasielastic Scattering from
  the Reaction $^3$He$^\uparrow(e,e^\prime)$}},\ }\href
  {https://doi.org/10.1103/PhysRevLett.115.172502} {\bibfield  {journal}
  {\bibinfo  {journal} {Phys. Rev. Lett.}\ }\textbf {\bibinfo {volume} {115}},\
  \bibinfo {pages} {172502} (\bibinfo {year} {2015})},\ \Eprint
  {https://arxiv.org/abs/1502.02636} {arXiv:1502.02636 [nucl-ex]} \BibitemShut
  {NoStop}%
\bibitem [{\citenamefont {Long}\ \emph {et~al.}(2019)\citenamefont {Long} \emph
  {et~al.}}]{Long:2019iig}%
  \BibitemOpen
  \bibfield  {author} {\bibinfo {author} {\bibfnamefont {E.}~\bibnamefont
  {Long}} \emph {et~al.},\ }\bibfield  {title} {\bibinfo {title} {{Measurement
  of the single-spin asymmetry $A_y^0$ in quasi-elastic
  $^3$He$^\uparrow$($e,e'n$) scattering at $0.4 < Q^2 < 1.0$ GeV$/c^2$}},\
  }\href {https://doi.org/10.1016/j.physletb.2019.134875} {\bibfield  {journal}
  {\bibinfo  {journal} {Phys. Lett. B}\ }\textbf {\bibinfo {volume} {797}},\
  \bibinfo {pages} {134875} (\bibinfo {year} {2019})},\ \Eprint
  {https://arxiv.org/abs/1906.04075} {arXiv:1906.04075 [nucl-ex]} \BibitemShut
  {NoStop}%
\bibitem [{\citenamefont {Grauvogel}\ \emph {et~al.}(2021)\citenamefont
  {Grauvogel}, \citenamefont {Kutz},\ and\ \citenamefont
  {Schmidt}}]{Grauvogel:2021btg}%
  \BibitemOpen
  \bibfield  {author} {\bibinfo {author} {\bibfnamefont {G.~N.}\ \bibnamefont
  {Grauvogel}}, \bibinfo {author} {\bibfnamefont {T.}~\bibnamefont {Kutz}},\
  and\ \bibinfo {author} {\bibfnamefont {A.}~\bibnamefont {Schmidt}},\
  }\bibfield  {title} {\bibinfo {title} {{Target-normal single spin asymmetries
  measured with positrons}},\ }\href
  {https://doi.org/10.1140/epja/s10050-021-00531-7} {\bibfield  {journal}
  {\bibinfo  {journal} {Eur. Phys. J. A}\ }\textbf {\bibinfo {volume} {57}},\
  \bibinfo {pages} {213} (\bibinfo {year} {2021})},\ \Eprint
  {https://arxiv.org/abs/2103.05205} {arXiv:2103.05205 [nucl-ex]} \BibitemShut
  {NoStop}%
\bibitem [{\citenamefont {Carlson}\ and\ \citenamefont
  {Vanderhaeghen}(2007)}]{Carlson:2007sp}%
  \BibitemOpen
  \bibfield  {author} {\bibinfo {author} {\bibfnamefont {C.~E.}\ \bibnamefont
  {Carlson}}\ and\ \bibinfo {author} {\bibfnamefont {M.}~\bibnamefont
  {Vanderhaeghen}},\ }\bibfield  {title} {\bibinfo {title} {{Two-Photon Physics
  in Hadronic Processes}},\ }\href
  {https://doi.org/10.1146/annurev.nucl.57.090506.123116} {\bibfield  {journal}
  {\bibinfo  {journal} {Ann. Rev. Nucl. Part. Sci.}\ }\textbf {\bibinfo
  {volume} {57}},\ \bibinfo {pages} {171} (\bibinfo {year} {2007})},\ \Eprint
  {https://arxiv.org/abs/hep-ph/0701272} {arXiv:hep-ph/0701272} \BibitemShut
  {NoStop}%
\bibitem [{\citenamefont {'t~Hooft}(1974)}]{tHooft:1973alw}%
  \BibitemOpen
  \bibfield  {author} {\bibinfo {author} {\bibfnamefont {G.}~\bibnamefont
  {'t~Hooft}},\ }\bibfield  {title} {\bibinfo {title} {{A Planar Diagram Theory
  for Strong Interactions}},\ }\href
  {https://doi.org/10.1016/0550-3213(74)90154-0} {\bibfield  {journal}
  {\bibinfo  {journal} {Nucl. Phys. B}\ }\textbf {\bibinfo {volume} {72}},\
  \bibinfo {pages} {461} (\bibinfo {year} {1974})}\BibitemShut {NoStop}%
\bibitem [{\citenamefont {Witten}(1979)}]{Witten:1979kh}%
  \BibitemOpen
  \bibfield  {author} {\bibinfo {author} {\bibfnamefont {E.}~\bibnamefont
  {Witten}},\ }\bibfield  {title} {\bibinfo {title} {{Baryons in the $1/N$
  Expansion}},\ }\href {https://doi.org/10.1016/0550-3213(79)90232-3}
  {\bibfield  {journal} {\bibinfo  {journal} {Nucl. Phys. B}\ }\textbf
  {\bibinfo {volume} {160}},\ \bibinfo {pages} {57} (\bibinfo {year}
  {1979})}\BibitemShut {NoStop}%
\bibitem [{\citenamefont {Gervais}\ and\ \citenamefont
  {Sakita}(1984{\natexlab{a}})}]{Gervais:1983wq}%
  \BibitemOpen
  \bibfield  {author} {\bibinfo {author} {\bibfnamefont {J.-L.}\ \bibnamefont
  {Gervais}}\ and\ \bibinfo {author} {\bibfnamefont {B.}~\bibnamefont
  {Sakita}},\ }\bibfield  {title} {\bibinfo {title} {{Large-$N$ QCD Baryon
  Dynamics: Exact Results from Its Relation to the Static Strong Coupling
  Theory}},\ }\href {https://doi.org/10.1103/PhysRevLett.52.87} {\bibfield
  {journal} {\bibinfo  {journal} {Phys. Rev. Lett.}\ }\textbf {\bibinfo
  {volume} {52}},\ \bibinfo {pages} {87} (\bibinfo {year}
  {1984}{\natexlab{a}})}\BibitemShut {NoStop}%
\bibitem [{\citenamefont {Gervais}\ and\ \citenamefont
  {Sakita}(1984{\natexlab{b}})}]{Gervais:1984rc}%
  \BibitemOpen
  \bibfield  {author} {\bibinfo {author} {\bibfnamefont {J.-L.}\ \bibnamefont
  {Gervais}}\ and\ \bibinfo {author} {\bibfnamefont {B.}~\bibnamefont
  {Sakita}},\ }\bibfield  {title} {\bibinfo {title} {{Large-$N$ Baryonic
  Soliton and Quarks}},\ }\href {https://doi.org/10.1103/PhysRevD.30.1795}
  {\bibfield  {journal} {\bibinfo  {journal} {Phys. Rev. D}\ }\textbf {\bibinfo
  {volume} {30}},\ \bibinfo {pages} {1795} (\bibinfo {year}
  {1984}{\natexlab{b}})}\BibitemShut {NoStop}%
\bibitem [{\citenamefont {Dashen}\ and\ \citenamefont
  {Manohar}(1993)}]{Dashen:1993as}%
  \BibitemOpen
  \bibfield  {author} {\bibinfo {author} {\bibfnamefont {R.~F.}\ \bibnamefont
  {Dashen}}\ and\ \bibinfo {author} {\bibfnamefont {A.~V.}\ \bibnamefont
  {Manohar}},\ }\bibfield  {title} {\bibinfo {title} {{Baryon-pion couplings
  from large-$N_c$ QCD}},\ }\href
  {https://doi.org/10.1016/0370-2693(93)91635-Z} {\bibfield  {journal}
  {\bibinfo  {journal} {Phys. Lett. B}\ }\textbf {\bibinfo {volume} {315}},\
  \bibinfo {pages} {425} (\bibinfo {year} {1993})},\ \Eprint
  {https://arxiv.org/abs/hep-ph/9307241} {arXiv:hep-ph/9307241} \BibitemShut
  {NoStop}%
\bibitem [{\citenamefont {Dashen}\ \emph {et~al.}(1994)\citenamefont {Dashen},
  \citenamefont {Jenkins},\ and\ \citenamefont {Manohar}}]{Dashen:1993jt}%
  \BibitemOpen
  \bibfield  {author} {\bibinfo {author} {\bibfnamefont {R.~F.}\ \bibnamefont
  {Dashen}}, \bibinfo {author} {\bibfnamefont {E.~E.}\ \bibnamefont
  {Jenkins}},\ and\ \bibinfo {author} {\bibfnamefont {A.~V.}\ \bibnamefont
  {Manohar}},\ }\bibfield  {title} {\bibinfo {title} {{$1/N_c$ expansion for
  baryons}},\ }\href {https://doi.org/10.1103/PhysRevD.51.2489} {\bibfield
  {journal} {\bibinfo  {journal} {Phys. Rev. D}\ }\textbf {\bibinfo {volume}
  {49}},\ \bibinfo {pages} {4713} (\bibinfo {year} {1994})},\ \bibinfo {note}
  {[Erratum: Phys.Rev.D 51, 2489 (1995)]},\ \Eprint
  {https://arxiv.org/abs/hep-ph/9310379} {arXiv:hep-ph/9310379} \BibitemShut
  {NoStop}%
\bibitem [{\citenamefont {Dashen}\ \emph {et~al.}(1995)\citenamefont {Dashen},
  \citenamefont {Jenkins},\ and\ \citenamefont {Manohar}}]{Dashen:1994qi}%
  \BibitemOpen
  \bibfield  {author} {\bibinfo {author} {\bibfnamefont {R.~F.}\ \bibnamefont
  {Dashen}}, \bibinfo {author} {\bibfnamefont {E.~E.}\ \bibnamefont
  {Jenkins}},\ and\ \bibinfo {author} {\bibfnamefont {A.~V.}\ \bibnamefont
  {Manohar}},\ }\bibfield  {title} {\bibinfo {title} {{Spin flavor structure of
  large $N_c$ baryons}},\ }\href {https://doi.org/10.1103/PhysRevD.51.3697}
  {\bibfield  {journal} {\bibinfo  {journal} {Phys. Rev. D}\ }\textbf {\bibinfo
  {volume} {51}},\ \bibinfo {pages} {3697} (\bibinfo {year} {1995})},\ \Eprint
  {https://arxiv.org/abs/hep-ph/9411234} {arXiv:hep-ph/9411234} \BibitemShut
  {NoStop}%
\bibitem [{\citenamefont {Jenkins}(1998)}]{Jenkins:1998wy}%
  \BibitemOpen
  \bibfield  {author} {\bibinfo {author} {\bibfnamefont {E.~E.}\ \bibnamefont
  {Jenkins}},\ }\bibfield  {title} {\bibinfo {title} {{Large-$N_c$ baryons}},\
  }\href {https://doi.org/10.1146/annurev.nucl.48.1.81} {\bibfield  {journal}
  {\bibinfo  {journal} {Ann. Rev. Nucl. Part. Sci.}\ }\textbf {\bibinfo
  {volume} {48}},\ \bibinfo {pages} {81} (\bibinfo {year} {1998})},\ \Eprint
  {https://arxiv.org/abs/hep-ph/9803349} {arXiv:hep-ph/9803349} \BibitemShut
  {NoStop}%
\bibitem [{\citenamefont {Fernando}\ and\ \citenamefont
  {Goity}(2020)}]{Fernando:2019upo}%
  \BibitemOpen
  \bibfield  {author} {\bibinfo {author} {\bibfnamefont {I.~P.}\ \bibnamefont
  {Fernando}}\ and\ \bibinfo {author} {\bibfnamefont {J.~L.}\ \bibnamefont
  {Goity}},\ }\bibfield  {title} {\bibinfo {title} {{$SU(3)$ vector currents in
  baryon chiral perturbation theory combined with the $1/N_c$ expansion}},\
  }\href {https://doi.org/10.1103/PhysRevD.101.054026} {\bibfield  {journal}
  {\bibinfo  {journal} {Phys. Rev. D}\ }\textbf {\bibinfo {volume} {101}},\
  \bibinfo {pages} {054026} (\bibinfo {year} {2020})},\ \Eprint
  {https://arxiv.org/abs/1911.00987} {arXiv:1911.00987 [hep-ph]} \BibitemShut
  {NoStop}%
\bibitem [{\citenamefont {Jenkins}\ \emph {et~al.}(2002)\citenamefont
  {Jenkins}, \citenamefont {Ji},\ and\ \citenamefont
  {Manohar}}]{Jenkins:2002rj}%
  \BibitemOpen
  \bibfield  {author} {\bibinfo {author} {\bibfnamefont {E.~E.}\ \bibnamefont
  {Jenkins}}, \bibinfo {author} {\bibfnamefont {X.}~\bibnamefont {Ji}},\ and\
  \bibinfo {author} {\bibfnamefont {A.~V.}\ \bibnamefont {Manohar}},\
  }\bibfield  {title} {\bibinfo {title} {{$\Delta \rightarrow N \gamma$ in
  large-$N_c$ QCD}},\ }\href {https://doi.org/10.1103/PhysRevLett.89.242001}
  {\bibfield  {journal} {\bibinfo  {journal} {Phys. Rev. Lett.}\ }\textbf
  {\bibinfo {volume} {89}},\ \bibinfo {pages} {242001} (\bibinfo {year}
  {2002})},\ \Eprint {https://arxiv.org/abs/hep-ph/0207092}
  {arXiv:hep-ph/0207092} \BibitemShut {NoStop}%
\bibitem [{\citenamefont {Dai}\ \emph {et~al.}(1996)\citenamefont {Dai},
  \citenamefont {Dashen}, \citenamefont {Jenkins},\ and\ \citenamefont
  {Manohar}}]{Dai:1995zg}%
  \BibitemOpen
  \bibfield  {author} {\bibinfo {author} {\bibfnamefont {J.}~\bibnamefont
  {Dai}}, \bibinfo {author} {\bibfnamefont {R.~F.}\ \bibnamefont {Dashen}},
  \bibinfo {author} {\bibfnamefont {E.~E.}\ \bibnamefont {Jenkins}},\ and\
  \bibinfo {author} {\bibfnamefont {A.~V.}\ \bibnamefont {Manohar}},\
  }\bibfield  {title} {\bibinfo {title} {{Flavor symmetry breaking in the
  $1/N_c$ expansion}},\ }\href {https://doi.org/10.1103/PhysRevD.53.273}
  {\bibfield  {journal} {\bibinfo  {journal} {Phys. Rev. D}\ }\textbf {\bibinfo
  {volume} {53}},\ \bibinfo {pages} {273} (\bibinfo {year} {1996})},\ \Eprint
  {https://arxiv.org/abs/hep-ph/9506273} {arXiv:hep-ph/9506273} \BibitemShut
  {NoStop}%
\bibitem [{\citenamefont {Goity}(2005)}]{Goity:2004pw}%
  \BibitemOpen
  \bibfield  {author} {\bibinfo {author} {\bibfnamefont {J.~L.}\ \bibnamefont
  {Goity}},\ }\bibfield  {title} {\bibinfo {title} {{$1/N_c$ countings in
  baryons}},\ }\href {https://doi.org/10.1134/1.1903092} {\bibfield  {journal}
  {\bibinfo  {journal} {Phys. Atom. Nucl.}\ }\textbf {\bibinfo {volume} {68}},\
  \bibinfo {pages} {624} (\bibinfo {year} {2005})},\ \Eprint
  {https://arxiv.org/abs/hep-ph/0405304} {arXiv:hep-ph/0405304} \BibitemShut
  {NoStop}%
\bibitem [{\citenamefont {Calle~Cordon}\ and\ \citenamefont
  {Goity}(2013)}]{CalleCordon:2012xz}%
  \BibitemOpen
  \bibfield  {author} {\bibinfo {author} {\bibfnamefont {A.}~\bibnamefont
  {Calle~Cordon}}\ and\ \bibinfo {author} {\bibfnamefont {J.~L.}\ \bibnamefont
  {Goity}},\ }\bibfield  {title} {\bibinfo {title} {{Baryon Masses and Axial
  Couplings in the Combined 1/$N_c$ and Chiral Expansions}},\ }\href
  {https://doi.org/10.1103/PhysRevD.87.016019} {\bibfield  {journal} {\bibinfo
  {journal} {Phys. Rev. D}\ }\textbf {\bibinfo {volume} {87}},\ \bibinfo
  {pages} {016019} (\bibinfo {year} {2013})},\ \Eprint
  {https://arxiv.org/abs/1210.2364} {arXiv:1210.2364 [nucl-th]} \BibitemShut
  {NoStop}%
\bibitem [{\citenamefont {Scoccola}\ \emph {et~al.}(2008)\citenamefont
  {Scoccola}, \citenamefont {Goity},\ and\ \citenamefont
  {Matagne}}]{Scoccola:2007sn}%
  \BibitemOpen
  \bibfield  {author} {\bibinfo {author} {\bibfnamefont {N.~N.}\ \bibnamefont
  {Scoccola}}, \bibinfo {author} {\bibfnamefont {J.~L.}\ \bibnamefont
  {Goity}},\ and\ \bibinfo {author} {\bibfnamefont {N.}~\bibnamefont
  {Matagne}},\ }\bibfield  {title} {\bibinfo {title} {{Analysis of Negative
  Parity Baryon Photoproduction Amplitudes in the $1/N_c$ Expansion}},\ }\href
  {https://doi.org/10.1016/j.physletb.2008.03.056} {\bibfield  {journal}
  {\bibinfo  {journal} {Phys. Lett. B}\ }\textbf {\bibinfo {volume} {663}},\
  \bibinfo {pages} {222} (\bibinfo {year} {2008})},\ \Eprint
  {https://arxiv.org/abs/0711.4203} {arXiv:0711.4203 [hep-ph]} \BibitemShut
  {NoStop}%
\bibitem [{\citenamefont {Goity}\ and\ \citenamefont
  {Scoccola}(2007)}]{Goity:2007ft}%
  \BibitemOpen
  \bibfield  {author} {\bibinfo {author} {\bibfnamefont {J.~L.}\ \bibnamefont
  {Goity}}\ and\ \bibinfo {author} {\bibfnamefont {N.~N.}\ \bibnamefont
  {Scoccola}},\ }\bibfield  {title} {\bibinfo {title} {{Photo-production of
  Positive Parity Excited Baryons in the $1/N_c$ Expansion of QCD}},\ }\href
  {https://doi.org/10.1103/PhysRevLett.99.062002} {\bibfield  {journal}
  {\bibinfo  {journal} {Phys. Rev. Lett.}\ }\textbf {\bibinfo {volume} {99}},\
  \bibinfo {pages} {062002} (\bibinfo {year} {2007})},\ \Eprint
  {https://arxiv.org/abs/hep-ph/0701244} {arXiv:hep-ph/0701244} \BibitemShut
  {NoStop}%
\bibitem [{\citenamefont {Goity}\ \emph {et~al.}(2022)\citenamefont {Goity},
  \citenamefont {Weiss},\ and\ \citenamefont {Willemyns}}]{Goity:2022yro}%
  \BibitemOpen
  \bibfield  {author} {\bibinfo {author} {\bibfnamefont {J.~L.}\ \bibnamefont
  {Goity}}, \bibinfo {author} {\bibfnamefont {C.}~\bibnamefont {Weiss}},\ and\
  \bibinfo {author} {\bibfnamefont {C.~T.}\ \bibnamefont {Willemyns}},\
  }\bibfield  {title} {\bibinfo {title} {{Target normal single-spin asymmetry
  in inclusive electron-nucleon scattering with two-photon exchange: Analysis
  using $1/N_c$ expansion}},\ }\href
  {https://doi.org/10.1016/j.physletb.2022.137580} {\bibfield  {journal}
  {\bibinfo  {journal} {Phys. Lett. B}\ }\textbf {\bibinfo {volume} {835}},\
  \bibinfo {pages} {137580} (\bibinfo {year} {2022})},\ \Eprint
  {https://arxiv.org/abs/2207.07588} {arXiv:2207.07588 [hep-ph]} \BibitemShut
  {NoStop}%
\bibitem [{\citenamefont {Christ}\ and\ \citenamefont
  {Lee}(1966)}]{Christ:1966zz}%
  \BibitemOpen
  \bibfield  {author} {\bibinfo {author} {\bibfnamefont {N.}~\bibnamefont
  {Christ}}\ and\ \bibinfo {author} {\bibfnamefont {T.~D.}\ \bibnamefont
  {Lee}},\ }\bibfield  {title} {\bibinfo {title} {{Possible Tests of $C_{st}$
  and $T_{st}$ Invariances in $l^{\pm} + N \rightarrow l^{\pm} + \Gamma$ and $A
  \rightarrow B + e^+ + e^-$}},\ }\href
  {https://doi.org/10.1103/PhysRev.143.1310} {\bibfield  {journal} {\bibinfo
  {journal} {Phys. Rev.}\ }\textbf {\bibinfo {volume} {143}},\ \bibinfo {pages}
  {1310} (\bibinfo {year} {1966})}\BibitemShut {NoStop}%
\bibitem [{\citenamefont {Low}(1958)}]{Low:1958sn}%
  \BibitemOpen
  \bibfield  {author} {\bibinfo {author} {\bibfnamefont {F.~E.}\ \bibnamefont
  {Low}},\ }\bibfield  {title} {\bibinfo {title} {{Bremsstrahlung of very
  low-energy quanta in elementary particle collisions}},\ }\href
  {https://doi.org/10.1103/PhysRev.110.974} {\bibfield  {journal} {\bibinfo
  {journal} {Phys. Rev.}\ }\textbf {\bibinfo {volume} {110}},\ \bibinfo {pages}
  {974} (\bibinfo {year} {1958})}\BibitemShut {NoStop}%
\bibitem [{\citenamefont {Gorchtein}\ \emph {et~al.}(2004)\citenamefont
  {Gorchtein}, \citenamefont {Guichon},\ and\ \citenamefont
  {Vanderhaeghen}}]{Gorchtein:2004ac}%
  \BibitemOpen
  \bibfield  {author} {\bibinfo {author} {\bibfnamefont {M.}~\bibnamefont
  {Gorchtein}}, \bibinfo {author} {\bibfnamefont {P.~A.~M.}\ \bibnamefont
  {Guichon}},\ and\ \bibinfo {author} {\bibfnamefont {M.}~\bibnamefont
  {Vanderhaeghen}},\ }\bibfield  {title} {\bibinfo {title} {{Beam normal spin
  asymmetry in elastic lepton-nucleon scattering}},\ }\href
  {https://doi.org/10.1016/j.nuclphysa.2004.06.008} {\bibfield  {journal}
  {\bibinfo  {journal} {Nucl. Phys. A}\ }\textbf {\bibinfo {volume} {741}},\
  \bibinfo {pages} {234} (\bibinfo {year} {2004})},\ \Eprint
  {https://arxiv.org/abs/hep-ph/0404206} {arXiv:hep-ph/0404206} \BibitemShut
  {NoStop}%
\bibitem [{\citenamefont {Afanasev}\ and\ \citenamefont
  {Merenkov}(2004)}]{Afanasev:2004pu}%
  \BibitemOpen
  \bibfield  {author} {\bibinfo {author} {\bibfnamefont {A.~V.}\ \bibnamefont
  {Afanasev}}\ and\ \bibinfo {author} {\bibfnamefont {N.~P.}\ \bibnamefont
  {Merenkov}},\ }\bibfield  {title} {\bibinfo {title} {{Collinear photon
  exchange in the beam normal polarization asymmetry of elastic electron-proton
  scattering}},\ }\href {https://doi.org/10.1016/j.physletb.2004.08.023}
  {\bibfield  {journal} {\bibinfo  {journal} {Phys. Lett. B}\ }\textbf
  {\bibinfo {volume} {599}},\ \bibinfo {pages} {48} (\bibinfo {year} {2004})},\
  \Eprint {https://arxiv.org/abs/hep-ph/0407167} {arXiv:hep-ph/0407167}
  \BibitemShut {NoStop}%
\bibitem [{\citenamefont {Gorchtein}(2006)}]{Gorchtein:2005yz}%
  \BibitemOpen
  \bibfield  {author} {\bibinfo {author} {\bibfnamefont {M.}~\bibnamefont
  {Gorchtein}},\ }\bibfield  {title} {\bibinfo {title} {{Beam normal spin
  asymmetry in the quasi-RCS approximation}},\ }\href
  {https://doi.org/10.1103/PhysRevC.73.055201} {\bibfield  {journal} {\bibinfo
  {journal} {Phys. Rev. C}\ }\textbf {\bibinfo {volume} {73}},\ \bibinfo
  {pages} {055201} (\bibinfo {year} {2006})},\ \Eprint
  {https://arxiv.org/abs/hep-ph/0512105} {arXiv:hep-ph/0512105} \BibitemShut
  {NoStop}%
\bibitem [{\citenamefont {Carlson}\ \emph {et~al.}(2017)\citenamefont
  {Carlson}, \citenamefont {Pasquini}, \citenamefont {Pauk},\ and\
  \citenamefont {Vanderhaeghen}}]{Carlson:2017lys}%
  \BibitemOpen
  \bibfield  {author} {\bibinfo {author} {\bibfnamefont {C.~E.}\ \bibnamefont
  {Carlson}}, \bibinfo {author} {\bibfnamefont {B.}~\bibnamefont {Pasquini}},
  \bibinfo {author} {\bibfnamefont {V.}~\bibnamefont {Pauk}},\ and\ \bibinfo
  {author} {\bibfnamefont {M.}~\bibnamefont {Vanderhaeghen}},\ }\bibfield
  {title} {\bibinfo {title} {{Beam normal spin asymmetry for the $e p \to e
  \Delta(1232)$ process}},\ }\href {https://doi.org/10.1103/PhysRevD.96.113010}
  {\bibfield  {journal} {\bibinfo  {journal} {Phys. Rev. D}\ }\textbf {\bibinfo
  {volume} {96}},\ \bibinfo {pages} {113010} (\bibinfo {year} {2017})},\
  \Eprint {https://arxiv.org/abs/1708.05316} {arXiv:1708.05316 [hep-ph]}
  \BibitemShut {NoStop}%
\bibitem [{\citenamefont {Koshchii}\ and\ \citenamefont
  {Afanasev}(2019)}]{Koshchii:2019mgv}%
  \BibitemOpen
  \bibfield  {author} {\bibinfo {author} {\bibfnamefont {O.}~\bibnamefont
  {Koshchii}}\ and\ \bibinfo {author} {\bibfnamefont {A.}~\bibnamefont
  {Afanasev}},\ }\bibfield  {title} {\bibinfo {title} {{Lepton mass effects for
  beam-normal single-spin asymmetry in elastic muon-proton scattering}},\
  }\href {https://doi.org/10.1103/PhysRevD.100.096020} {\bibfield  {journal}
  {\bibinfo  {journal} {Phys. Rev. D}\ }\textbf {\bibinfo {volume} {100}},\
  \bibinfo {pages} {096020} (\bibinfo {year} {2019})},\ \Eprint
  {https://arxiv.org/abs/1905.10217} {arXiv:1905.10217 [hep-ph]} \BibitemShut
  {NoStop}%
\end{thebibliography}%
\end{document}